\documentclass[preprintnumbers,10pt,a4paper,superscriptaddress,aps,pre,twocolumn]{revtex4-2}

\usepackage{physics}
\usepackage{graphicx}
\usepackage{xcolor}
\usepackage[hidelinks]{hyperref} % this and the next two have to be in this order
\usepackage[capitalise]{cleveref} % automatic referencing

\crefname{plural_equation}{Eqs.}{Eqs.}
\Crefname{plural_equation}{Equations}{Equations}
\creflabelformat{plural_equation}{(#2\textup{#1}#3)}

\DeclareMathOperator{\sign}{sign}

\DeclareMathOperator{\cov}{Cov}
\DeclareMathOperator{\variance}{Var}

\newcommand{\conj}[1]{\overline{#1}}
\newcommand{\outlier}{\lambda_\text{outlier}}
\newcommand{\outliers}[1]{\lambda_\text{outlier}^{(#1)}}
\newcommand{\edge}{\lambda_\text{edge}}
\newcommand{\PF}{\lambda_\text{PF}}
\newcommand{\eigval}[1]{\lambda_{#1}}
\newcommand{\bulk}{\lambda_\text{bulk}}

\newcommand{\avg}[1]{\left\langle #1 \right\rangle}
\newcommand{\wh}[1]{\widehat{#1}}

\newcommand{\id}{{\openone}}

\newcommand{\nocontentsline}[3]{}
\let\oldaddcontentsline=\addcontentsline
\let\addcontentsline=\nocontentsline

\begin{document}

\title{Eigenvalue spectra of finely structured random matrices}
    
\author{Lyle Poley}
\affiliation{Theoretical Physics, Department of Physics and Astronomy, School of Natural Sciences, The University of Manchester, Manchester M13 9PL, UK}
\author{Tobias Galla}
\affiliation{Instituto de F{\' i}sica Interdisciplinar y Sistemas Complejos IFISC (CSIC-UIB), 07122 Palma de Mallorca, Spain}
% \affiliation{Theoretical Physics, Department of Physics and Astronomy, School of Natural Sciences, The University of Manchester, Manchester M13 9PL, UK}
\author{Joseph W. Baron}
\affiliation{Laboratoire de Physique de l’Ecole Normale Sup\'{e}rieure, ENS, Universit\'{e} PSL, CNRS, Sorbonne Universit\'{e}, Universit\'{e} de Paris, F-75005 Paris, France}

\date{\today}
\begin{abstract}
Random matrix theory allows for the deduction of stability criteria for complex systems using only a summary knowledge of the statistics of the interactions between components. As such, results like the well-known elliptical law are applicable in a myriad of different contexts. However, it is often assumed that all components of the complex system in question are statistically equivalent, which is unrealistic in many applications. Here, we introduce the concept of a finely-structured random matrix. These are random matrices with element-specific statistics, which can be used to model systems in which the individual components are statistically distinct. By supposing that the degree of `fine structure' in the matrix is small, we arrive at a succinct `modified' elliptical law. We demonstrate the direct applicability of our results to the niche and cascade models in theoretical ecology, as well as a model of a neural network, and a directed network with arbitrary degree distribution. The simple closed form of our central results allow us to draw broad qualitative conclusions about the effect of fine structure on stability.
\end{abstract}

\maketitle

\section{Introduction}
% introduce random matrices
Random matrix theory (RMT) is a branch of mathematics and physics concerned with the properties and applications of large matrices whose entries have given statistics. Early applications that were important for the development of RMT include invariant theory \cite{hurwitz1897erzeugung,forresterLogGasesRandomMatrices2010,diaconisHurwitzOriginsRandom2016}, numerical analysis \cite{vonneumannNumericalInvertingMatrices1947,edelmanEigenvaluesConditionNumbers1988,highamAccuracyStabilityNumerical2002}, the study of covariance matrices \cite{wishartGENERALISEDPRODUCTMOMENT1928a,marcenkoDISTRIBUTIONEIGENVALUESSETS1967,fisherSamplingDistributionStatistics1939,hsuDistributionRootsCertain1939,royPStatisticsGeneralisationsAnalysis1939,girshickSamplingTheoryRoots1939}, and nuclear physics \cite{wignerCharacteristicVectorsBordered1955,dysonStatisticalTheoryEnergy1962,mehtaRandomMatrices2004,dysonBrownianMotionModelEigenvalues1962,dysonDynamicsDisorderedLinear1953}. Since its inception, RMT has become an area of study in its own right \cite{taoRandomMatricesCircular2008a,taoTopicsRandomMatrix2012,edelmanRandomMatrixTheory2005,akemannOxfordHandbookRandom2015} and has proven to be an invaluable tool in the study of disordered systems, with applications in spin glass physics \cite{mezard1987,auffingerRandomMatricesComplexity2013,crisantiSphericalpspinInteractionSpin1992}, neural networks \cite{kuczalaEigenvalueSpectraLarge2016,sternChaosHeterogeneousNeural2014,sompolinskyChaosRandomNeural1988,rajanEigenvalueSpectraRandom2006,sommersSpectrumLargeRandom1988,hopfieldNeuralNetworksPhysical1982,coolenTheoryNeuralInformation2005}, and theoretical ecology \cite{allesinaStabilityCriteriaComplex2012,mayWillLargeComplex1972,MAYBOOK,stoneFeasibilityStabilityLarge2018,allesinaStabilityComplexityRelationship2015,baronDispersalinducedInstabilityComplex2020,landiComplexityStabilityEcological2018} to name a few.

%Why is RMT so useful? self averaging and universal
The broad applicability of RMT is in part due to two key properties shared by many random matrix ensembles: the self averaging property and universality. Many ensembles are `self averaging', meaning that the eigenvalue spectrum of a single realization approaches a deterministic (non-random) limit with increasing matrix size. The term `universality' \cite{woodUniversalityCircularLaw2012,taoRandomMatricesUniversality2015} indicates that this limit does not depend on the intricacies of the specific distribution from which the matrix elements are drawn, and only depends on a small number of its characteristics such as the first and second moments. Hence, in many cases we can do away with the specifics of a given system and focus instead on its statistical properties.

One central result of RMT is the celebrated elliptical law \cite{sommersSpectrumLargeRandom1988,girkoEllipticLaw1986}, which applies in a universal manner to a range of self-averaging random matrix ensembles, can be summarized as follows. If the elements of a large matrix are drawn in diagonally opposite pairs, independently and from the same distribution, then the eigenvalues will be confined to an ellipse in the complex plane, with the possibility of a single outlier. The shape of this ellipse and the location of the outlier eigenvalue are dependent only on the mean and variance of the elements, and on the correlation coefficient between diagonally opposite elements. If this large matrix represents the Jacobian matrix of some dynamical system, then the elliptical law tells us that we do not need to know the precise values of all the matrix elements in order to determine the stability of the system. Instead, we need only know the \textit{statistical} properties of the interactions between the various components of the system. This observation allows for very general qualitative deductions about which characteristics of a complex system permit stability \cite{allesinaStabilityCriteriaComplex2012}. 

%The limitations of the elliptic law -- motivating our study
However, for the elliptical law to be applicable, it is necessary that all transpose pairs of elements be statistically equivalent. In recent applications, particularly in ecology and neural networks, this assumption breaks down. There is often a hierarchy, or ordering, to the elements which cannot be ignored \cite{allesinaPredictingStabilityLarge2015,barbierGenericAssembltPatterns2018,schubertCellTypeSpecificCircuits2003,yoshimuraExcitatoryCorticalNeurons2005,levySpatialProfileExcitatory2012,gravel2016stability,barabas2017self,aljadeffTransitionChaosRandom2015,kuczalaEigenvalueSpectraLarge2016,aljadeffChaosHeterogeneousNeural2014,barbierFingerprintsHighDimensionalCoexistence2021}. As a result, many random matrix models have been developed to describe the various structures specific to the system in question, which result in modifications to the usual elliptical law. In this work, we aim to tie together a large subset of these models.

%Our result as a unification of previous results with all the advantages of the beloved elliptic law
To this end, we introduce the concept of a `finely-structured' random matrix (FSRM), which has elements with statistics that vary depending on the position in the matrix. By considering this fine structure as a perturbation to an elliptical random matrix (i.e. a random matrix for which the elliptic law applies \cite{sommersSpectrumLargeRandom1988,girkoEllipticLaw1986,orourkeLowRankPerturbations2014}), we derive a correction to the elliptical law that depends only on a few summary statistics of the model. A similar correction for the associated outlier eigenvalue can also be determined.

The results we derive are simple and compact, so they are easily applied to many previously studied random matrix models. Indeed, we show that a wide variety of models can be understood as `finely structured' upon careful analysis, and we demonstrate that our formulae accurately predict the eigenvalue spectra (and consequently the stability) of these systems. Ultimately, we use the modified elliptic law to draw qualitative conclusions about which aspects of fine structure affect the stability of disordered systems.

The rest of this work is laid out as follows. First, in \cref{section:matrix_ensemble}, we define more precisely what we mean by a finely-structured random matrix. In \cref{section:spectrum_approximation} we outline the derivation of the equations that determine the support of the spectrum of a general FSRM, and we use a perturbative approach to derive the modified elliptical law and associated outlier eigenvalue [\cref{eq:bulk_approx,eq:outlier_approx}], which approximate this support. Then, in \cref{section:directed_network,section:exact_solution,section:neural_networks,section:niche_model} we demonstrate how the modified elliptical law can be applied to a number of random matrix models. Specifically, in \cref{section:exact_solution} we provide an exact solution for the cascade model from theoretical ecology \cite{allesinaPredictingStabilityLarge2015,ahmadianPropertiesNetworksPartially2015a}, and show that this solution reduces to the modified elliptical law when the level of fine structure is small. In \cref{section:directed_network}, we approximate the eigenvalue spectra of elliptical matrices with dense and random network structure. In \cref{section:neural_networks}, we apply our results to a toy neural network model inspired by Ref.~\cite{kuczalaEigenvalueSpectraLarge2016}. Finally, in \cref{section:niche_model} we show how the `niche model' \cite{allesinaStabilityCriteriaComplex2012,williamsSimpleRulesYield2000} can be understood as a finely-structured system and predict ecosystem stability in this model. We conclude by discussing possible extensions of the methods and results put forward in this work. 

\section{Finely-structured random matrices}\label{section:matrix_ensemble}
Our focus is a particular ensemble of finely-structured random matrices (FSRMs), which includes many existing generalizations to the elliptical ensemble of Refs.~\cite{girkoEllipticLaw1986,sommersSpectrumLargeRandom1988} as special cases. Specifically, we consider FSRMs $\vb J$ of size $N \times N$, whose elements have the following statistics
\begin{align}
    \avg{J_{ij}}_J &= \frac{u_{ij}}{N}, \nonumber\\
    \variance_J\qty(J_{ij}) &= \frac{s_{ij}}{N}, \nonumber\\
    \cov_J\qty(J_{ij}, J_{ji}) &= \frac{t_{ij}}{N},\label[plural_equation]{eq:fsrm_stats}
\end{align}
where angular brackets $\avg{\cdot}_J,\variance_J(\cdot)$ and $\cov_J(\cdot)$ denote the average, variance and covariance over realizations of the random matrix $\vb J$ respectively. We note that if the higher moments of $J_{ij}$ drop off sufficiently quickly with $N$, then the results we obtain will be universal in the sense that we only need to know the moments in \cref{eq:fsrm_stats}, and not the full joint probability distribution of the elements of $\vb J$, to find the eigenvalue spectrum [see \cref{sm:section:The_bulk_spectrum} of the Supplemental Material (SM)].

We emphasize that the coefficients $u_{ij}, s_{ij}$ and $t_{ij}$ are not random variables, but are instead fixed parameters of the model. We assume that all of $u_{ij}, s_{ij}$ and $t_{ij}$ are of order $N^0$. The factors of $N$ in \cref{eq:fsrm_stats} then ensure that we obtain a sensible large-$N$ limit for the spectrum of $\vb J$ \cite{mezard1987}. The definitions of $\vb s$ and $\vb t$ also imply that we must have $s_{ij} \geq 0$ and $t_{ij} = t_{ji}$.

Similar random matrix ensembles have been studied in a number of different contexts \cite{kuczalaEigenvalueSpectraLarge2016,aljadeffEigenvaluesBlockStructured2015,aljadeffChaosHeterogeneousNeural2014,cookNonHermitianRandomMatrices2018,cookNonHermitianRandomMatrices2020,cookLowerBoundsSmallest2018,ahmadianPropertiesNetworksPartially2015a,ajankiUniversalityGeneralWignertype2017,ajankiStabilityMatrixDyson2019,altLocationSpectrumKronecker2018,altLocalLawRandom2017}. However, as far as we are aware, none of these previous studies allow for non-zero $\vb u$ and $\vb t$ simultaneously, nor do they derive the central result of this paper, the modified elliptical law and associated outlier eigenvalue that we present in \cref{section:spectrum_approximation}.

% recovering the elliptical law, what remains true
The FSRMs that we study are a generalization of the elliptical ensemble in the following sense. If the statistics of the $(i,j)$-element do not depend on $i$ and $j$ (written $u_{ij} = \mu, s_{ij} = \sigma^2$ and $t_{ij} = \gamma\sigma^2$), then the random matrix $\vb J$ is an elliptical random matrix of the type in \cite{orourkeLowRankPerturbations2014}. In this case, we say that the statistics of the FSRM $\vb J$ have no fine structure. In the more general case where the statistics of $\vb J$ do depend on position in the matrix, we find that the majority of the eigenvalues are still confined to a bulk region, which is generally not an ellipse, and there are ${\cal O}(N^0)$ outlier eigenvalues due to the non-zero mean values $u_{ij}$. An example is shown in \cref{fig:HT_spectrum}~(a). We see a `star' shaped bulk region containing the majority of the eigenvalues, as well as four isolated outlier eigenvalues.
%%
%%
%%
%%
%%
% The main result of the paper, a general formula for the boundary of the eigenvalue spectrum of a finely-structured matrix.
\section{Fine-structure corrections to the Elliptical law and associated outlier}
\label{section:spectrum_approximation}
\subsection{Preliminaries and definitions}
\label{section:spectrum_approximation_prelims}
As is common in the study of disordered systems, we assume that the spectra of the large random matrices that we study are self-averaging \cite{mezard1987}. That is, we will assume that the eigenvalue spectrum, averaged over realizations of $\vb J$, is equivalent to the spectrum of a single realization of the matrix for large $N$. We verify this assumption in panel (a) of \cref{fig:DR_networks,fig:HT_spectrum,fig:neural_networks,fig:NM_spectrum}, in which our analytical predictions, which describe ensemble averaged properties, are compared to single realizations of the FSRM models of interest.

Similarly to the random matrices studied in Refs.~\cite{orourkeLowRankPerturbations2014,baronDispersalinducedInstabilityComplex2020,taoOutliersSpectrumIid2014}, we write the spectrum of $\vb J$ as the sum of the bulk spectrum and the isolated outlier eigenvalues as follows
\begin{align}
    \rho_J(z) &= \avg{\frac{1}{N}\sum_{i=1}^{N}\delta\Big(z - \eigval{i}[\vb J]\Big)}_J,\nonumber \\
    &= \rho_\text{bulk}(z) + \frac{1}{N}\sum_k \delta\qty(z - \outliers{k}),
\end{align}
where $\eigval{i}[\vb J]$ are the eigenvalues of $\vb J$ indexed by $i$, $\rho_\text{bulk}(z)$ is the continuous bulk part of the spectrum at $z = x + iy$, and $\outliers{k}$ are the outlier eigenvalues, indexed by $k$. 

For a general random matrix $\vb J$, the spectrum $\rho_J(z)$ can be derived from the disorder-averaged resolvent matrix (see Refs. \cite{sommersSpectrumLargeRandom1988,benaych-georgesEigenvaluesEigenvectorsFinite2010,orourkeLowRankPerturbations2014}), defined by
\begin{align}
    \vb G(z) \equiv \avg{\qty(z\vb I - \vb J)^{-1}}_J,\label{eq:resolvent_definition}
\end{align}
where $\vb I$ is the $N \times N$ identity matrix. Similarly to Refs. \cite{sommersSpectrumLargeRandom1988,baronDispersalinducedInstabilityComplex2020,baron_directed_2022,brayEvidenceMasslessModes1979,kuczalaEigenvalueSpectraLarge2016}, the resolvent matrix is  diagonal in the limit $N\to \infty$, and we write $G_i(z)$ for the diagonal element $[\vb G(z)]_{ii}$. In \cref{sm:section:The_bulk_spectrum,sm:section:self_consistent_equations} of the SM, we find a set of equations that self-consistently determine the $\{G_i(z)\}$ for the FSRM ensemble defined in \cref{eq:fsrm_stats} using well-known replica methods \cite{sommersSpectrumLargeRandom1988,baronDispersalinducedInstabilityComplex2020,haakeStatisticsComplexLevels1992,anandStabilityDynamicalProperties2009}. We then describe how these expressions for the $\{G_i(z)\}$ can, in principle, be solved and used to obtain the spectrum of $\vb J$. 

% hard to get explicit solutions, so we approximate
The equations relating the $\{G_i(z)\}$, $\rho_J(z)$ and the statistics of $\vb J$ are complicated. Non-trivial choices of the matrix statistics for which $\rho_J(z)$ can be found explicitly are difficult to come by. Further, when an explicit solution exists, it can be rather complex (see \cref{section:exact_solution}, and \cref{sm:section:explicit_solutions_for} of the SM for examples), and this complexity can obfuscate the qualitative effect of fine structure on stability. 

% The main result of the paper, a general formula for the boundary of the eigenvalue spectrum of a finely-structured matrix.
Therefore, we proceed by treating the fine structure as a perturbation, seeking a fine-structure correction to the elliptical law and outlier eigenvalue which can be interpreted qualitatively. We write the statistics in \cref{eq:fsrm_stats} as a sum of index-independent and index-dependent parts
\begin{align}
    u_{ij} &= \mu + u^{(1)}_{ij}, \nonumber\\
    s_{ij} &= \sigma^2 + s_{ij}^{(1)}, \nonumber\\
    t_{ij} &= \gamma \sigma^2 + t^{(1)}_{ij},\label[plural_equation]{eq:approx_stats}
\end{align}
where the `zeroth-order', or elliptical, statistics are (in the limit of large $N$)
\begin{align}
    \mu &= \frac{1}{N^2}\sum_{ij}u_{ij}, \nonumber\\
    \sigma^2 &= \frac{1}{N^2}\sum_{ij}s_{ij}, \nonumber\\
    \gamma &= \frac{1}{\sigma^2 N^2}\sum_{ij}t_{ij},\label[plural_equation]{eq:avg_params}
\end{align}
and where we call the matrices $\vb u^{(1)}, \vb s^{(1)}, \vb t^{(1)}$ the fine-structure parts of $\vb u, \vb s, \vb t$ respectively.  We also define the following `first-order' (or fine-structure) parameters (again to be understood in the large-$N$ limit),
\begin{align}
    R &\equiv \frac{1}{\sigma^4N^3}\sum_{ijk}\frac{1}{2}\qty[s^{(1)}_{ij} + s^{(1)}_{ji}]t^{(1)}_{jk},\nonumber\\
    S &\equiv \frac{1}{\sigma^4N^3}\sum_{ijk}s^{(1)}_{ij}s^{(1)}_{jk}, \nonumber\\
    T &\equiv \frac{1}{\sigma^4N^3}\sum_{ijk}t^{(1)}_{ij}t^{(1)}_{jk}, \nonumber\\
    U &\equiv \frac{1}{\mu^2N^3}\sum_{ijk}u^{(1)}_{ij}u^{(1)}_{jk},\nonumber \\
    V &\equiv \frac{1}{\mu\sigma^2 N^3}\sum_{ijk}\frac12\qty[u^{(1)}_{ij} + u^{(1)}_{ji}]t^{(1)}_{jk}.\label[plural_equation]{eq:fs_params}
\end{align}

As we will show, these fully characterize the fine-structure correction to the elliptical law and outlier eigenvalue when the amount of fine structure is small. We can interpret these new parameters by noticing that $R, S, T, U$ and $V$ are directly related to the row and column sums of the statistics of $\vb J$. For example, if we define the $i$-th row sum of $\vb t$ as $t_i \equiv \sum_j t_{ij}/N$, then we have $T = \sum_i(t_i - \gamma\sigma^2)^2/N\sigma^4$. Hence, up to a pre-factor, $T$ is the variance of the row sums of the matrix $\vb t$.

If the fine-structure contributions are all zero, then the spectrum of $\vb J$ is simply the well-known elliptical law (see \cref{sm:section:elliptical_law} of the SM) \cite{sommersSpectrumLargeRandom1988,orourkeLowRankPerturbations2014}. That is, the majority of the eigenvalues are contained in the ellipse
\begin{align}
    \qty(\frac{x}{1 + \gamma})^2 + \qty(\frac{y}{1 - \gamma})^2 = \sigma^2,\label{eq:ellipse0}
\end{align}
and, provided $|\mu| > \sigma$, there is a single outlier eigenvalue located at (see \cite{orourkeLowRankPerturbations2014,benaych-georgesEigenvaluesEigenvectorsFinite2010,edwardsEigenvalueSpectrumLarge1976}),
\begin{align}
    \outlier = \mu + \frac{\gamma \sigma^2}{\mu}.\label{eq:outlier0}
\end{align}
\subsection{Modified elliptical law}
\label{subsection:modified_elliptical_law}
% how to do the approximation
The fine-structure correction to the support of the spectrum of $\vb J$ can be computed by introducing a small perturbation parameter $\epsilon$ which measures the extent to which the statistics of $\vb J$ are finely structured. We assume that all elements of $\vb u^{(1)},\ \vb s^{(1)}$ and $\vb t^{(1)}$ are proportional to this small quantity $\epsilon$ so that the fine structure parameters in \cref{eq:fs_params} are proportional to $\epsilon^2$. We then find that the leading-order correction to the elliptical law is not of order $\epsilon$, but is in fact of order $\epsilon^2$. Here, we only present the final result. For a full and detailed derivation of all results up to and including \cref{eq:edge_approx}, we refer the reader to \cref{sm:section:approximate_self_consistent_equations,sm:section:Modified_elliptical_law} of the SM. All relationships derived in this section are understood to be accurate up to second order in $\epsilon$.

By supposing that the diagonal elements of the resolvent $G_i(z)$ have the following small-$\epsilon$ expansion
\begin{align}
    G_i(z) = G^{(0)}_i(z) + G^{(1)}_i(z)\epsilon + G^{(2)}_i(z)\epsilon^2+\order{\epsilon^3}, \label{eq:G_small_epsilon_expansion}
\end{align}
we can find an expression for the trace of the resolvent $\sum_iG_i(z)/N$ which is accurate to second order in $\epsilon$. Using standard methods (see e.g. Ref.~\cite{sommersSpectrumLargeRandom1988}), the trace of the resolvent can then be related to the eigenvalue density in the complex plane, and we can thus determine the support of the eigenvalue spectrum.

The support of the bulk spectrum can be expressed in terms of the elliptical parameters $\sigma^2, \gamma$, as well as the fine-structure parameters $R, S$ and $T$. In the end, we find the following deformed ellipse
\begin{align}
    \qty(\frac{x}{a})^2 + \qty(\frac{y}{b})^2 = \sigma^2 - \frac{4c}{\sigma^2}\qty(\frac{x}{a})^2\qty(\frac{y}{b})^2,\label{eq:bulk_approx}
\end{align}
with
\addtocounter{equation}{-1}
\begin{subequations}
\begin{align}
    a &= 1 + \gamma + \frac{1}{2}(1 - \gamma)(S + T + 4R) + 2T, \nonumber\\
    b &= 1 - \gamma + \frac{1}{2}(1 + \gamma)(S + T - 4R) + 2T, \nonumber\\
    c &= 8\frac{T - R\gamma}{1 - \gamma^2}. 
\end{align}
\end{subequations}
When $S, T, R \to 0$, we recover the usual elliptical law in \cref{eq:ellipse0}.

In \cref{fig:DR_networks,fig:neural_networks,fig:NM_spectrum}, rather than plot \cref{eq:bulk_approx} directly by solving for $x$ or $y$, we use the following parametric representation
\begin{align}
    x(\theta) &= a \sigma \cos(\theta)\qty[1 - c\sin[2](\theta)], \nonumber\\
    y(\theta) &= b \sigma \sin(\theta)\qty[1 - c\cos[2](\theta)].\label[plural_equation]{eq:parametric_approx}
\end{align}
For a demonstration of the equivalence of this parametric form to \cref{eq:bulk_approx}, see \cref{sm:section:parametric_form_of_MEL} of the SM.

The quantity $a$ determines the location of the rightmost edge of the bulk region $\edge$, and hence stability. More precisely, we have
\begin{align}
    \frac{\edge}{\sigma} = 1 + \gamma + \frac{1}{2}(1 - \gamma)(S + T + 4R) + 2T. \label{eq:edge_approx}
\end{align}
\subsection{Outlier eigenvalue}
Within a calculation to second order in $\epsilon$, we find at most one outlier eigenvalue in the spectrum of an FSRM. The outlier can be expressed in terms of the elliptical parameters $\mu, \sigma^2, \gamma$, as well as the fine-structure parameters $T, U$ and $V$ as follows (see \cref{sm:section:approx_outlier} in the SM)
\begin{multline}
    \frac{\outlier}{\mu} = 1 + \frac{\gamma \sigma^2}{\mu^2} \\[1mm]
    + \qty(1 - \frac{\gamma \sigma^2}{\mu^2})\qty(U + \frac{2\sigma^2V}{\mu^2}) + \frac{2 \sigma^4 T}{\mu^4}. \label{eq:outlier_approx}
\end{multline}
Similarly to how an elliptical matrix has no outlier if $|\mu|\leq \sigma$, the finely structured matrix under consideration only has an outlier at \cref{eq:outlier_approx} if the following condition holds
\begin{align}
    |\mu| > \sigma \qty[ 1 + \frac{1}{2}\qty\Big(S + T + 4R) - \qty(U + \frac{2\sigma^2 V}{\mu^2})]. \label{eq:outlier_condition}
\end{align}
In the case of an equality in \cref{eq:outlier_condition}, we have $\outlier = \edge$, and the outlier eigenvalue is `absorbed' into the bulk spectrum. Note that for $T, U, V \to 0$, we recover the outlier for elliptical matrices in \cref{eq:outlier0}.

\cref{eq:bulk_approx,eq:outlier_approx,eq:edge_approx} are the central results of this paper. Given an arbitrary finely-structured random matrix, \cref{eq:bulk_approx,eq:outlier_approx} provide an explicit and direct approximation for the support of the spectrum of that matrix. The fine-structure correction to the leading eigenvalue of a general FSRM is given by the maximum of $\edge$ and $\outlier$. We can therefore make general statements about the stability of systems for which the Jacobian matrix is finely-structured using \cref{eq:outlier_approx,eq:edge_approx} (when the strength of the fine structure is small).
\subsection{Stability}
\label{section:small_fs_stability}
%\cref{eq:edge_approx,eq:outlier_approx} give the first order fine-structure correction to the stability of systems with element specific statistics. 
In this section, we use \cref{eq:edge_approx,eq:outlier_approx} to understand in more detail what kinds of fine structure promote (in)stability. That is, if the matrix $\mathbf{J} -\mathbf{\id}$ is the Jacobian matrix of a dynamical system linearized about its fixed point (as it is for the models in \cref{section:exact_solution,section:directed_network,section:neural_networks,section:niche_model}), we see that if any of the eigenvalues of $\mathbf{J}$ have a real part that exceeds $1$, the fixed point is unstable to perturbations. Thus, by understanding how the fine structure affects the rightmost eigenvalue of $\mathbf{J}$, we gain insight into the effect of fine structure on stability. All mathematical details in this section can be found in \cref{sm:section:effect_f_fs_on_stability} of the SM.

In our analysis we use the following matrix semi-norm \cite{conwayHilbertSpaces2007}
\begin{align}
    \|\vb A\| \equiv \sqrt{\frac{1}{N^3}\sum_i\qty\bigg(\sum_jA_{ij})^2},
\end{align}
which satisfies $\|\vb A\| \geq 0 $, and where $\|\vb A\| = 0$ if and only if the row sums of $\vb A$ are all equal to zero. We also define the following quantities
\begin{align}
    S_s &= \frac{1}{4\sigma^4}\left\|\vb s^{(1)} + \vb s^{(1)T}\right\|^2, \nonumber\\
    S_a &= \frac{1}{4\sigma^4}\left\|\vb s^{(1)} - \vb s^{(1)T}\right\|^2, \nonumber\\
    U_s &= \frac{1}{4\mu^2}\left\|\vb u^{(1)} + \vb u^{(1)T}\right\|^2, \nonumber\\
    U_a &= \frac{1}{4\mu^2}\left\|\vb u^{(1)} - \vb u^{(1)T}\right\|^2,\label[plural_equation]{eq:sym_antisym_params}
\end{align}
where $\vb s^{(1)T}$ denotes the transpose of the matrix $\vb s^{(1)}$, and similarly for $\vb u^{(1)T}$

As $\|\cdot \|$ is a matrix semi-norm, we can interpret $S_s$ as quantifying the amount of fine structure present in the matrix $(\vb s + \vb s^T)/2$, the symmetric part of $\vb s$. We also interpret $S_a$ as quantifying the amount of fine structure in the antisymmetric part of $\vb s$, and similarly for $U_s$ and $U_a$. Because $T = \|\vb t^{(1)}\|^2/\sigma^4$ [see \cref{eq:fs_params}] and $\vb t^{(1)T} = \vb t^{(1)}$, we interpret the parameter $T$ as directly quantifying the amount of (symmetric) fine structure in the matrix $\vb t$.
 
The fine-structure parameters $R, S, U$ and $V$ are related to $S_s, S_a, U_s, U_a$ and $T$ via 
\begin{align}
    R &= \rho_1 \sqrt{S_s T},\nonumber\\
    S &= S_s - S_a, \nonumber \\
    U &= U_s - U_a, \nonumber \\
    V &= \rho_2 \sqrt{U_s T}. \label[plural_equation]{eq:fs_params_sym_antisym}
\end{align}
The coefficients $\rho_1$ and $\rho_2$, defined in \cref{sm:eq:rho_12_definition} in the SM, are numbers that vary in the range $[-1, 1]$. Their precise values are immaterial to the arguments that follow. 

Substituting \cref{eq:fs_params_sym_antisym} into \cref{eq:edge_approx}, the following deductions of the effect of fine-structure on the bulk edge $\edge$ are drawn. We find that increased antisymmetric fine structure in the matrix $\vb s$ (as measured by $S_a$) always decreases the value of $\edge$. That is 
\begin{align}
    \pdv{\edge}{S_a} &\leq 0.\label{eq:fs_stability_Sa}
\end{align}
If $\gamma \geq -1/3$, then symmetric fine structure in the matrices $\vb t$ and $\vb s$ increases the value of $\edge$ 
\begin{align}
    \edge\qty(S_s, T) &\geq \edge\qty(S_s=0, T),\nonumber\\
    \edge\qty(S_s, T) &\geq \edge\qty(S_s, T=0),\label[plural_equation]{eq:edge_stability_sym}
\end{align}
If $\gamma < -1/3$, then symmetric fine structure in the matrices $\vb t$ and $\vb s$ can be stabilizing or destabilizing.

We can make similar deductions for the outlier eigenvalue by substituting \cref{eq:fs_params_sym_antisym} into \cref{eq:outlier_approx}. Using \cref{eq:outlier_condition}, we see that \cref{eq:outlier_approx} only corresponds to an outlier eigenvalue to the right of the bulk spectrum if $\mu/\sigma > 1 + \delta$, where $\delta$ goes to zero as the fine-structure correction goes to zero. Hence, to leading order in the fine structure, the factor of $(1 - \gamma\sigma^2/\mu^2)$ in \cref{eq:outlier_approx} is positive. Using this observation, we find that antisymmetric fine structure in the matrix $\vb u$ decreases the outlier eigenvalue
\begin{align}
    \pdv{\outlier}{U_a} \leq 0,\label{eq:fs_stability_Ua}
\end{align}
and the parameters $U_s$ and $T$ increase the outlier eigenvalue, for any value of $\gamma$
\begin{align}
    \outlier\qty(U_s, T) &\geq \outlier\qty(U_s=0, T),\nonumber\\
    \outlier\qty(U_s, T) &\geq \outlier\qty(U_s, T=0).\label[plural_equation]{eq:outlier_stability_sym}
\end{align}

We can therefore draw broad qualitative conclusions on the effect of fine structure on stability. \Cref{eq:fs_stability_Sa,eq:fs_stability_Ua} tell us that antisymmetric fine structure in the statistics of an FSRM is always a stabilizing influence, and \cref{eq:outlier_stability_sym,eq:edge_stability_sym} tell us that symmetry in the fine structure of the statistics of a FSRM largely promotes instability. Indeed, for symmetric fine structure to stabilize a given system, it is necessary, but not sufficient, that $\gamma$ lies in the range $[-1, -1/3]$ and that the inequality in \cref{eq:outlier_condition} is violated.

\section{Summary of the examples in the following sections}
The remainder of this paper is concerned with the application of the results for the fine-structure corrections to the eigenvalue spectrum presented thus far [\cref{eq:outlier_approx,eq:bulk_approx}]. The purpose of these examples is to make clear what constitutes a finely-structured random matrix in context, and to demonstrate the breadth of the applicability of the modified elliptical law.

\begin{figure*}
    \includegraphics[width=\textwidth]{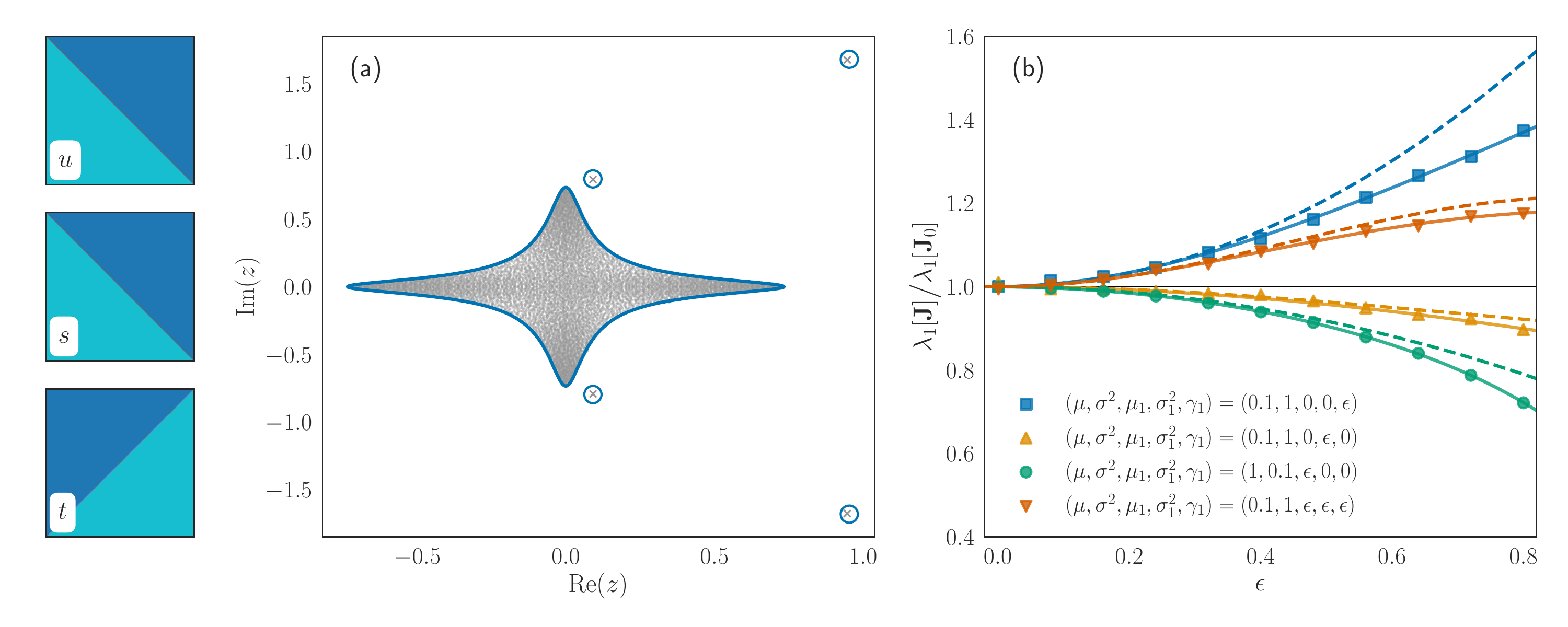}
    \caption{Support of the spectrum of FSRMs constructed according to the cascade model. $(u)$, $(s)$ and $(t)$: heatmaps of the matrices $\vb u$, $\vb s$ and $\vb t$ as defined in \cref{eq:cascade_stats}. (a): The bulk and outliers of a single $8000 \times 8000$ cascade model FSRM with $\mu = 2.5, \mu_1 = 3.5, \sigma = 0.5, \sigma_1 = 0.1, \gamma = 0, \gamma_1 = 1$. The blue solid lines and circles are the exact analytical prediction for the boundary of the bulk and outliers [given in \cref{sm:section:cascade_model} of the SM]. (b): The leading eigenvalue $\eigval{1}[\vb J]$ of a cascade model FSRM with $\gamma = 0.2$ and all other model parameters as indicated. We also divide the leading eigenvalue by $\eigval{1}[\vb J_0]$, the leading eigenvalue of a corresponding elliptical random matrix with statistics $\mu$, $\sigma$ and $\gamma$ to isolate the effect of fine structure. For the blue, orange and red curves (squares and triangles), $\eigval{1}[\vb J_0]=\sigma(1 + \gamma)$ because the bulk edge is the leading eigenvalue. For the green curve (circles), $\eigval{1}[\vb J_0]=\mu + \gamma\sigma^2/\mu$ because the outlier is the leading eigenvalue. Solid lines are the exact solution [\cref{sm:section:cascade_model}] and dashed lines are the fine-structure corrections to the leading eigenvalue given in \cref{eq:cascade_edge_approx,eq:cascade_approx_outlier}. Each marker is the result of diagonalizing a single $8000\times 8000$ instance of the matrix.}
    \label{fig:HT_spectrum}
\end{figure*}

In the first example (\cref{section:exact_solution}) we study a generalization of the cascade model from theoretical ecology \cite{allesinaPredictingStabilityLarge2015,aljadeffLowdimensionalDynamicsStructured2016}. This is a simple example of an FSRM ensemble, for which there is an exact solution, which we derive in \cref{sm:section:cascade_model} of the SM. We evaluate the fine-structure correction to the elliptical law for this model using the modified elliptical law, and verify that this correction produces the same result as an expansion of the exact solution for small fine structure. An instance of this model, and our solution, is shown in \cref{fig:HT_spectrum}. 

\begin{figure*}
    \includegraphics[width=\textwidth]{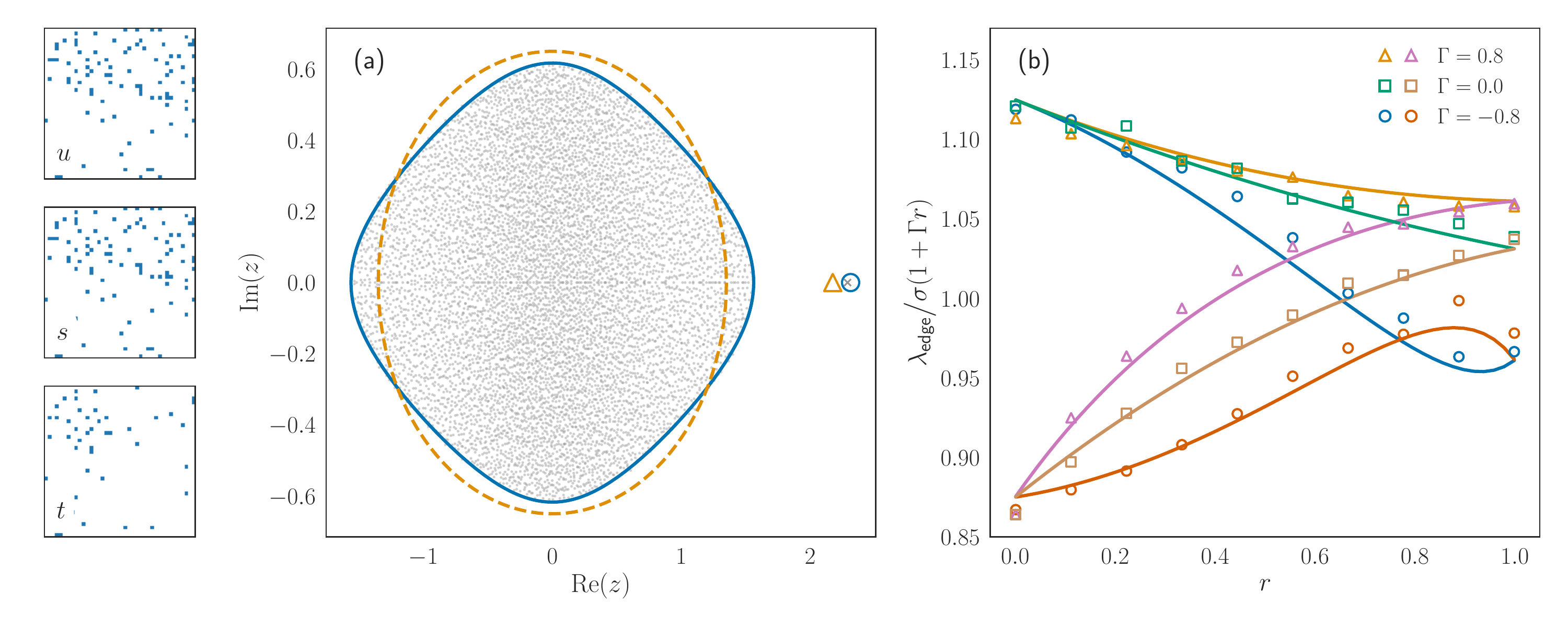}
    \caption{The spectrum of a directed network with statistics as in \cref{eq:dr_stats}. $(u), (s), (t)$: Heatmaps of the statistics of $\vb J = \vb A \circ \vb K$ for a single realization of the network $\vb A$ with $N = 40$, coloured squares indicate the presence of a link and white squares indicate no link. (a) Eigenvalues of a single $8000\times 8000$ realization of $\vb J$. Blue solid line and circle are the fine-structure correction to the eigenvalue spectrum of a directed network with $\mu=2$, $\sigma=1$, $\Gamma=0.7$, orange dashed line and triangle show the elliptical law with no fine-structure correction. Half of the nodes have expected degrees $(k^\leftrightarrow, k^\text{in}, k^\text{out}) = (325, 300, 300)$ and the other half have expected degrees $(75, 100, 100)$. (b) Ratio of the bulk edge eigenvalue of $\vb J$ to $\sigma(1 + \Gamma r)$, the bulk edge without the fine-structure correction. The top three curves when $r=0$ correspond to random matrices with positive degree correlations $\rho$, the bottom three curves correspond to ensembles with the same degree distributions with the links rearranged so that the degree correlation is negative, demonstrating that increasing $\rho$ leads to destabilization. When $r=1$, the degree heterogeneity $h^2$ controls the size of the fine-structure correction, and it stabilizes/destabilizes depending on the value of $\Gamma$. The network has the same dichotomous structure as that used in panel (a) and $\mu=0.5$, $\sigma=1$ and $\Gamma$ as indicated. Markers are the result of diagonalizing a single $8000\times 8000$ realization of the matrix. We verify \cref{eq:network_outlier}, the fine-structure correction to the outlier eigenvalue, in \cref{sm:fig:network_outlier} of the SM.}
    \label{fig:DR_networks}
\end{figure*}

In a second example (\cref{section:directed_network}), for which there is no simple exact solution, we show that directed complex networks with dense random network structure can also be treated as FSRMs. We find that the amount of fine structure present depends on three statistical properties of the network, and we derive the fine-structure correction to the elliptical law, which is valid when these statistics are small. The results can be seen in \cref{fig:DR_networks}. We also note that our fine-structure corrections reproduce the exact result of Ref.~\cite{neriLinearStabilityAnalysis2020} when the network has no undirected links, and the formulae reproduce the results of \cite{baron_directed_2022} when the network has no directed links.

The third example (\cref{section:neural_networks}) considers a square grid of interacting neurons arranged in physical space \cite{kuczalaEigenvalueSpectraLarge2016}. The interaction between any two neurons depends continuously on the physical distance between them in the grid. We use \cref{eq:outlier_approx,eq:bulk_approx} to compute the fine-structure correction to the spectrum of random matrices with statistics modelling the interacting neurons, arriving at a simple closed-form expression, where previous works had to find the spectrum of similar models numerically. Our approximation is compared with numerical results in this case in \cref{fig:neural_networks}.

In the final example (\cref{section:niche_model}), we consider a more sophisticated model in theoretical ecology. We use our central results to provide a simple approximation for the eigenvalue spectrum of a random matrix constructed according to the niche model \cite{williamsSimpleRulesYield2000,allesinaStabilityCriteriaComplex2012}, for which precise analytical results have so far been difficult to come by. In the case of the niche model, there is no limit of any model parameter one could take to recover the conventional elliptical law. Despite this, our fine-structure correction still works well (an example is shown in \cref{fig:NM_spectrum}).
\section{The Cascade model of complex ecosystems}
\label{section:exact_solution}

\subsection{Model Definition}
Following Robert May \cite{mayWillLargeComplex1972,MAYBOOK}, we consider a set of ordinary differential equations that govern the time evolution of a set of species abundances $x_i$,
\begin{align}
    \dot x_i = f_i\qty(x_1, x_2, \dots, x_N).
\end{align}
We imagine that this system is linearized about a fixed point such that
\begin{align}
    \delta{\dot x}_i = -\delta x_i + \sum_{j\neq i} J_{ij} \delta x_j,
\end{align}
where $\delta x_i$ is the deviation of $x_i$ from its fixed point value. The diagonal elements of the Jacobian are set to $-1$ so that the system without interactions would be stable. We   wish to understand what kinds of interactions (or Jacobian matrix elements $J_{ij}$) permit stability. If the largest real part of any eigenvalue of $\vb J$ is greater than one, then the ecological equilibrium will be unstable.

May's original work makes use of the circular law \cite{girko1985circular,baiCircularLaw1997,taoRandomMatricesUniversality2015}, a special case of the elliptical law with zero mean ($\mu = 0$) and uncorrelated matrix entries ($\gamma = 0$). More recent work \cite{tangCorrelationInteractionStrengths2014,Galla_2018,buninEcologicalCommunitiesLotkavolterra2017}, has given an ecological interpretation to the parameters $\mu$ and $\gamma$, associating them respectively with the average strength of interspecies interaction and with the proportion of interspecies interactions that are of predator prey type $p$ (the proportion of links satisfying $J_{ij}J_{ji} < 0$) through $\gamma = \cos(\pi p)$ (see also Ref.~\cite{poleyGeneralizedLotkaVolterraModel2023}). Here, and later in \cref{section:niche_model}, we also incorporate the possibility of a hierarchy amongst species \cite{poleyGeneralizedLotkaVolterraModel2023,allesinaPredictingStabilityLarge2015}, which can be interpreted as fine structure.

We consider the FSRM ensemble with statistics given by
\begin{align}
    u_{ij} &=
    \mu + \mu_1 \times
    \begin{cases}
        1, & i < j \\
        -1, & i > j
    \end{cases}
    , \nonumber\\
    s_{ij} &= \sigma^2 + \sigma_1^2 \times
    \begin{cases}
        1, & i < j \\
        -1, & i > j
    \end{cases}
    , \nonumber\\
    t_{ij} &= \gamma\sigma^2
     + \gamma_1\sigma^2\times
    \begin{cases}
        1, & i + j < N \\
        -1, & i + j > N
    \end{cases}
    ,\label[plural_equation]{eq:cascade_stats}
\end{align}
where $\vb u, \vb s, \vb t$ are defined in \cref{eq:fsrm_stats} [the structure of these matrices is illustrated in panels $(u), (s)$ and $(t)$ of \cref{fig:HT_spectrum}]. If $\gamma_1 = 0$, then we recover the cascade model of Refs.~\cite{allesinaPredictingStabilityLarge2015,aljadeffLowdimensionalDynamicsStructured2016}. The triangular structure of $\vb u$ reflects the hierarchy amongst species (see Refs.~\cite{poleyGeneralizedLotkaVolterraModel2023,allesinaPredictingStabilityLarge2015} for more in-depth discussion). The generalization presented here allows for $t_{ij}$ to be index-dependent. In writing down \cref{eq:cascade_stats} we have anticipated that $\mu, \sigma^2$ and $\gamma$ in \cref{eq:cascade_stats} will coincide with the quantities defined in \cref{eq:avg_params}. 
\subsection{Eigenvalue spectrum}
To find the fine-structure correction to the elliptical law due to the statistics in \cref{eq:cascade_stats}, we suppose that $\mu_1$, $\sigma_1^2$ and $\gamma_1$ are proportional to a small parameter~$\epsilon$. We make the following substitutions,
\begin{align}
    \mu_1 \to \epsilon \mu_1, \hspace{0.5cm}
    %, 
    \sigma_1^2 \to \epsilon \sigma_1^2, \hspace{0.5cm}
    \gamma_1 \to \epsilon \gamma_1.\label{eq:cascade_small_epsilon_replacements}
\end{align}
If $\epsilon = 0$ then the statistics \cref{eq:cascade_stats} reduce to the elliptical law with parameters $\mu, \sigma^2$ and $\gamma$ [see \cref{eq:avg_params}]. 

Let us now find the correction due to the fine structure in the cascade model. The calculation of the fine-structure parameters $R, S, T, U$ and $V$ [using \cref{eq:fs_params}] is detailed in \cref{sm:section:MEL_cascade_boundary} of the SM. We find $R = V = 0$, and the non-zero terms are
\begin{gather}
    S = -\frac{\epsilon^2\sigma_1^4}{3\sigma^4}, \hspace{5mm} T = \frac{\epsilon^2\gamma_1^2}{3},\hspace{5mm} U = -\frac{\epsilon^2\mu_1^2}{3\mu^2}. \label{eq:cascade_fs_corrections}
\end{gather}
Substituting the corrections from \cref{eq:cascade_fs_corrections} into \cref{eq:bulk_approx,eq:edge_approx,eq:outlier_approx}, we obtain the fine-structure correction to the spectrum of an FSRM with cascade model statistics. We can see how the addition of fine structure affects stability by looking at the correction to the bulk edge [\cref{eq:edge_approx}]
\begin{align}
    \frac{\edge}{\sigma} &= 1 + \gamma + \frac{\epsilon^2}{6}\qty[\qty(5 - \gamma)\gamma_1^2 - \qty(1 - \gamma)\frac{\sigma_1^4}{\sigma^4}], \label{eq:cascade_edge_approx}
\end{align}
and the fine-structure correction to the outlier eigenvalue [\cref{eq:outlier_approx}]
\begin{multline}
    \frac{\outlier}{\mu} = 1 + \frac{\gamma\sigma^2}{\mu^2} + \frac{\epsilon^2}{3}\qty[\frac{2\sigma^4}{\mu^4}\gamma_1^2 - \qty(1 - \frac{\gamma\sigma^2}{\mu^2})\frac{\mu_1^2}{\mu^2}].\label{eq:cascade_approx_outlier}
\end{multline}
Recalling $|\gamma|\leq1$, we see that $\edge$ increases if $\gamma_1$ is increased and decreases if $\sigma_1$ increases. Further, \cref{eq:outlier_condition} tells us that \cref{eq:cascade_approx_outlier} only corresponds to an outlier eigenvalue if $\mu/\sigma > 1 + \order{\epsilon^2}$. Hence, when \cref{eq:cascade_approx_outlier} is valid, the term $\epsilon^2(1 - \gamma\sigma^2/\mu^2)$ is positive (to second order in $\epsilon$) so we can conclude that $\gamma_1$ increases, and $\mu_1$ decreases the value of $\outlier$. As $\sigma_1, \mu_1$ encode the amount of asymmetry in the fine-structured parts of the matrices $\vb s$ and $\vb u$, and as $\gamma_1$ encodes the amount of symmetrical fine structure in the matrix $\vb t$, our findings here are consistent with the discussion in \cref{section:small_fs_stability}. The above stated dependence of stability on the various model parameters is also confirmed in \cref{fig:HT_spectrum}~(b).

For FSRMs with statistics as in \cref{eq:cascade_stats}, there exist exact expressions for the support of the bulk region and outlier eigenvalues. The solution is derived in \cref{sm:section:cascade_model} of the SM, and verified in panels (a) and (b) of \cref{fig:HT_spectrum}. In \cref{sm:section:verification_of_MEL} of the SM, we also demonstrate that expanding the exact solution in small $\epsilon$ leads to the same result as the modified elliptical law, thus verifying \cref{eq:bulk_approx,eq:edge_approx,eq:outlier_approx} analytically.
\section{Directed complex networks as fine structure}
\label{section:directed_network}
\subsection{Model definition}
As a second example, we consider the eigenvalue spectra of matrices that represent weighted and directed complex networks. Here, the network structure is the source of fine structure in the model.

In particular, we consider matrices of the form $[\vb J]_{ij} = [\vb A \circ \vb K]_{ij} = A_{ij}K_{ij}$, where $\vb A$ is the adjacency matrix of the network ($A_{ij} = 1$ if a link exists going from node $j$ to node $i$ and is zero otherwise), $\vb K$ encodes the weights of the edges, which can be positive or negative, and $\circ$ denotes the element-wise product of matrices. Here, $\vb K$ is an elliptical random matrix with statistics
\begin{align}
    \avg{K_{ij}}_K &= \frac{\mu}{p}, \nonumber\\
    \variance_K(K_{ij}) &= \frac{\sigma^2}{p}, \nonumber\\
    \cov_K\qty(K_{ij}, K_{ji}) &= \frac{\Gamma\sigma^2}{p},  \label[plural_equation]{eq:dr_non_zero_stats}  
\end{align}
where $p = \sum_{ij}A_{ij}/N$ is the average degree of nodes in the network. The scaling with $p$ of the above statistics ensures a sensible dense limit $p\to\infty$, with $p/N$ held constant as $N\to\infty$. As the statistics of $\vb K$ contain no fine structure, our corrections to the elliptical law and outlier eigenvalue will only depend on properties of the network. 

The network is constructed as follows: Each pair of nodes $i$ and $j$ is joined with a directed edge $i \to j$, another directed edge $j \to i$, a reciprocal edge $i \leftrightarrow j$, or no edge with probabilities $P_{j \gets i}, P_{i \gets j}, P_{i\leftrightarrow j}$ and $P_{i \not\leftrightarrow j}$ respectively ($P_{j \gets i}+P_{i \gets j}+P_{i\leftrightarrow j}+P_{i \not\leftrightarrow j}=1$). These events are taken to be mutually exclusive, so our definitions are such that we cannot obtain an undirected link from two directed links. We also define the reciprocity (the ratio of undirected links to total links) \cite{garlaschelliPatternsLinkReciprocity2004,newmanEmailNetworksSpread2002,wassermanSocialNetworkAnalysis1994,squartiniReciprocityWeightedNetworks2013} of the network
\begin{align}
    r = \frac{1}{pN}\sum_{ij}P_{i\leftrightarrow j},\label{eq:reciprocity}
\end{align}
as well as the mean undirected, exclusively-in and exclusively-out degrees of each node
\begin{align}
    k^\leftrightarrow_i &= \sum_j P_{i\leftrightarrow j}, \nonumber\\
    k^\text{in}_i &= \sum_j P_{i\gets j}, \nonumber\\
    k^\text{out}_i &= \sum_j P_{j\gets i}.\label[plural_equation]{eq:dr_degree_distributions}
\end{align}
The relations $\sum_ik^\leftrightarrow_i = prN$, $\sum_ik^\text{in}_i = (1-r)pN$ and $\sum_ik^\text{out}_i = (1-r)pN$ follow from the definitions. Note that even if the underlying network is completely undirected ($r = 0$), the weighted network $\vb A \circ \vb K$ can still be considered directed because of the presence of positive and negative weights (see, for example, Ref.~\cite{baron_directed_2022}).

In \cref{sm:section:directed_networks} of the SM, we show that $\vb J = \vb A \circ \vb K$ (the random matrix of interest) has the same spectrum as a fully-connected FSRM with statistics
\begin{align}
    u_{ij} &= \frac{N}{p}\big(P_{i\leftrightarrow j} + P_{i \gets j}\big)\mu, \nonumber\\
    s_{ij} &= \frac{N}{p}\big(P_{i\leftrightarrow j} + P_{i \gets j}\big)\sigma^2, \nonumber\\
    t_{ij} &= \frac{N}{p}P_{i\leftrightarrow j}\Gamma\sigma^2. \label[plural_equation]{eq:dr_stats}
\end{align}
Hence, network structure can be interpreted as a manifestation of fine structure, and we can approximate the spectrum of $\vb A \circ \vb K$ using the modified elliptical law.
\subsection{Eigenvalue spectrum}
To find the fine-structure correction to the spectrum we first compute the elliptical parameters [\cref{eq:avg_params}] from the statistics \cref{eq:dr_stats}. The definitions of $\mu$ and $\sigma^2$ in \cref{eq:dr_stats} coincide with those in \cref{eq:avg_params}, and we find $\gamma=\Gamma r$. 

The fine-structure parameters $R, S, T, U, V$ are computed directly using \cref{eq:fs_params}, and they depend on three additional statistical properties of the network. These are the variance of the undirected degrees $h^2$, otherwise known as the degree heterogeneity \cite{baron_directed_2022}, the correlation coefficient between exclusively undirected degrees and exclusively directed degrees $\tau$, and the correlation coefficient between exclusively in and out-degrees $\rho$, otherwise known as the degree correlation coefficient \cite{neriLinearStabilityAnalysis2020}. Explicitly, these statistics are
\begin{align}
    h^2 &= \sum_i\frac{\qty[k^\leftrightarrow_i - rp]^2}{Np^2}, \nonumber\\
    \tau &= \sum_i\frac{\qty[k^\leftrightarrow_i - rp]\qty[k^\text{in}_i + k^\text{out}_i - 2(1 - r)p]}{2Np^2}, \nonumber\\
    \rho &= \sum_i\frac{\qty[k^\text{in}_i - (1 - r)p]\qty[k^\text{out}_i - (1 - r)p]}{Np^2}.\label[plural_equation]{eq:network_stats}
\end{align}
Written in terms of $h^2, \tau$ and $\rho$, the fine-structure parameters are then
\begin{align}
    R &= \Gamma \qty(\tau + h^2), \nonumber\\
    S &= \rho + 2\tau + h^2, \nonumber\\
    T &= \Gamma^2 h^2, \nonumber\\
    U &= \rho + 2\tau + h^2, \nonumber\\
    V &= \Gamma \qty(\tau + h^2).\label[plural_equation]{eq:network_fs_params}
\end{align}
The fine-structure correction to the spectrum of $\vb J$ due to the network is found by substituting the values of $R, S, T, U$ and $V$ into \cref{eq:bulk_approx,eq:outlier_approx}. This approximation is valid provided $h^2, \tau$, and $\rho$ are all small. For details of the calculation, see \cref{sm:section:directed_networks} of the SM.

We can analyse the effect of network structure on stability by looking at the fine-structure correction to the bulk edge [see \cref{eq:edge_approx}],
\begin{multline}
    \frac{\edge}{\sigma} = 1 + \Gamma r \\ + \frac12h^2\qty\Big[1 + (4 - r)\Gamma + (5 - 4r)\Gamma^2 - r\Gamma^3] \\
         + \frac{1}{2}(1 - \Gamma r)\qty\Big[2\qty(1 + 2\Gamma)\tau + \rho], \label{eq:network_edge}
\end{multline}
and the fine-structure correction to the outlier eigenvalue [see \cref{eq:outlier_approx}],
\begin{multline}
    \frac{\outlier}{\mu} = 1 + \frac{\Gamma\sigma^2}{\mu^2} r \\ + h^2\qty[1 + (2 - r)\frac{\Gamma\sigma^2}{\mu^2} + 2(1 - r)\frac{\Gamma^2\sigma^4}{\mu^4}]\\
    + \qty(1 - \frac{\Gamma\sigma^2}{\mu^2}r)\qty[2\qty(1 + \frac{\Gamma\sigma^2}{\mu^2})\tau + \rho].\label{eq:network_outlier}
\end{multline}
Examining \cref{eq:network_outlier,eq:network_edge}, and recalling that there is no outlier if $\mu/\sigma < 1 + \delta$, where $\delta$ goes to zero as the fine structure goes to zero, we find that the network heterogeneity $h^2$ always increases the value of the outlier eigenvalue, and increases the value of $\edge$ if $\Gamma > 1 - \sqrt{2}$ or if $r < 0.73$ [as can be seen by examining the factor multiplying $h^2$ in \cref{eq:network_edge}]. We also observe that $\edge$ and $\outlier$ always increase with increasing correlation coefficient $\rho$. Given that a greater values of $h^2$ and $\rho$ both connote greater degrees of symmetric fine structure, we see that our results agree with the general remarks made in \cref{section:small_fs_stability}.
\begin{figure*}
    \includegraphics[width=\textwidth]{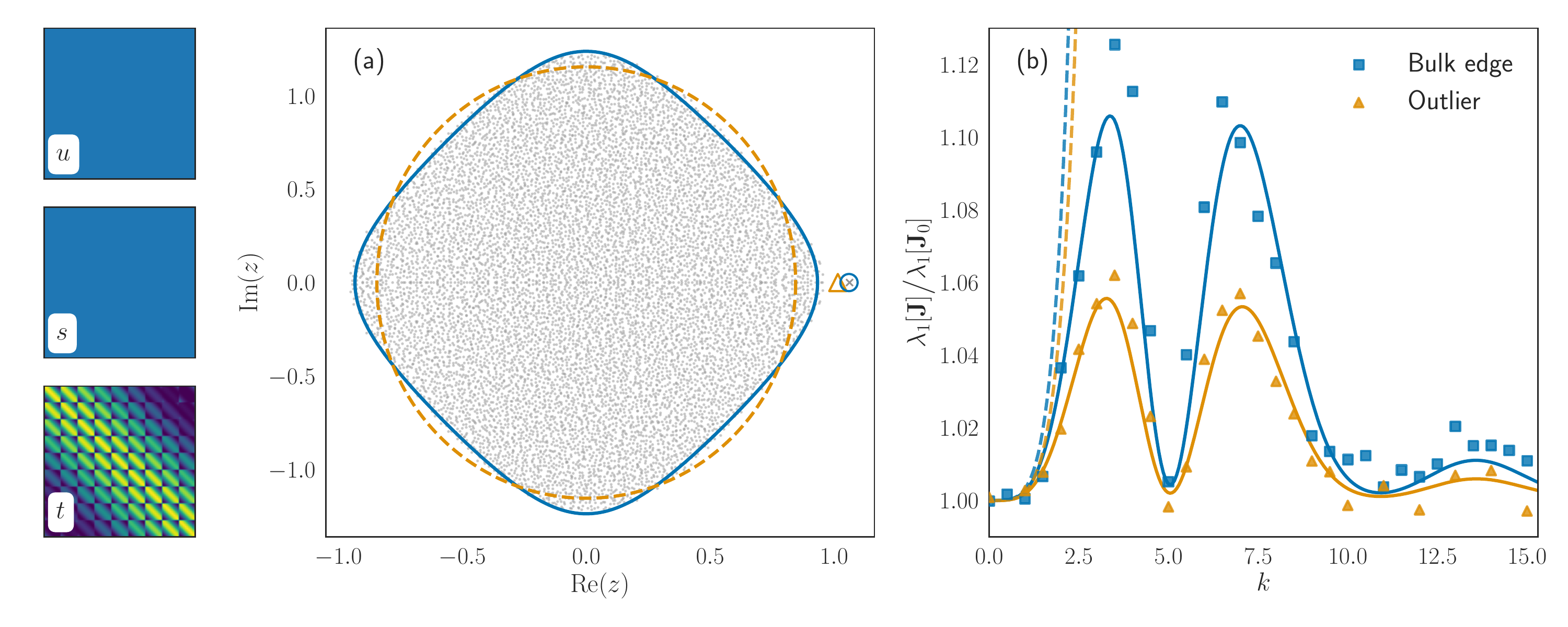}
    \caption{Spectrum of a FSRM with statistics as in \cref{eq:neural_stats}. (u), (s) and (t): Heatmaps of the matrices $\vb u$, $\vb s$ and $\vb t$ in \cref{eq:neural_stats}. (a) Eigenvalues of a single $8000\times 8000$ realization of the matrix. Blue solid line and circle are the fine-structure correction to the eigenvalue spectrum [\cref{eq:bulk_approx,eq:outlier_approx}], orange dashed line and triangle show the elliptical law with no fine-structure correction. Parameters are $\mu=1.1, \sigma=1, \Gamma=1$ and $k=3$. (b) Blue curves (squares): ratio of the fine-structure prediction for the bulk edge of $\vb J$ to the bulk edge of an elliptical matrix with no fine structure. Orange curves (triangles): as for the blue curves but for the outlier eigenvalue. Solid curves indicate the fine-structure correction [\cref{eq:outlier_approx,eq:bulk_approx}], dashed lines indicate the small-$k$ correction [\cref{eq:neural_stability_small_k}]. Blue (squares) markers are the result of diagonalizing a single $8000\times 8000$ realization of the matrix, orange (triangles) markers are the result of diagonalizing $100$ instances of $4000\times 4000$ realizations of the matrix and averaging. Parameters are $\mu = 1.1$, $\sigma = 1$ and $\Gamma = 1$, with $k$ as indicated.}
    \label{fig:neural_networks}
\end{figure*}

In \cref{sm:section:directed_networks} of the SM, we further confirm our results by demonstrating that our fine-structure correction [\cref{eq:network_edge,eq:network_outlier}] reduces to the approximate results derived in Ref.~\cite{baron_directed_2022} when the underlying network is exclusively directed ($r = 1$), and reduces to the exact results of Ref.~\cite{neriLinearStabilityAnalysis2020} when the underlying network is exclusively undirected ($r = 0$).
\section{Fine structure in a toy neural network model}
\label{section:neural_networks}
\subsection{Model Definition}
In the study of neural networks, firing-rate models are used to investigate the dynamical interaction of the neurons. The activation of the $i$-th neuron $x_i$ is often taken to follow dynamical rules of the type  \cite{sompolinskyChaosRandomNeural1988,aljadeffTransitionChaosRandom2015}
\begin{align}
    \dot x_i = -x_i + \sum_j J_{ij}\tanh(x_i),
\end{align}
where $J_{ij}$ dictates the presence and strength of an effect of neuron $j$ onto the activation of neuron $i$. Similarly to May's analysis of complex ecosystems, when the leading eigenvalue of $\vb J$ is greater than $1$, the state $x_i = 0$ becomes unstable, leading to a qualitative change in the dynamics. 

Inspired by a toy model proposed in \cite{kuczalaEigenvalueSpectraLarge2016}, we consider $N = n^2$ neurons regularly spaced on a finite square grid (with no periodic boundary conditions), and with correlated interactions. We suppose that the degree to which the interaction strengths $J_{ij}$ and $J_{ji}$ between any two neurons are correlated is a function of the physical distance between them in the grid \cite{smithSpatialTemporalScales2008,rosenbaumSpatialStructureCorrelated2017}. For the purposes of illustration of the method, we consider an FSRM ensemble with statistics that depend in a simple way on the distances between neurons
\begin{align}
    u_{ij} &= \mu, \nonumber\\
    s_{ij} &= \sigma^2, \nonumber\\
    t_{ij} &= \Gamma\sigma^2\cos(k|\vb r_i - \vb r_j|), \label[plural_equation]{eq:neural_stats}   
\end{align}
where $\vb r_i = (x_i, y_i)$ is the position of the $i$-th neuron in the grid and $k$ is a parameter controlling the scale over which correlations in the interaction of one neuron with other neurons varies. We use a lattice spacing of $1/n$, so that the total size of the grid is $1\times 1$.

We suppose that the neurons are labelled $1, 2, \dots, n$ in the top row, $n + 1, n + 2, \dots, 2n$ in the second row, and so on, so that there are $n^2$ neurons in total. The matrix $\mathbf{t}$ therefore has size $n^2\times n^2$. An example with $n = 9$ is shown in \cref{fig:neural_networks} (so the matrix of statistics $t_{ij}$ has size $81\times 81$). The block-like structure is due to the specific labelling of the neurons; there is a discontinuous jump in the function $|\vb r_i - \vb r_j|$ at the end of each row. Ultimately, the labelling of the neurons has no effect on the spectrum of $\vb J$.

\subsection{Eigenvalue spectrum}
If the range parameter $k$ is zero, then we recover the usual elliptical law, so it is fair to assume that the approximation works if $k\ll 1$. However, as \cref{fig:neural_networks}~(b) demonstrates, our approximation works well for most values of $k$, not just small values. This is because the fine-structure correction to the elliptical law is small for all values of $k$.

Assuming that the number of neurons in the grid is large, we can compute the elliptical and fine-structure parameters using \cref{eq:avg_params,eq:fs_params}. The mean and variance are $\mu$ and $\sigma^2$ respectively, and the correlation coefficient is
\begin{align}
    \gamma &= \Gamma F_1(k), \label{eq:neural_gamma_calc}
\end{align}
where the $F_1(k)$ is an integral over the coordinates of the grid
\begin{align}
    F_1(k) &= \int\dd{\vb r}\dd{\vb r'}\cos(k|\vb r - \vb r'|),\nonumber\\
    &= \int_0^1\dd{x}\dd{y}A(k, x, y), \label{eq:NN_F1}
\end{align}
with
\begin{align}
    A(k, x, y) \equiv \int_0^1\dd{x'}\dd{y'}\cos(k\sqrt{(x - x')^2 + (y - y')^2}). \label{eq:NN_A}
\end{align}
Details of the derivation of \cref{eq:neural_gamma_calc}, as well as of all other calculations in this section, can be found in \cref{sm:section:neural_network} of the SM. From \cref{eq:neural_stats}, we see that the matrices $\vb u$ and $\vb s$ have no fine structure. Hence, the parameter $T$ is the only non-zero fine-structure parameter
\begin{align}
    T &= \Gamma^2 F_2(k),\label{eq:neural_T_calc}
\end{align}
where the function $F_2(k)$ is also an integral over the spatial coordinates of the grid 
\begin{align}
    F_2(k) \equiv \int_0^1 \dd{x}\dd{y}\qty[A(k, x, y) - F_1(k)]^2.
\end{align}
Plugging the values of $\mu, \sigma^2, \gamma$ and $T$ into \cref{eq:outlier_approx,eq:edge_approx,eq:bulk_approx} gives the fine-structure correction for FSRMs with statistics as in \cref{eq:neural_stats}.

When $k = 0$, the statistics in \cref{eq:neural_stats} reduce to those of an elliptical random matrix. Therefore, the amount of fine structure present in the system is guaranteed to be small if $k$ is small. Expanding \cref{eq:edge_approx,eq:outlier_approx} in powers of $k$ gives us some insight into the (de)stabilizing effect of the spatial dependence of interactions between neurons
\begin{align}
    \frac{\edge}{\sigma\qty[1 + \Gamma F_1(k)]} &= 1 + \frac{\qty(5 - \Gamma)}{720}\frac{\Gamma^2k^4}{1 + \Gamma} + \order{k^6},  \nonumber\\
    \frac{\mu\outlier}{\mu^2 + \Gamma\sigma^2F_1(k)} &= 1 + \frac{1}{180}\frac{\Gamma^2\sigma^4k^4}{\mu^2\qty\big(\mu^2 + \Gamma\sigma^2)} + \order{k^6}.\label[plural_equation]{eq:neural_stability_small_k}
\end{align}
We have divided by the zeroth order factors in the fine structure to highlight the relative effect of the fine structure on the spectrum. Our approximation for the bulk and outlier are confirmed in \cref{fig:neural_networks}. In particular, panel (b) reveals that the fine structure present in the statistics \cref{eq:neural_stats} is always destabilizing, which is consistent with the general condition in \cref{eq:edge_stability_sym}.
\begin{figure*}
    \includegraphics[width=\textwidth]{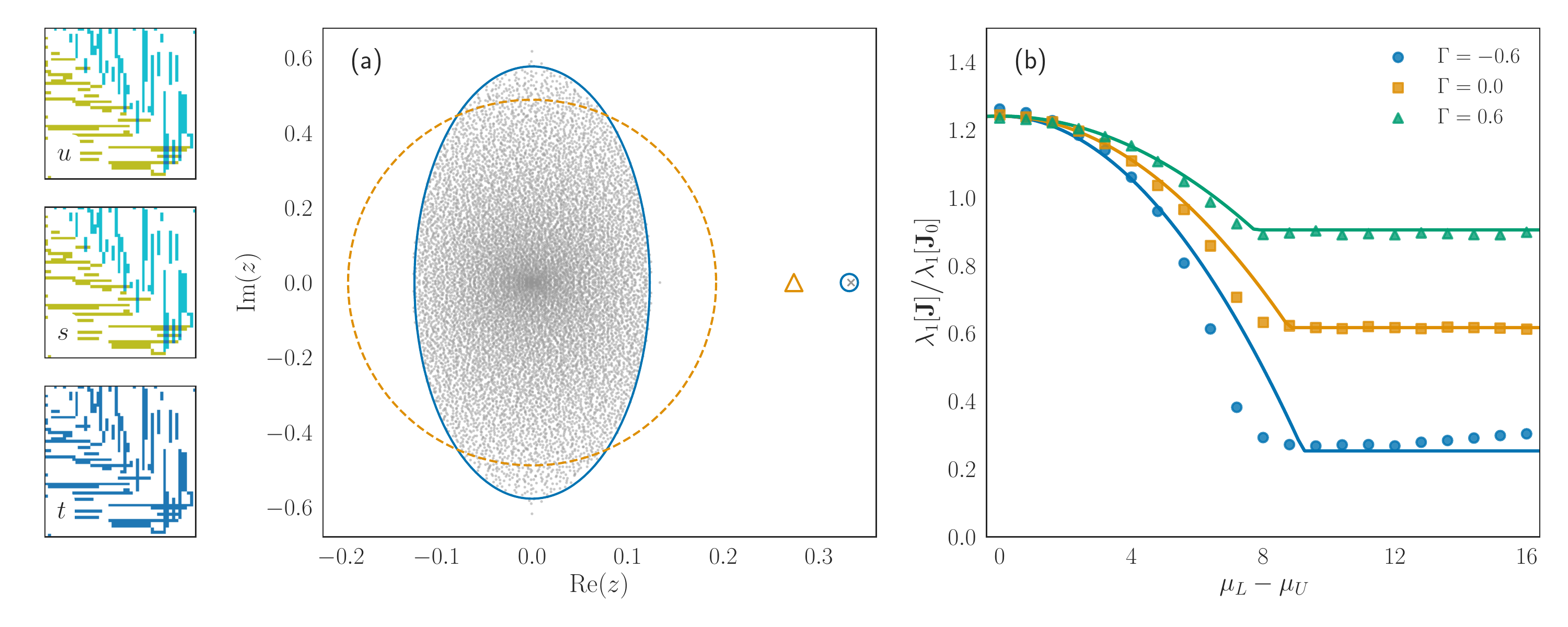}
    \caption{Spectrum of an FSRM constructed according to the niche model [\cref{eq:niche_stats}]. (u), (s), (t): Heatmaps of the statistics $\vb u, \vb s, \vb t$ with $N = 50$. Colour indicates the strength and presence of a link, and white indicates no link in the network. (a) Predictions from \cref{eq:bulk_approx,eq:outlier_approx} are shown in blue (solid line and circle), and the elliptical approximation to the spectrum is shown in orange (dashed line and triangle). Grey markers are the result of exact diagonalization of a $12000\times12000$ realization of the matrix with $\mu_L=3, \mu_U=5, \sigma_L=0.6, \sigma_U=1.4, \gamma=-0.6$ and $d = 0.05$. (b) The ratio of the maximum eigenvalue of a niche model matrix to that of an elliptical matrix with no fine-structure correction. Lines indicate the result of the fine-structure correction [\cref{eq:outlier_approx,eq:bulk_approx}], markers are computed from the average of $10$ realizations of $4000\times 4000$ matrices. The remaining system parameters are $\mu_L + \mu_U=8, \sigma_L=1.2, \sigma_U=0.8, d=0.1$, and the parameters $\Gamma$ and $\mu_L - \mu_U$ are as indicated in the figure.}
    \label{fig:NM_spectrum}
\end{figure*}
\section{The niche model of complex ecosystems}
\label{section:niche_model}
\subsection{Model definition}
\label{section:niche_model_definition}
Our final example builds on the same basic random-matrix framework of Robert May, as described in \cref{section:exact_solution}, but involves a more intricate interaction network.

The niche model is primarily a model of ecological network structure \cite{williamsSimpleRulesYield2000}, which approximates many statistical features of real ecological networks \cite{allesinaStabilityComplexityRelationship2015,williamsStabilizationChaoticNonpermanent2004}. However, due to its complexity, it is a daunting task to find an exact analytical expression for the eigenvalue spectra of the associated interaction matrices. Here we show how \cref{eq:outlier_approx,eq:bulk_approx} deliver a simple approximate solution to this problem. 

First, we describe the steps for the construction of a niche model network. To construct a realization of the network, we draw three random numbers for each of the $N$ species in the system, the `niche value' $\eta_i$, `niche range' $d_i$, and `niche centre' $c_i$ \cite{williamsSimpleRulesYield2000,allesinaGeneralModelFood2008}, whose distributions are described below. These quantities determine where in the hierarchy a species lies, how many species it interacts with, and where in the hierarchy the species it interacts with are, respectively. The network structure is encoded in the matrix $\vb E$, with $E_{ij} = 1$ if $c_i - d_i/2 < \eta_j < c_i + d_i/2$ and $E_{ij} = 0$ otherwise.

\begin{figure}
    \includegraphics[width=0.45\textwidth]{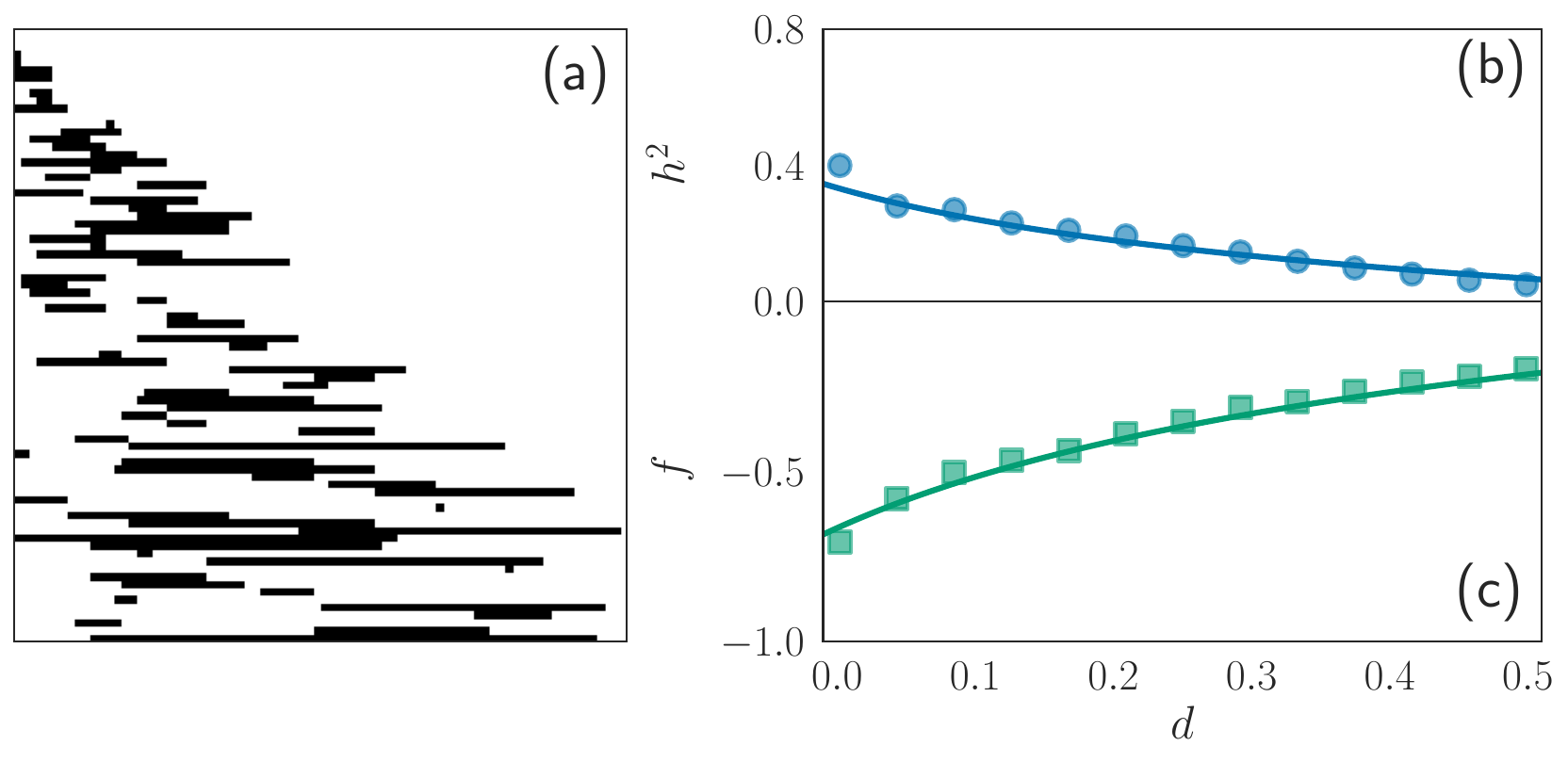}
    \caption{Structure and statistics of the matrix $\vb E$ in the niche model. (a) A single realization of the matrix $\vb E$ with $N = 80$ and $d = 0.15$, constructed according to rules (1) to (4) in \cref{section:niche_model_definition}. Panel (b): Solid line shows the degree heterogeneity ($h^2$) as given in \cref{eq:niche_network_stats_expression}, symbols are from simulations. Panel (c): Solid line shows the quantity $f$ in \cref{eq:niche_network_stats_definition}, and the symbols are again from computer generated networks. Simulation data in (b) and (c) are from a single realization of $\vb E$ with $N = 2000$.}
    \label{fig:NM_adjacency_matrices}
\end{figure}

Following Ref.~\cite{williamsSimpleRulesYield2000}, the niche value, range and centre of species $i$ are defined by the following steps:
\begin{enumerate}
    \item The niche value $\eta_i$ is a uniform random variable between $0$ and $1$. Species are re-labelled so that $\eta_1 < \eta_2 < \dots < \eta_N$. Species with high niche value are at the top of the hierarchy, and species with low niche value are at the bottom.
    \item The niche range is $d_i = \eta_i X_i$, where $X_i$ is a beta distributed random variable sampled independently for each species from the distribution $p(X_i) = \qty(\frac{1}{2d} - 1) (1 - X_i)^{\frac{1}{2d} - 2}$. The value of the parameter $d$ is the same for all species, and it is equal to the average niche range $d = \sum_id_i/N$ when $N$ is large. $d$ takes values in the interval $\qty[0, \frac{1}{2}]$.
    \item The niche centre $c_i$ is uniformly sampled in the range $[d_i/2, \eta_i]$ if $\eta_i + d_i/2 < 1$, and sampled in the range $[d_i/2, 1 - d_i/2]$ if $\eta_i + d_i/2 \geq 1$. By definition, the niche centre of a species is always smaller than that species' niche value, reflecting the idea that species generally `look down' the hierarchy for food.
\end{enumerate}
The structure of $\vb E$, and therefore of the niche model network, varies solely due to the average niche range $d$. A single realization of $\vb E$ is illustrated in \cref{fig:NM_adjacency_matrices}(a). 

Given a realization of the network $\vb E$, a random matrix constructed according to the niche model has elements $J_{ij} = E_{ij}K^L_{ij} + E_{ji}K^U_{ji}$, where $K^L_{ij}$ are independent random variables with mean $\mu_L/N$ and variance $\sigma_L^2/N$, and where $K^U_{ij}$ are also independent random variables, with mean $\mu_U/N$ and variance $\sigma_U^2/N$. The elements of $\vb K^L$ and $\vb K^U$ are also correlated via $N\cov_K(K^L_{ij}, K^U_{ij}) = \Gamma\sigma_L\sigma_U$.

For a fixed $\vb E$, the matrix $\vb J$ is therefore an FSRM, with statistics
\begin{align}
    u_{ij} &= \mu_L E_{ij} + \mu_U E_{ji}, \nonumber\\
    s_{ij} &= \sigma^2_L E_{ij} + \sigma^2_U E_{ji}, \nonumber\\
    t_{ij} &= \Gamma\sigma_L\sigma_U\qty(E_{ij} + E_{ji}). \label[plural_equation]{eq:niche_stats}
\end{align}
Elements of the adjacency matrix $A_{ij}$ are equal to one if either of $E_{ij}$ or $E_{ji}$ are equal to one. Otherwise, $A_{ij}=0$. We also assume that $\mu_L > \mu_U$, which implies that if $E_{ij} = 1$ and $E_{ji}=0$, then species $i$ benefits more from the relationship than species $j$ on average. If $E_{ij} = E_{ji} = 1$, then on average species $i$ and species $j$ benefit an equal amount from their mutual relationship. See Refs.~\cite{allesinaGeneralModelFood2008,williamsSimpleRulesYield2000} for discussion and motivation of the model.

In the following section we will compute the fine-structure correction using \cref{eq:niche_stats}, that is, we compute the fine-structure parameters for a fixed instance of the network. However, as we will see, the fine-structure correction depends only on a few summary statistics of $\{E_{ij}\}$.

\subsection{Eigenvalue spectrum}
\label{section:approximate_adjacency_matrix}
We now proceed to calculate the zeroth-order parameters in the fine structure [using \cref{eq:avg_params}]. They are
\begin{align}
    \mu &= d\qty(\mu_L + \mu_U), \nonumber\\
    \sigma^2 &= d\qty(\sigma_L^2 + \sigma_U^2), \nonumber\\
    \gamma &= \frac{2\Gamma\sigma_L\sigma_U}{\sigma_L^2 + \sigma_U^2}.\label[plural_equation]{eq:niche_elliptical_params}
\end{align}

The fine-structure correction [\cref{eq:fs_params}] can be expressed in terms of two additional properties of the underlying niche model network: the degree heterogeneity of the niche model network with adjacency matrix $\vb A$, and a property encoding the amount of asymmetry in the matrix $\vb E$
\begin{align}
    h^2 &\equiv \frac{1}{4d^2N^3}\sum_{ijk}\qty\Big(A_{ij}A_{jk} - 4d^2),\nonumber\\
    f &\equiv \frac{1}{4d^2N^3}\sum_{ijk}\qty\Big(E_{ij} - E_{ji})\qty\Big(E_{jk} - E_{kj}).\label[plural_equation]{eq:niche_network_stats_definition}
\end{align}
We also introduce the following parameters, defined in terms of $\mu_L, \mu_U, \sigma_L, \sigma_U$, which measure the amount of asymmetry in the interspecies interaction strengths and variances
\begin{align}
    \mu_1 &= d(\mu_L - \mu_U), \nonumber\\
    \sigma_1^2 &= d(\sigma_L^2 - \sigma_U^2).\label[plural_equation]{eq:niche_mu_1_sigma_1}
\end{align}

In terms of the aforementioned quantities, the fine-structure parameters read [see \cref{sm:section:niche_model} for details of the calculation]
\begin{align}
    R &= \gamma h^2, \nonumber\\
    S &= h^2 + \frac{\sigma_1^4}{\sigma^4}f, \nonumber\\
    T &= \gamma^2h^2, \nonumber\\
    U &= h^2 + \frac{\mu_1^2}{\mu^2}f, \nonumber\\
    V &= \gamma h^2. \label[plural_equation]{eq:niche_fs_params}
\end{align}
Therefore, the fine-structure correction to the elliptical law and associated outlier eigenvalue only depend on three characteristics of the niche model network: $d, h^2$ and $f$, and not on any particular instance of $\vb E$. If $\sigma_1 = \mu_1 = 0$, then the niche model is a special case of the FSRMs in \cref{section:directed_network} where the adjacency matrix of the underlying network is symmetric ($r=1$). In this case \cref{eq:network_fs_params} and \cref{eq:niche_fs_params} give the same results.

Plugging the elliptical [\cref{eq:niche_elliptical_params}] and fine-structure [\cref{eq:niche_fs_params}] parameters into \cref{eq:bulk_approx,eq:outlier_approx} gives the fine-structure correction to the support of the spectrum of a random matrix constructed according to the niche model. In particular, we can analyze the combined effect of the network and hierarchical interactions on stability. We find the following correction to the bulk edge 
\begin{align}
    \frac{\edge}{\sigma} =& \qty(1 + \gamma)\qty[1 + \frac{1}{2}\qty(1 + 2\gamma - \gamma^2)h^2]\nonumber \\
    &+ \frac{1}{2}\qty(1 - \gamma)\frac{\sigma_1^4}{\sigma^4}f, \label{eq:niche_edge}
\end{align}
and the correction to the outlier eigenvalue is
\begin{multline}
    \frac{\outlier}{\mu} = \qty(1 + \frac{\gamma\sigma^2}{\mu^2})\qty(1 + h^2) 
    + \qty(1 - \frac{\gamma\sigma^2}{\mu^2})\frac{\mu_1^2}{\mu^2}f.\label{eq:niche_outlier}
\end{multline}
We note that the quantity $f$ is always negative. Recalling the restriction $|\gamma| \leq 1$, and that there is no outlier if $\mu/\sigma \leq 1 + \delta$, where $\delta$ vanishes as the fine structure vanishes, we see that increased asymmetry in both the interaction strengths (i.e. increased $\mu_1$ and $\sigma_1^2$) and network structure (i.e. more negative $f$) reduce $\edge$ and $\outlier$. Hence, increased asymmetry is always stabilizing. The effect of degree heterogeneity is more complicated. If there are no asymmetrical interactions ($\mu_1 = \sigma_1 = 0$), then larger degree heterogeneity always increases the value of $\outlier$ and increases $\edge$ provided $\gamma > 1 - \sqrt{2}$. If there are asymmetrical interactions, then increased degree heterogeneity can be stabilizing or destabilizing, as it increases the amount of symmetrical and asymmetrical fine structure in the matrices $\vb u$ and $\vb s$.

As the network structure ultimately depends only on the average niche range $d$, the parameters $h^2$ and $f$ must be functions of $d$ only. In \cref{sm:section:niche_model} of the SM, we find approximate expressions for $h^2$ and $f$, valid when $d$ is small, but not so small as to make the network sparse
\begin{align}
    h^2 &\approx \frac{2 - d - 2d^2}{6(1 + 3d)},\nonumber\\
    f &\approx - h^2 - \frac{1}{3(1 + d)^2}. \label[plural_equation]{eq:niche_network_stats_expression}
\end{align}
\Cref{eq:niche_network_stats_expression} are compared to the measured values [obtained from generating $\vb E$ according to rules (1) to (4)] in \cref{fig:NM_adjacency_matrices}. We find good numerical agreement for all possible values of $d$, not just small ones.

The validity of our results, as well as our assertions on the effect of $\mu_1$ on stability, are verified in \cref{fig:NM_spectrum} for varying strength of hierarchy between species, where we see that increasing $\mu_1$ is indeed stabilizing. The validity of our assertions on the effect of the degree heterogeneity of the network are verified in \cref{sm:fig:niche_varying_d} of the SM. The performance of our approximations is surprisingly good considering that the size of the fine-structure correction can be as large as $\approx 70\%$ of the uncorrected value. \Cref{fig:NM_adjacency_matrices} (b) also reveals that the degree heterogeneity is always non-zero, and has a minimum value of approximately $0.1$. Hence, any combination of the model parameters leads to a non-vanishing fine-structure correction.

\section{Conclusions}
The elliptical law carries information about the stability of a densely-connected disordered system in which all components are statistically equivalent. However, in many cases, we have some knowledge about the system's structure and want to know how this feature affects its stability. In this work, we have provided an explicit formula for a large subset of such models that answers this question, so long as the effect of the additional structure is small.

Further, our modified elliptical law reveals that antisymmetry in the statistics (as opposed to the elements themselves) of an FSRM is stabilizing, and symmetry is usually destabilizing. We have seen how this observation manifests itself in a number of examples. For instance, by interpreting hierarchy in an ecosystem as antisymmetric fine structure in the statistics of the associated random-matrix models, we are able to conclude that hierarchy is a stabilizing factor. Also, by interpreting the degree heterogeneity of a network as symmetric fine structure, we can conclude that degree heterogeneity is usually a destabilizing factor.

There are several possible avenues for future work. We have treated fine structure as a perturbation to the elliptic law, but there are other aspects which could be treated similarly. For example, the corrections to the elliptic law due to a sparse network structure \cite{baronPathIntegralApproach2023}, or additional correlations \cite{aceitunoUniversalHypotrochoidicLaw2019, baron_generalised_2022} have been addressed in this way. Combining such corrections using a perturbative approach is interesting prospect for future work. Alternatively, we could consider the fine-structure corrections to the stability criteria of some non-linear dynamical system with an FSRM as its interaction matrix, such as the generalized Lotka-Volterra equations \cite{poleyGeneralizedLotkaVolterraModel2023}. 

A further interesting possibility for future work would be to drop the restriction that all mean values of the matrix elements, $\avg{J_{ij}}_J$, scale with $1/N$, instead allowing for some to be of order $1$, perhaps by imposing conditions similar to those in Ref.~\cite{rogersUniversalSumProduct2010}. We could then consider systems where agents interact with some agents very strongly, and with others randomly. For example, many ecological models include self-regulation of species (see, e.g., Refs.~\cite{baronDispersalinducedInstabilityComplex2020,barabas2017self,deangelisStabilityConnectanceFood1975,vanaltenaFoodWebStability2016,barbierGenericAssembltPatterns2018}), which could be incorporated into an FSRM for which the diagonal elements of $\vb u$ are of order $1$, rather than of order $1/N$.

Finally, we remark that one can compute all parameters necessary for the modified elliptical law with just one instance of the FSRM $\vb J$. This opens up the possibility of applications of the modified elliptical law to situations with real-world data, in a similar spirit to uses of the standard elliptical law \cite{coyteEcologyMicrobiomeNetworks2015}.

\acknowledgments
Financial support has been received from the Agencia Estatal de Investigación (AEI, MCI, Spain) MCIN/AEI/10.13039/501100011033 and Fondo Europeo de Desarrollo Regional (FEDER, UE) under Project APASOS (PID2021-122256NB-C21/C22) and the María de Maeztu Program for units of Excellence in R\&D, grant CEX2021-001164-M. LP acknowledges funding by the Engineering and Physical Sciences Research Council UK, grant number EP/T517823/1. JWB is supported by grants from the Simons Foundation (\#454935 Giulio Biroli).

\newpage

\setcounter{section}{0}		%Resets section numbers
\setcounter{page}{1}		%Resets page numbers
\setcounter{equation}{0}	%Resets equation numbers
\setcounter{figure}{0}		%Resets figure numbers
\setcounter{table}{0}		%Resets table numbers
\renewcommand{\thesection}{S\arabic{section}} 		%Adds "S" before arabic number of sections
\renewcommand{\thepage}{S\arabic{page}} 			%Adds "S" before arabic number of pages
\renewcommand{\theequation}{S\arabic{equation}}  	%Adds "S" before arabic number of equations
\renewcommand{\thefigure}{S\arabic{figure}}  		%Adds "S" before arabic number of figures
\renewcommand{\thetable}{S\arabic{table}}  

\begin{widetext}
{\begin{center}{\Large ------ Supplemental Material ------}\end{center}}
\setlength{\parskip}{-0.6pt}
\setlength{\parindent}{0pt}
\tableofcontents

\setlength{\parskip}{8pt}
\setlength{\parindent}{0pt}
% this line marks the end of the contents that wont be included in the contents. All sections below this command are included in the table of contents
\let\addcontentsline=\oldaddcontentsline
\let\nocontentsline=\oldaddcontentsline

\section*{Overview}

This document contains additional details of the calculations presented in the main text, as well as additional figures and examples of FSRM eigenvalue spectra.

In \cref{sm:section:The_bulk_spectrum}, we compute the disorder average of the resolvent matrix of the $N\times N$ finely-structured random matrix (FSRM) $\vb J$ [\cref{eq:resolvent_definition} in the main text] by means of a saddle point approximation for large $N$ of the so-called eigenvalue potential \cite{sommersSpectrumLargeRandom1988}. Given that the disorder-averaged resolvent matrix is diagonal, we then derive a set of self-consistent equations that relate the diagonal elements of the resolvent matrix, the statistics of the FSRM ensemble, and the support of the spectrum of $\vb J$ in \cref{sm:section:self_consistent_equations}.

In \cref{sm:section:approximate_self_consistent_equations}, we expand the self-consistent equations derived in \cref{sm:section:self_consistent_equations} to second order in a small parameter $\epsilon$, which measures the amount of fine structure present in the system. We then solve these simplified equations explicitly in \cref{sm:section:Modified_elliptical_law}, deriving our modified elliptical law and outlier eigenvalue as consequences [\cref{eq:bulk_approx,eq:outlier_approx} in the main text].

In \cref{sm:section:effect_f_fs_on_stability}, we use the modified elliptical law and outlier eigenvalue to justify the claims made in \cref{section:small_fs_stability} of the main text. In particular, we demonstrate that antisymmetric fine structure in the statistics of a FSRM generally stabilizes the system, and symmetric fine structure largely destabilizes the system unless $\gamma < -1/3$. If $\gamma < -1/3$ then symmetric fine structure can stabilize or destabilize the system.

In \cref{sm:section:explicit_solutions_for}, we analyze a number of FSRM ensembles for which the self-consistent equations for the support of the spectrum can be solved exactly. Specifically, in \cref{sm:section:the_case_T_0} we present an exact solution in the case where the fine-structure parameter $T = 0$. This class of FSRM models includes the elliptical law, the cascade model of Ref.~\cite{allesinaPredictingStabilityLarge2015}, and the directed networks model of Ref.~\cite{neriLinearStabilityAnalysis2020} as special cases. In \cref{sm:section:cascade_model} we find the support of the spectrum for random matrices constructed according to the cascade model as presented in \cref{section:exact_solution} of the main text.

In \cref{sm:section:verification_of_MEL}, we demonstrate that the modified elliptical law gives the same result as a direct expansion of the exact result in the case of the cascade model of \cref{section:exact_solution} in the main text, verifying our small-fine-structure correction analytically. 

Finally, in \cref{sm:section:approximate_solutions_for} we detail the computation of the elliptical and fine-structure parameters for the models presented in \cref{section:directed_network,section:niche_model,section:neural_networks} of the main text. Specifically, we compute the parameters $\mu, \sigma, \gamma, R, S, T, U$ and $V$ as defined in \cref{eq:avg_params,eq:fs_params} of the main text for a directed network, a neural network model and for the niche model from theoretical ecology. 
\newpage

\section{The bulk spectrum of \texorpdfstring{$J$}{}}
\label{sm:section:The_bulk_spectrum}
\subsection{The eigenvalue potential}
To find the spectrum of an $N\times N$ finely-structured matrix $\vb J$ with element specific statistics, we calculate the disorder averaged resolvent matrix, given by \cref{eq:resolvent_definition} in the main text. It is well-known that the density of the bulk region can be derived from the disorder averaged resolvent of $\vb J$ via [see Ref.~\cite{feinbergNonHermiteanRandomMatrix1997} for detailed derivations of \cref{sm:eq:G_to_rho,sm:eq:eigenvalue_potential,sm:eq:phi_to_G}]
\begin{align}
    \rho_\text{bulk}(z) = \frac{1}{\pi N}\overline{\partial}_z\Tr \vb G(z), \label{sm:eq:G_to_rho}
\end{align}
where $\overline{\partial}_z = (\partial_x + i\partial_y)/2$ and $\Tr$ denotes the trace.

From \cref{sm:eq:G_to_rho}, we see that the eigenvalue density is non-zero when $\Tr \vb G(z)$ is a non-analytic function of $z$. Hence, the support of $\rho_\text{bulk}(z)$ is the boundary of the region in the complex plane for which $\Tr \vb G(z)$ is an analytic function. Following \cite{sommersSpectrumLargeRandom1988}, the problem of computing $\Tr \vb G(z)$ can be tackled by considering the eigenvalue potential
\begin{align}
    \Phi(z, \conj z) \equiv -\frac{1}{N}\avg{\ln \det \qty(\conj z\vb I - \vb J^\dagger)\qty(z\vb I - \vb J)}_J, \label{sm:eq:eigenvalue_potential}
\end{align}
where $\conj z$ denotes the complex conjugate of $z$. The potential is related to the resolvent matrix through
\begin{align}
    \partial_z\Phi(z, \conj z) = \Tr \vb G(z). \label{sm:eq:phi_to_G}
\end{align}
In the large-$N$ limit, we will compute $\Phi(z, \conj z)$ using the replica trick together with a saddle point approximation \cite{sommersSpectrumLargeRandom1988,stratonovichMethodCalculatingQuantum1957a,hubbardCalculationPartitionFunctions1959,haakeStatisticsComplexLevels1992,baron_generalised_2022}. The result is ultimately a set of self-consistent equations which determine the resolvent matrix $\vb G(z)$, as well as spectral density $\rho_\text{bulk}(z)$.

For the calculation of the eigenvalue potential it is convenient to write the FSRM $\vb J$ [with statistics given by \cref{eq:fsrm_stats} in the main text] as the sum of a mean and a fluctuating part
\begin{align}
    J_{ij} = \frac{u_{ij}}{N} + \sqrt{\frac{s_{ij}}{N}}w_{ij},\label{sm:eq:J_statistics}
\end{align}
where the mean zero random variables $w_{ij}$ satisfy
\begin{align}
    \avg{w_{ij}^2}_J &= 1, \nonumber\\
    \avg{w_{ij}w_{ji}}_J &= \gamma_{ij}.
\end{align}
The correlation parameters $\gamma_{ij}$ are related to the covariance parameters $t_{ij}$, defined in \cref{eq:fsrm_stats} in the main text, via $t_{ij} = \gamma_{ij}\sigma_{ij}\sigma_{ji}$.

Similarly to \cite{baronDispersalinducedInstabilityComplex2020,sommersSpectrumLargeRandom1988,haakeStatisticsComplexLevels1992}, we evaluate the eigenvalue potential $\Phi(z, \conj z)$ using replicas. This involves using the identity $\ln x = \lim_{n\to 0}\qty(x^n - 1)/n$, but assuming that $n$ is an integer. This reduces the problem of computing the disorder average of a logarithm to the problem of computing the disorder average of the 'replicas' $\avg{x^n}_J$. 

In previous works, it has been shown that the replicas decouple \cite{sommersSpectrumLargeRandom1988,baronDispersalinducedInstabilityComplex2020,haakeStatisticsComplexLevels1992,baron_directed_2022}, that is, that $\avg{x^n}_J = \avg{x}^n_J$. Hence, if the replicas decouple, then the average of the logarithm is the logarithm of the average. In this work, we assume that the replicas decouple, so that 
\begin{align}
    \Phi(z, \conj z) = -\frac{1}{N}\ln \avg{\det \qty(\conj z\vb I - \vb J^\dagger)\qty(z\vb I - \vb J)}_J, \label{eq:eigenvalue_potential}
\end{align}
We will also follow \cite{baron_directed_2022,orourkeLowRankPerturbations2014,taoOutliersSpectrumIid2014,benaych-georgesEigenvaluesEigenvectorsFinite2010} and assume that if the rank of the matrix $\vb u$ is finite as $N$ increases, then it has an $\order{1/N}$ contribution to the bulk spectrum and can therefore be set to zero for the derivation of the bulk spectrum. Whilst the elements $u_{ij}$ turn out to have no effect on the bulk spectrum, we will see in section \cref{sm:section:outliers} that they are important in determining the location of possible outlier eigenvalues.
\subsection{Disorder average}
To carry out the average of the determinant in \cref{eq:eigenvalue_potential} over realizations of $\vb J$, we first express it as the following Gaussian integral over variables $p_{i = 1, \dots, N}$
\begin{align}
    \exp[N\Phi(z)] &= \avg{\int \prod_i \frac{\dd[2]{p_i}}{\pi} \exp[- \sum_{ij} \conj{p_i} (\conj z \delta_{ij} - J_{ji})(z \delta_{ij} - J_{ij})p_j]}_J, \nonumber\\
    &= \avg{\int \prod_{ij} \frac{\dd[2]{p_i}\dd[2]{q_j}}{\pi^2} \exp[- \sum_{i}\conj{q_i} q_i + i\sum_{ij} \conj{p_i} (\conj z \delta_{ij} - J_{ji})q_j + \conj{q_i}(z \delta_{ij} - J_{ij})p_j]}_J,
\end{align}
where the second equality follows from a Hubbard Stratonovich transformation \cite{hubbardCalculationPartitionFunctions1959,stratonovichMethodCalculatingQuantum1957a}. Substituting \cref{sm:eq:J_statistics}, the ensemble average of the disordered part of $\vb J$ can be computed by taylor expanding the exponent in powers of $1/N$. One can then re-exponentiate the result, assuming that higher moments decay sufficiently fast with the system size. Focusing on the parts of the exponent which contain the disordered terms (terms containing the random variable $w_{ij}$) we have
\begin{align}
    \Bigg\langle\exp \Bigg[-\frac{i}{\sqrt{N}} \sum_{ij}\conj{p_i}\sqrt{s_{ji}}w_{ji}q_j + \conj{q_i}\sqrt{s_{ij}}w_{ij}p_i\Bigg]\Bigg\rangle _J  
    &\approx 1 - \frac{1}{2N}\avg{\qty(\sum_{ij}\conj{p_i}\sqrt{s_{ji}}w_{ji}q_j + \conj{q_i}\sqrt{s_{ij}}w_{ij}p_i)^2}_J, \nonumber \\
    &= 1 - \frac{1}{2N} \sum_{ij}s_{ij} \qty(q_i\conj{p_j} + \conj{q_i}p_j)^2 + t_{ij}\qty(p_i\conj{q_j} + \conj{p_i}q_j)\qty(p_j\conj{q_i} + \conj{p_j}q_i), \nonumber \\
    &\approx \exp[ - \frac{1}{2N} \sum_{ij}s_{ij} \qty(q_i\conj{p_j} + \conj{q_i}p_j)^2 + t_{ij}\qty(p_i\conj{q_j} + \conj{p_i}q_j)\qty(p_j\conj{q_i} + \conj{p_j}q_i)].\label{sm:eq:disorder_average}
\end{align}
Now that we have averaged over realizations of $\vb J$, our expression for the eigenvalue potential $\Phi(z, \conj z)$ reads
\begin{align}
    \exp[N\Phi(z)] = & \int \prod_{ij} \frac{\dd[2]{p_i}\dd[2]{q_j}}{\pi^2} \exp[ - \sum_{i}\conj{q_i} q_i
    -i\sum_{i}\qty(\conj{p_i}q_i\conj z + p_i\conj{q_i} z)]\nonumber\\
    &\times\exp[ - \frac{1}{2N} \sum_{ij}s_{ij} \qty(q_i\conj{p_j} + \conj{q_i}p_j)^2 + t_{ij}\qty(p_i\conj{q_j} + \conj{p_i}q_j)\qty(p_j\conj{q_i} + \conj{p_j}q_i)]. \label{sm:eq:eval_potential_1}
\end{align}
We can perform the integral in \cref{sm:eq:eval_potential_1} for large $N$ by transforming the integrand into a form amenable to a saddle point approximation. The transformation involves introducing the following parameters
\begin{align}
    P_i &= \conj{p_i}p_i,  &Q_i &= \conj{q_i}q_i, \nonumber\\
    R_i &= \conj{p_i}q_i,  &R_i^* &= p_i \conj{q_i}.\label[plural_equation]{sm:eq:macro_parameters}
\end{align}
Following \cite{baron_directed_2022,sommersSpectrumLargeRandom1988,edwardsEigenvalueSpectrumLarge1976}, we neglect terms involving $\qty(p_i)^2, \qty(q_i)^2, p_iq_i$ and $\conj{p_i}\conj{q_i}$ as these do not ultimately contribute to the integral to leading order in $N$. We introduce $P_i, Q_i, R_i, R^*_i$ into \cref{sm:eq:eval_potential_1} by inserting Dirac deltas in their complex exponential form
\begin{align}
    \delta\qty(P_i - \conj{p_i}p_i) &\propto \int d\wh P_i \exp[i\wh P_i\qty(P_i - \conj{p_i}p_i)],\nonumber\\
    \delta\qty(Q_i - \conj{q_i}q_i) &\propto \int d\wh Q_i \exp[i\wh Q_i\qty(Q_i - \conj{q_i}q_i)],\nonumber\\ 
    \delta\qty(R_i - \conj{p_i}q_i) &\propto \int \dd[2]{\wh R_i} \exp[i\wh R^*_i\qty(R_i - \conj{p_i}q_i) + i\wh R_i\qty(R^*_i - p_i\conj{q_i})],
\end{align}
where we have ignored the pre-factors involved in expressing the deltas as complex exponentials because they only contribute a constant factor to the eigenvalue potential, which does not affect the spectrum.

With the complex exponentials inserted, our expression for the eigenvalue potential now reads
\begin{align}
    &\exp[N\Phi(z, \conj z)] = \int \prod_i \dd{P_i}\dd{\wh P_i} \dd{Q_i}\dd{\wh Q_i} \dd[2]{R_i}\dd[2]{\wh R_i} \nonumber\\
    &\times\exp[N\qty{-\frac{1}{N}\sum_i Q_i - \frac{1}{N^2}\sum_{ij}\qty(s_{ij} P_j Q_i + \frac 12 t_{ij} \qty(R_iR_j + R_i^*R_j^*)) - \frac{i}{N}\sum_i\qty(\conj z R_i + z R_i^*)}] \nonumber\\
    &\times\exp[N\qty{\frac{i}{N}\sum_i\qty(\wh P_iP_i + \wh Q_iQ_i + \wh R_iR_i^* + \wh R_i^*R_i)}] \nonumber\\
    % %
    &\times\exp[N\sum_i\ln\qty{\int\frac{\dd[2]{p_i}\dd[2]{q_i}}{\pi^2}\exp(i\conj{p_i}p_i~ \wh P_i + \conj{q_i}q_i~ \wh Q_i + \conj{p_i}q_i~ \wh R_i^*  + p_i\conj{q_i}~ \wh R_i )}].\label{sm:eq:eval_potential_2}
\end{align}
The integral in the final term is a standard Gaussian integral, it can be explicitly evaluated
\begin{align}
    \int\frac{\dd[2]{p_i}\dd[2]{q_i}}{\pi^2}\exp(i\conj{p_i}p_i~ \wh P_i + \conj{q_i}q_i~ \wh Q_i + \conj{p_i}q_i~ \wh R_i^*  + p_i\conj{q_i}~ \wh R_i ) = \frac{1}{\wh R_i \wh R^*_i - \wh P_i \wh Q_i} \equiv K_i.
\end{align}
As the integrand in \cref{sm:eq:eval_potential_2} is now of order $\exp(N)$, we can evaluate the integrals over the parameters $P_i, Q_i, R_i, R^*_i$ and their hatted counterparts with a saddle point approximation. For the hatted parameters, we arrive at the following saddle point equations 
\begin{align}
    iP_i &= -\wh Q_iK_i, \nonumber\\
    iQ_i &= -\wh P_iK_i, \nonumber\\
    iR_i &= \wh R_iK_i, \nonumber\\
    iR_i^* &= \wh R_i^*K_i, \label[plural_equation]{sm:eq:saddle_PQRR*}
\end{align}
and for the unhatted parameters
\begin{align}
    i\wh P_i &= \frac{1}{N}\sum_j Q_j s_{ji}, \nonumber\\
    i \wh Q_i &= 1 + \frac{1}{N}\sum_j s_{ij} P_j, \nonumber\\
    i\wh R_i &= \frac{1}{N}\sum_j t_{ij} R_j^* + i z, \nonumber\\
    i\wh R_i^* &= \frac{1}{N}\sum_j t_{ij} R_j + i \conj z.\label[plural_equation]{sm:eq:saddle_PQRR*_hat}
\end{align}
Using \cref{sm:eq:saddle_PQRR*}, we find 
\begin{align}
    K = P_iQ_i - R^*_iR_i.
\end{align}
Eliminating the hatted parameters from \cref{sm:eq:saddle_PQRR*_hat,sm:eq:saddle_PQRR*}, we arrive at.
\begin{align}
    Q_i &= \frac{1}{N}\sum_j Q_js_{ji} K_i,\label{sm:eq:Q} \\
    P_i &= \qty(1 + \frac{1}{N}\sum_js_{ij}P_j)K_i, \label{sm:eq:P}\\
    R_i &= \qty(iz + \frac{1}{N}\sum_jt_{ij}R^*_j)K_i, \label{sm:eq:R1}\\
    R_i^* &= \qty(i\conj z + \frac{1}{N}\sum_jt_{ij}R_j)K_i.\label{sm:eq:R2}
\end{align}
Similarly to Refs.~\cite{baron_directed_2022,baronDispersalinducedInstabilityComplex2020}, we can relate the saddle point values of $R_i, R^*_i$ to the diagonal elements of the resolvent matrix $G_i(z)$. First, we suppose that the eigenvalue potential depended not just on one value of $z$, but on a diagonal matrix $\vb z$ of complex variables with elements $z_i\delta_{ij}$
\begin{align}
    \Phi(\vb z) \equiv -\frac{1}{N}\avg{\ln \det \qty(\vb{z} - \vb J)^\dagger\qty(\vb{z} - \vb J)}_J. \label{eq:eigenvalue_potential_alternative}
\end{align}
Following the derivation of \cref{sm:eq:eval_potential_2} with this alternative definition of the eigenvalue potential, the only difference is that we replace $z \to z_i$ and $\conj z \to \conj z_i$ throughout. Comparing the partial derivative of the alternative potential with respect to $z_i$ in \cref{eq:eigenvalue_potential_alternative,sm:eq:eval_potential_2} gives 
\begin{align}
    \partial_{z_i} \Phi(\vb z) = \frac{1}{N}G_i(z) = \frac{1}{N}iR^*_i, \nonumber\\
    \partial_{\conj z_i} \Phi(\vb z) = \frac{1}{N}\conj G_i(z) = \frac{1}{N}iR_i. \label[plural_equation]{sm:eq:G_to_RR*}
\end{align}
As the eigenvalue potential of the FSRM $\vb J$ could be determined from this alternative eigenvalue potential with $z_i = z$, we conclude that the relations in \cref{sm:eq:G_to_RR*} hold for our original eigenvalue potential.

Eliminating $R_i$ and $R_i^*$ in \cref{sm:eq:R1,sm:eq:R2} for $G_i$, we arrive at the following system of equations for the diagonal elements of the resolvent matrix 
\begin{align}
    \overline{G_i} = \qty(z - \frac{1}{N}\sum_jt_{ij}G_j)K_i, \label{sm:eq:G_general}
\end{align}
with $K_i = P_iQ_i + |G_i|^2$. Hence, to find the spectrum of the matrix $\vb J$, one must first simultaneously solve \cref{sm:eq:Q,sm:eq:P,sm:eq:G_general} for the diagonal resolvent elements $G_i$. The spectral density can then be computed according to \cref{sm:eq:G_to_rho}.
\section{Self-consistent equations for the support of the spectrum}\label{sm:section:self_consistent_equations}
In this section we show how \cref{sm:eq:Q,sm:eq:P,sm:eq:G_general} simplify if we are only interested in knowing the boundary of the support of the bulk spectrum, and not the density of eigenvalues inside. We then derive a similar condition which determines the outlier eigenvalues.
\subsection{Boundary of the bulk spectrum}
\label{sm:section:boundary_of_the_bulk_spectrum}
There are two solutions to \cref{sm:eq:Q,sm:eq:P,sm:eq:G_general}, corresponding to $Q_i = 0$ and $Q_i \neq 0$. When $Q_i = 0$, we have $K_i = |G_i|^2$ and \cref{sm:eq:G_general} reduces to
\begin{align}
    \frac{1}{G_i(z)} &= z - \frac{1}{N}\sum_jt_{ij}G_j(z).
\end{align}
Hence, the diagonal elements of the resolvent are functions of $z$ only, and not of $\conj z$. $G_i$ is therefore an analytic function and the spectral density is zero. Therefore, the solution for which $Q_i = 0$ corresponds to the region of the complex plane outside the bulk spectrum. Conversely, if $Q_i \neq 0$, the trace of the resolvent matrix is a non-analytic function as \cref{sm:eq:G_general} contains both $G_i$ and its complex conjugate. Therefore, the eigenvalue density is non-zero, and we are in the bulk region of the spectrum of $\vb J$. 

Consequently, to find the support of the bulk spectrum we seek the set of points $\bulk$ where the two solutions to \cref{sm:eq:P,sm:eq:Q,sm:eq:G_general} meet. The solutions are distinguished by whether $Q_i$ is positive or zero, so we write $Q_i \to \delta Q_i$ for small positive $\delta > 0$ and expand to leading order in $\delta$, finding
\begin{align}
    Q_i &= |G_i(\bulk)|^2\frac{1}{N}\sum_jQ_j s_{ji}, \label{sm:eq:Q2}\\
    P_i &=  |G_i(z)|^2\qty(1 + \frac{1}{N}\sum_js_{ij}P_j) , \label{sm:eq:P2}\\
    \frac{1}{G_i(z)} &= z - \frac{1}{N}\sum_jt_{ij}G_j(z),\label{sm:eq:resolvent_block}
\end{align}
Both \cref{sm:eq:P2,sm:eq:resolvent_block} are valid outside the bulk region, including on the boundary, and \cref{sm:eq:Q2} is valid only on the bulk boundary. \Cref{sm:eq:resolvent_block} determines $G_i(z)$ outside the bulk region.

Because $Q_i\geq 0$ [see \cref{sm:eq:macro_parameters}], \cref{sm:eq:Q2} implies that the vector $[\vb Q]_i  = Q_i$ is the left Perron-Frobenius (PF) eigenvector of the matrix $[\vb s\vb G(\bulk)\conj{\vb G}(\bulk)]_{ij}/N = s_{ij}\qty|G_{j}(\bulk)|^2/N$ with eigenvalue $1$. We can also manipulate \cref{sm:eq:P2} into an eigenvalue equation. By dividing \cref{sm:eq:P2} by the positive quantity $p \equiv \frac{1}{N}\sum_iP_i$ and defining $B_i' = P_i/p$, we obtain
\begin{align}
    B'_i = |G_i(z)|^2\frac{1}{N}\sum_j\qty(p + s_{ij})B'_j.\label{sm:eq:bulk_block_2}
\end{align}
Now, because the vector $\vb B'$ has all positive entries [see \cref{sm:eq:macro_parameters}], it must be the right PF eigenvector of the positive matrix with elements $[\vb G(z)\conj{\vb G}(z)\qty(p\vb 1 + \vb s)]_{ij}/N = |G_i(z)|^2\qty(p + s_{ij})/N$ for some positive number $p$, with eigenvalue $1$. Writing $\PF\qty[\vb M]$ for the PF eigenvalue of a matrix $\vb M$ with all entries positive, we can write \cref{sm:eq:Q2,sm:eq:P2} as 
\begin{align}
    \PF\qty[\frac{1}{N}\vb s \vb G\qty(\bulk)\conj{\vb G}\qty(\bulk)] = 1,\label{sm:eq:bulk_boundary}\\
    \PF\qty[\frac{1}{N}\qty(p\vb 1 + \vb s)\vb G(z) \conj{\vb G}(z)] = 1,\label{sm:eq:outlier_ellimination_0}
\end{align}
where $\vb 1$ is an $N\times N$ matrix of ones. It is a known fact that if any one of a positive matrices elements increases, then so too does the PF eigenvalue of that matrix \cite{cohenDerivativesSpectralRadius1978,friedlandInequalitiesSpectralRadius1975,deutschDerivativesPerronRoot1984}. Therefore, the RHS of \cref{sm:eq:outlier_ellimination_0} is an increasing function of $p$. It has a minimum when $p = 0$ with value $\PF\qty[\vb s \vb G(z) \conj{\vb G}(z)/N]$ and a maximum value of $\infty$. We can therefore combine \cref{sm:eq:bulk_boundary,sm:eq:outlier_ellimination_0} into a single condition. Points $z$ outside the bulk spectrum all satisfy
\begin{align}
    \PF\qty[\frac{1}{N}\vb s \vb G\qty(z)\conj{\vb G}\qty(z)] \leq 1, \label{sm:eq:outlier_ellimination_2}
\end{align}
with equality only on the boundary of the bulk spectrum.

\subsection{Outliers}
\label{sm:section:outliers}
So far, we have derived conditions determining the diagonal elements of the resolvent matrix outside the bulk region, as well as the boundary of the bulk region itself. In our derivation of these conditions, the matrix $\vb u$ played no part and was effectively set to $0$ \cite{taoOutliersSpectrumIid2014,baronDispersalinducedInstabilityComplex2020}. Outlier eigenvalues are isolated eigenvalues of the FSRM $\vb J$ which lie outside the bulk spectrum. Writing $\vb J$ as sum of its deterministic and random part as in \cref{sm:eq:J_statistics}, the outlier eigenvalues $\outliers{k}$ are defined by 
\begin{align}
    \det[\outliers{k}\vb I - \vb J_0 - \frac{1}{N}\vb{u}] = 0,\label{sm:eq:outlier_0}
\end{align}
where $\vb J_0$ is equal to the random matrix $\vb J$ with the mean values $\vb u$ set to zero, $[\vb J_0]_{ij} = w_{ij}\sqrt{s_{ij}/N}$. Since $\outliers{k}$ is outside the bulk region, it is not an eigenvalue of $\vb J_0$. Therefore, the matrix $\outliers{k}\vb I - \vb J_0$ is invertible, and its inverse is the resolvent $[\vb G(z)]_{ij} = \delta_{ij}G_i(z)$, where $G_i(z)$ are the diagonal elements of the resolvent matrix of $\vb J$. We note that in this step we have assumed that the resolvent is self-averaging property of the FSRM assemble to equate a single realization of the resolvent of $\vb J$ with its disorder average. That is, we have assumed that
\begin{align}
    \qty(\outliers{k}\vb I - \vb J_0)^{-1} \approx \avg{\qty(\outliers{k}\vb I - \vb J_0)^{-1}}_J = \vb G\qty(\outliers{k}).
\end{align}
Multiplying \cref{sm:eq:outlier_0} by $\det[\vb G]$, we obtain
\begin{align}
    \det[\vb I - \frac{1}{N} \vb u\vb G\qty(\outliers{k})] = 0.\label{sm:eq:outliers}
\end{align}
Hence, to find outlier eigenvalues, we first solve \cref{sm:eq:resolvent_block} for the diagonal elements of the resolvent matrix $G_i(z)$, which is valid for all $z$ outside the bulk of the eigenvalue spectrum. Then, we substitute $G_i(z)$ into \cref{sm:eq:outliers} and enumerate all solutions. Finally, we must check that the solutions to \cref{sm:eq:outliers} actually lie outside the bulk spectrum, which we can do with \cref{sm:eq:outlier_ellimination_2}.
\subsection{Summary}
\label{sm:section:self_consistent_equations_summary}
We now summarize and repeat the relevant equations derived in \cref{sm:section:self_consistent_equations,sm:section:The_bulk_spectrum}. These equations self-consistently determine the boundary of the bulk spectrum and outlier eigenvalues of a FSRM with statistics given by \cref{eq:fsrm_stats} in the main text, we repeat them here
\begin{align}
    \avg{J_{ij}}_J &= \frac{u_{ij}}{N}, \nonumber\\
    \variance_J\qty(J_{ij}) &= \frac{s_{ij}}{N}, \nonumber\\
    \cov_J\qty(J_{ij}, J_{ji}) &= \frac{t_{ij}}{N}.
\end{align}
First, we find the diagonal elements of the resolvent matrix (valid for complex numbers $z$ outside the bulk spectrum) from the following equation 
\begin{align}
    \frac{1}{G_i(z)} = z - \frac{1}{N}\sum_jt_{ij}G_j(z). \label{sm:eq:resolvent_general}
\end{align}
The boundary of the bulk spectrum comprises the set of solutions $\bulk$ to the following equation
\begin{align}
    \PF\qty[\frac{1}{N}\vb s \vb G\qty(\bulk)\conj{\vb G}\qty(\bulk)] = 1,\label{sm:eq:bulk_boundary_general}
\end{align}
where $[\vb G(z)]_{ij} = \delta_{ij}G_i(z)$ is the diagonal resolvent matrix, the overline indicates element wise complex conjugation, and $\PF[\vb M]$ is the Perron-Frobenius eigenvalue of the matrix $\vb M$ with positive entries. The outlier eigenvalues are given by the simultaneous solutions $\outliers{k}$ of the following equation and inequality
\begin{align}
    \det\qty[\vb I - \frac{1}{N}\vb u\vb G\qty(\outliers{k})] = 0,\label{sm:eq:outliers_general}\\
    \PF\qty[\frac{1}{N}\vb s \vb G\qty(\outliers{k})\conj{\vb G}\qty(\outliers{k})] < 1.\label{sm:eq:outlier_ellimination_general}
\end{align}
For explicit examples of \cref{sm:eq:outliers_general,sm:eq:resolvent_general,sm:eq:bulk_boundary_general,sm:eq:outlier_ellimination_general}, including a derivation of the standard elliptical law, see \cref{sm:section:explicit_solutions_for}.
\section{Approximate self-consistent equations for the support of the spectrum}
\label{sm:section:approximate_self_consistent_equations}
\subsection{Approximate statistics of an FSRM}
In this section, we derive an explicit formula for the support of the eigenvalue spectrum of a general FSRM when there is a small amount of fine structure. Specifically, we suppose that the statistics of an FSRM $J_{ij}$, given by \cref{eq:approx_stats} in the main text can be written as 
    \begin{align}
        u_{ij} &= \mu + \epsilon u^{(1)}_{ij}, \nonumber\\
        s_{ij} &= \sigma^2 + \epsilon s^{(1)}_{ij}, \nonumber\\
        t_{ij} &= \gamma \sigma^2 + \epsilon t^{(1)}_{ij}, \label[plural_equation]{sm:eq:approx_stats}
    \end{align}
where the zeroth-order statistics in the fine structure are given by \cref{eq:avg_params} in the main text, which we repeat here
    \begin{align}
        \mu &= \frac{1}{N^2}\sum_{ij}u_{ij}, \nonumber\\
        \sigma^2 &= \frac{1}{N^2}\sum_{ij}s_{ij}, \nonumber\\
        \gamma &= \frac{1}{\sigma^2 N^2}\sum_{ij}t_{ij}. \label[plural_equation]{sm:eq:avg_params}
    \end{align}
By summing over the indices $i$ and $j$ in \cref{sm:eq:approx_stats}, we see that the sum over all elements of the fine-structure parts of the statistics must sum to zero. That is, the fine-structure part of the statistics satisfy
\begin{align}
    \sum_{ij}u^{(1)}_{ij} = \sum_{ij}s^{(1)}_{ij} = \sum_{ij}t^{(1)}_{ij} = 0.
\end{align}
We also recall the definitions of the fine-structure corrections [\cref{eq:fs_params} in the main text]
\begin{align}
    R &\equiv \frac{1}{\sigma^4N^3}\sum_{ijk}\frac{1}{2}\qty[s^{(1)}_{ij} + s^{(1)}_{ji}]t^{(1)}_{jk},\nonumber\\
    S &\equiv \frac{1}{\sigma^4N^3}\sum_{ijk}s^{(1)}_{ij}s^{(1)}_{jk}, \nonumber\\
    T &\equiv \frac{1}{\sigma^4N^3}\sum_{ijk}t^{(1)}_{ij}t^{(1)}_{jk}, \nonumber\\
    U &\equiv \frac{1}{\mu^2N^3}\sum_{ijk}u^{(1)}_{ij}u^{(1)}_{jk},\nonumber \\
    V &\equiv \frac{1}{\sigma^2\mu N^3}\sum_{ijk}\frac12\qty[u^{(1)}_{ij} + u^{(1)}_{ji}]t^{(1)}_{jk}.\label{sm:eq:fs_params}
\end{align}

Our aim is to find an explicit solution to the self-consistent equations summarized in \cref{sm:section:self_consistent_equations_summary}, valid to second order in $\epsilon$. On substituting the statistics [\cref{sm:eq:approx_stats}] into the self-consistent equations for determining the support of the bulk spectrum and outlier eigenvalues [\cref{sm:eq:resolvent_general,sm:eq:bulk_boundary_general,sm:eq:outliers_general}], we find 
\begin{align}
    \frac{1}{G_i(z)} &= z - \frac{1}{N}\sum_j\qty(\gamma\sigma^2 + \epsilon t^{(1)}_{ij})G_j(z), \label{sm:eq:resolvent_perturbative}\\
    1 &= \PF\qty[\frac{1}{N}\vb G\qty(\bulk) \conj{\vb G}\qty(\bulk)\qty(\sigma^2\vb 1 + \epsilon \vb s^{(1)})], \label{sm:eq:bulk_boundary_perturbative_0}\\
    0 &= \det[\vb I - \frac{1}{N}\vb G\qty(\outliers{k})\qty(\mu\vb 1 + \epsilon \vb u^{(1)})],\label{sm:eq:outliers_perturbative_0}\\
    1 &> \PF\qty[\frac{1}{N}\vb G\qty(\outliers{k}) \conj{\vb G}\qty(\outliers{k})\qty(\sigma^2\vb 1 + \epsilon \vb s^{(1)})], \label{sm:eq:outlier_ellimination_perturbative_0}
\end{align}
where $\vb 1$ is an $N\times N$ matrix of ones. We will not explicitly expand \cref{sm:eq:outlier_ellimination_perturbative_0}, as the steps are nearly identical to that of \cref{sm:eq:bulk_boundary_perturbative_0}.

\cref{sm:eq:bulk_boundary_perturbative_0,sm:eq:outliers_perturbative_0} can both be written as eigenvalue equations, we write them both in the equivalent matrix-vector form 
\begin{align}
    B_i &= |G_i(\bulk)|^2\frac{1}{N}\sum_{j}\qty(\sigma^2 + \epsilon s^{(1)}_{ij}) B_j, \label{sm:eq:bulk_boundary_perturbative}\\
    C_i &= G_i\qty(\outliers{k})\frac{1}{N}\sum_j\qty(\mu + \epsilon u^{(1)}_{ij})C_j,\label{sm:eq:outliers_perturbative}
\end{align}
where $\vb B$ is the right PF eigenvector of the positive matrix $\frac{1}{N}|\vb G(\bulk)|^2\qty(\sigma^2\vb 1 + \epsilon \vb s^{(1)})$, and where $\vb C$ is an eigenvector of the matrix $\frac{1}{N}|\vb G(z)|^2\qty(\sigma^2\vb 1 + \epsilon \vb s^{(1)})$. Both vectors $\vb B$ and $\vb C$ correspond to an eigenvalue of $1$. We note that \cref{sm:eq:Q2} is essentially the same as \cref{sm:eq:bulk_boundary_perturbative}, but is written in terms of the left eigenvector $\vb Q$.

To proceed with the derivation of the fine-structure corrections to the elliptical law and outlier eigenvalue, we suppose that the unknown variables $G_i, z, B_i$ and $C_i$ can be expanded in powers of $\epsilon$
\begin{align}
    G_i &= G^{(0)}_i + \epsilon G^{(1)}_i + \epsilon^2 G^{(2)}_i,\label{sm:eq:G_second_order}\\
    z &= z_0 + z_1 \epsilon + z_2 \epsilon^2,\label{sm:eq:z_second_order}\\
    B_i &= B^{(0)}_i + \epsilon B^{(1)}_i + \epsilon^2 B^{(2)}_i,\label{sm:eq:B_second_order}\\
    C_i &= C^{(0)}_i + \epsilon C^{(1)}_i + \epsilon^2 C^{(2)}_i. \label{sm:eq:C_second_order}  
\end{align}
In the remainder of this section, we will expand \cref{sm:eq:outliers_perturbative,sm:eq:resolvent_perturbative,sm:eq:bulk_boundary_perturbative} to second order in $\epsilon$. We will derive our modified elliptical law and outlier eigenvalue [\cref{eq:bulk_approx,eq:outlier_approx} in the main text] as the simultaneous solution to these expanded equations. 

In the remainder of this section we stop writing the explicit $z$ dependence for the resolvent, and will frequently use the abbreviations $G_0 = \sum_i G^{(0)}_i/N, G_1 = \sum_i G^{(1)}_i/N, G_2 = \sum_i G^{(2)}_i/N$ and similarly for $B_i$ and $C_i$.
\subsection{Second order expansion of \texorpdfstring{\cref{sm:eq:resolvent_general}}{}}\
\label{sm:section:expansion_of_resolvent}
On substituting \cref{sm:eq:G_second_order,sm:eq:z_second_order} into \cref{sm:eq:resolvent_perturbative}, we expand to second order in $\epsilon$ and equate terms. We proceed order by order in $\epsilon$, substituting the solution for the zeroth order equations in to the first order equations and so on. To order $\epsilon^0$ we find
\begin{align}
    z_0 &= \frac{1}{G^{(0)}_i} + \gamma \sigma^2G_0.
\end{align}
Hence, the zeroth order approximation to the diagonal elements of the resolvent, $G^{(0)}_{i}$, have no $i$ dependence and $G^{(0)}_i = G_0$. We may therefore write $z_0$ as
\begin{align}
    z_0 = \frac{1}{G_0} + \gamma \sigma^2G_0.
\end{align}
Equating terms of order $\epsilon$, we obtain 
\begin{align}
    z_1 &= \gamma \sigma^2G_1 - \frac{G^{(1)}_i}{G_0^2} + \frac{G_0}{N}\sum_jt^{(1)}_{ij}.
\end{align}
By summing over the index $i$, we find 
\begin{align}
    z_1 &= G_1\qty(\gamma \sigma^2 - \frac{1}{G_0^2}).
\end{align}
We also find the following useful expression for $G^{(1)}_i$, which we will use repeatedly throughout this section
\begin{align}
    G^{(1)}_i &= G_1 + \frac{G_0^3}{N}\sum_jt^{(1)}_{ij}. \label{sm:eq:G1i}
\end{align}
Equating terms of order $\epsilon^2$ gives
\begin{align}
    z_2 &= G_2\gamma\sigma^2 - \frac{G^{(2)}_i}{G_0^2} + \frac{\qty(G^{(1)}_i)^2}{G_0^3} + \frac{1}{N}\sum_jt^{(1)}_{ij}G^{(1)}_j,\nonumber \\
    &= G_2\gamma\sigma^2 - \frac{G^{(2)}_i}{G_0^2} + \frac{G_1^2}{G_0^3} + \frac{G_0^3}{N^2}\sum_{jk}\qty(t^{(1)}_{ij}t^{(1)}_{jk} + t^{(1)}_{ij}t^{(1)}_{ik}) + \qty(2G_0^3 + G_1)\frac{1}{N}\sum_jt^{(1)}_{ij},
\end{align}
where we have used \cref{sm:eq:G1i} to go from the first to the second line. Again, summing over the index $i$, we find
\begin{align}
    z_2 = G_2\qty(\gamma \sigma^2 - \frac{1}{G_0^2}) + \frac{G_1^2}{G_0^3} + 2\sigma^4 T G_0^3.
\end{align}
Finally, we compute $z = z_0 + \epsilon z_1 + \epsilon^2 z_2$ and arrive at the following compact expression for $G$, valid to second order in $\epsilon$
\begin{align}
    \frac{1}{G(z)} = z - \gamma \sigma^2 G(z) - 2 T\sigma^4 G(z)^3 \epsilon^2.\label{sm:eq:G_approx}
\end{align}
Hence, to second order in $\epsilon$, we can replace the $N$ equations in \cref{sm:eq:resolvent_perturbative} for each diagonal element of the resolvent $G_i(z)$ with a single equation for the trace $G(z)$. 
\subsection{Second order expansion of \texorpdfstring{\cref{sm:eq:bulk_boundary_general}}{}}
\label{sm:section:expansion_bulk}
We follow the same procedure as in the preceeding section for expanding \cref{sm:eq:bulk_boundary_perturbative} in small $\epsilon$, by substituting \cref{sm:eq:G_second_order,sm:eq:B_second_order} into \cref{sm:eq:bulk_boundary_perturbative} and equating term by term in powers of $\epsilon$. We also note that, from \cref{sm:section:expansion_of_resolvent}, we know $G_i^{(0)} = G_0$. To zeroth order we have  
\begin{align}
    B^{(0)}_i &= \frac{\sigma^2|G_0|^2}{N}\sum_jB^{(0)}_j.\label{sm:eq:B0}
\end{align}
As the elements $B_i$ must all be positive, by the Perron-Frobenius theorem the only solution to \cref{sm:eq:B0} requires that $\sigma^2|G_0|^2 = 1$ and $B^{(0)}_i = B_0$ for each $i$. Equating terms of order $\epsilon$, we find
\begin{align}
    B^{(1)}_i &= B_1 + B_0\frac{1}{\sigma^2N}\sum_j s^{(1)}_{ij} + 2\sigma^2B_0\Re{G_0^*G^{(1)}_i}.\label{sm:eq:B1}
\end{align}
On summing over the index $i$, we see that $G_1 = 0$. Finally, the second order equation is
\begin{align}
    B^{(2)}_i &= B_2 + \frac{1}{\sigma^2N}\sum_js^{(1)}_{ij}B^{(1)}_j + 2\sigma^2 \Re{G_0^*G^{(1)}_i}\qty[B_1 + \frac{B_0}{\sigma^2N}\sum_js^{(1)}_{ij}] + \sigma^2 B_0\qty[2\Re{G_0^*G^{(2)}_i} + 
    \qty|G^{(1)}_i|^2].
\end{align}
Summing over the index $i$ and substituting \cref{sm:eq:G1i,sm:eq:B1} gives
\begin{align}
    2\sigma^2\Re{G_0^*G_2} = -\qty\Big[T + 4\cos(2\varphi)R + S],
\end{align}
where we have written $\varphi$ for the argument of the summed resolvent $G = |G|e^{i\varphi}$. If we now compute the sum $|G|^2 = |G_0 + \epsilon G_1 + \epsilon^2 G_2|^2$, we arrive at 
\begin{align}
    \sigma^2|G(\bulk)|^2 = 1 - \epsilon^2\qty\Big[S + T + 4R \cos(2\varphi)]. \label{sm:eq:G_bulk_approx}
\end{align}
Hence, the values of the resolvent on the boundary of the bulk spectrum can be parametrized in terms of its complex argument $\varphi$ as 
\begin{align}
    G(\bulk) = \frac{1}{\sigma^2}\qty\bigg[1 - \epsilon^2\qty\Big[S + T + 4R \cos(2\varphi)]]e^{i\varphi}.
\end{align}
Finally, we remark that one could derive \cref{sm:eq:outlier_ellimination_perturbative_0} similarly to the expansion of \cref{sm:eq:bulk_boundary_perturbative_0} that we have performed here. The expansion of \cref{sm:eq:outlier_ellimination_perturbative_0} gives the following condition on outlier eigenvalues
\begin{align}
    \sigma^2\left |G\qty(\outliers{k})\right |^2 < 1 - \epsilon^2\qty\Big[S + T + 4R \cos(2\varphi^{(k)})],\label{sm:eq:G_outlier_ellimination_approx}
\end{align}
where $\varphi^{(k)}$ is the argument of the outlier eigenvalue $\outliers{k}$.
\subsection{Second order expansion of \texorpdfstring{\cref{sm:eq:outliers_general}}{}}
\label{sm:section:outlier_expansion}
Expanding \cref{sm:eq:outliers_perturbative} is very similar to the expansion of \cref{sm:eq:bulk_boundary_perturbative} detailed in \cref{sm:section:expansion_bulk}. The main difference between \cref{sm:eq:bulk_boundary_perturbative} and \cref{sm:eq:outliers_perturbative} is that, in \cref{sm:eq:bulk_boundary_perturbative}, the elements $B_i$ must all be positive, and so the Perron-Frobenius theorem guarantees that there is only one solution. In contrast, the elements $C_i$ in \cref{sm:eq:outliers_perturbative} do not need to be positive, and there are $N$ possible eigenvectors $\vb C$ for each possible outlier eigenvalue $\outliers{k}$. However, we will see that only one of these solutions satisfies the further condition given in \cref{sm:eq:G_outlier_ellimination_approx}, so in fact there is only one outlier eigenvalue.
 
Substituting \cref{sm:eq:G_second_order,sm:eq:C_second_order} into \cref{sm:eq:outliers_perturbative}, we have
\begin{align}
    C^{(0)}_i + \epsilon C^{(1)}_i + \epsilon^2 C^{(2)}_i = \qty(G_0 + \epsilon G^{(1)}_i + \epsilon^2 G^{(2)}_i)\frac{1}{N}\sum_j\qty(\mu + \epsilon u^{(1)}_{ij})\qty(C^{(0)}_i + \epsilon C^{(1)}_i + \epsilon^2 C^{(2)}_i).\label{sm:eq:Cm1}
\end{align}
To zeroth order, we have 
\begin{align}
    C^{(0)}_i = \frac{1}{N}\mu G_0\sum_jC^{(0)}_j.\label{sm:eq:C0}
\end{align}
One solution to \cref{sm:eq:C0} is to take $C^{(0)}_i = C_0$ and $uG_0 = 1$. There are an additional $N - 1$ solutions corresponding to a diverging value of $G_0$ and the $N - 1$ linearly independent vectors $\vb C^{(0)}$ such that $\sum_iC^{(0)}_i = 0$. However, for small $\epsilon$ we know from \cref{sm:section:expansion_bulk} that $|G_i(\bulk)|$ is $O(\epsilon^0)$. Therefore, as outlier eigenvalues must satisfy $|G_i(\outlier)| \leq |G_i(\bulk)|$ [\cref{sm:eq:G_outlier_ellimination_approx}], only the first solution to \cref{sm:eq:C0} has a chance of corresponding to an outlier eigenvalue when $\epsilon$ is small. For the remainder of this section we are only interested in the first solution.

Equating terms of order $\epsilon$ in the expansion of \cref{sm:eq:Cm1}, we find
\begin{align}
    C^{(1)}_i = C_1 + C_0\frac{1}{\mu N}\sum_ju^{(1)}_{ij} + \mu C_0 G^{(1)}_i,\label{sm:eq:C1}
\end{align}
summing over the index $i$, we find $G_1 = 0$. Finally, equating terms of order $\epsilon^2$ gives
\begin{align}
    C^{(2)}_i &= C_2 + \frac{1}{\mu N}\sum_ju^{(1)}_{ij}C^{(1)}_j + \mu  G^{(1)}_i\qty[C_1 + \frac{C_0}{\mu N}\sum_ju^{(1)}_{ij}] + \mu C_0 G^{(2)}_i. \label{sm:eq:C2}
\end{align}
On substituting \cref{sm:eq:C1,sm:eq:G1i} into \cref{sm:eq:C2} and summing over the index $i$, we arrive at the following expression for the resolvent, evaluated at the outlier eigenvalue
\begin{align}
    \mu G\qty(\outlier) = 1  - \epsilon^2\qty[U + \frac{2\sigma^2}{\mu^2}V].\label{sm:eq:G_outlier_approx}
\end{align}
\subsection{Summary}
To summarize, in this section we have expanded \cref{sm:eq:resolvent_general,sm:eq:bulk_boundary_general,sm:eq:outliers_general,sm:eq:outlier_ellimination_general} to second order in the degree of fine structure (as measured by the small parameter $\epsilon$). In doing so, we have found a set of self-consistent equations for determining the trace of the resolvent matrix [\cref{sm:eq:G_approx}], the boundary of the bulk spectrum [\cref{sm:eq:G_bulk_approx}], and the outlier eigenvalue [\cref{sm:eq:G_outlier_approx}]. We have also determined a criterion for testing whether a complex number is outside the bulk spectrum [\cref{sm:eq:G_outlier_ellimination_approx}]. This criterion allows us to verify whether the outlier eigenvalue, as determined from \cref{sm:eq:G_outlier_approx}, corresponds to an outlier eigenvalue.
\section{Modified elliptical law}
\label{sm:section:Modified_elliptical_law}
To find the support of the spectrum of a general FSRM exactly, we must simultaneously solve \cref{sm:eq:resolvent_general,sm:eq:bulk_boundary_general,sm:eq:outliers_general,sm:eq:outlier_ellimination_general} for the complex numbers $\bulk$ comprising the support of the bulk spectrum of $\vb J$, and for the isolated outlier eigenvalues $\outliers{k}$ outside the bulk spectrum. In \cref{sm:section:approximate_self_consistent_equations} we have seen that if the level of fine structure is small (as measured by the small parameter $\epsilon$), then the support of the spectrum is determined from the simultaneous solution of \cref{sm:eq:G_approx,sm:eq:G_bulk_approx,sm:eq:G_outlier_approx,sm:eq:G_outlier_ellimination_approx}. In this section we will show how the solution of these equations leads to our modified elliptical law [\cref{eq:bulk_approx} in the main text] and to the associated outlier eigenvalue [\cref{eq:outlier_approx} in the main text].
\subsection{Support of the bulk spectrum}
\label{sm:section:approx_bulk}
To find the support of the bulk spectrum, we must simultaneously solve \cref{sm:eq:G_bulk_approx,sm:eq:G_approx} to second order in $\epsilon$. Writing the resolvent in polar form $G(z) = |G(z)|e^{i\varphi(z)}$, we substitute \cref{sm:eq:G_bulk_approx} into \cref{sm:eq:G_approx} to obtain
\begin{align}
    \frac{\bulk(\varphi)}{\sigma} = e^{-i\varphi} + \gamma e^{i\varphi} + \frac{\epsilon^2}{2}\qty{4 T e^{3 i \varphi} + \qty[e^{-i\varphi} - \gamma e^{i\varphi}]\qty[T + S + 4\cos(2\varphi)R]}.\label{sm:eq:bulk_boundary_complex}
\end{align}
To second order in $\epsilon$, \cref{sm:eq:bulk_boundary_complex} is equivalent to the modified ellipse in \cref{eq:bulk_approx} of the main text. To show this, we write $\bulk(\varphi) = x(\varphi) + iy(\varphi)$ and take the real and imaginary parts of \cref{sm:eq:bulk_boundary_complex}
\begin{align}
    \frac{x(\varphi)}{\sigma} &= \qty(1 + \gamma) \cos(\varphi) + \frac{\epsilon^2}{2}\qty\Big{4T\cos(3\varphi) + [1 - \gamma]\qty[T + S + 4\cos(2\varphi)R]\cos(\varphi)}, \nonumber\\
    \frac{y(\varphi)}{\sigma} &= (-1 + \gamma) \sin(\varphi) + \frac{\epsilon^2}{2}\qty\Big{4T\sin(3\varphi) - [1 + \gamma]\qty[T + S + 4\cos(2\varphi)R]\sin(\varphi)}.
\end{align}
Note that, if we set $\epsilon = 0$, then we recover the usual elliptical law 
\begin{align}
    \frac{x^2}{(1 + \gamma)^2} + \frac{y^2}{(1 - \gamma)^2} = \sigma^2.
\end{align}
To find the order $\epsilon^2$ correction to the zeroth order ellipse, it is helpful to compute the following terms involving $x^2$ and $y^2$
\begin{align}
    \qty[\frac{x(\varphi)}{\sigma(1 + \gamma)}]^2 &= \cos[2](\varphi) + \frac{\epsilon^2}{1 + \gamma}\qty\Big{\qty\big[4T + (1 - \gamma)(T + S + 4R)]\cos[2](\varphi) - 2\qty\big[2T + (1 - \gamma)R]\sin[2](2\varphi)}, \nonumber\\
    \qty[\frac{y(\varphi)}{\sigma(1 - \gamma)}]^2 &= \sin[2](\varphi) + \frac{\epsilon^2}{1 - \gamma}\qty\Big{\qty\big[4T + (1 + \gamma)(T + S - 4R)]\sin[2](\varphi) - 2\qty\big[2T - (1 + \gamma)R]\sin[2](2\varphi)}.\label[plural_equation]{sm:eq:xy_sq}
\end{align}
By considering the product of \cref{sm:eq:xy_sq}, we find
\begin{align}
    \sin(2\varphi)^2 &=  \frac{4x(\varphi)^2y(\varphi)^2}{s^2(1 - \gamma^2)^2} + O\qty(\epsilon^2). \label{sm:eq:trig}
\end{align}
Substituting \cref{sm:eq:trig} into \cref{sm:eq:xy_sq} and adding up the resulting equations, we find
\begin{multline}
    \frac{x^2}{(1 + \gamma)^2}\qty{1 - \frac{\epsilon^2}{(1 + \gamma)}\qty[(1 - \gamma)(S + T + 4R) + 4T]} + \frac{y^2}{(1 - \gamma)^2}\qty[1 - \frac{\epsilon^2}{(1 - \gamma)}\qty[(1 + \gamma)(S + T - 4R) + 4T]]\\
    = \sigma^2 - \frac{32\epsilon^2(T - \gamma R)}{\sigma^2}\frac{x^2y^2}{\qty(1 - \gamma^2)^3}. \label{sm:eq:bulk_m1}
\end{multline}
Finally, we use the binomial identity $(1 + A\epsilon^2/2)^{-2} = 1 - A\epsilon^2 + O\qty(\epsilon^4)$ to move the terms of order $\epsilon^2$ on the left-hand side of \cref{sm:eq:bulk_m1} to the denominator, giving us \cref{eq:bulk_approx} from the main text 
\begin{align}
    \qty(\frac{x}{a})^2 + \qty(\frac{y}{b})^2 = \sigma^2 - \frac{4c\epsilon^2}{\sigma^2}\qty(\frac{x}{a})^2\qty(\frac{y}{b})^2,\label{sm:eq:bulk_approx}
\end{align}
with 
\addtocounter{equation}{-1}
\begin{subequations}
\begin{align}
    a &= 1 + \gamma + \frac{\epsilon^2}{2}\qty[(1 - \gamma)(S + T + 4R) + 4T], \nonumber\\
    b &= 1 - \gamma + \frac{\epsilon^2}{2}\qty[(1 + \gamma)(S + T - 4R) + 4T], \nonumber\\
    c &= 8\frac{T - R\gamma}{1 - \gamma^2}.
\end{align}
\end{subequations}
In the main text, we drop the factors of $\epsilon$ in our statement of the modified elliptical law, absorbing them into the fine-structure parameters $T, S$ and $R$. 
\subsubsection{Parametric form for the boundary of the bulk spectrum}
\label{sm:section:parametric_form_of_MEL}
In the main text we present the parametric equations \cref{eq:parametric_approx} as a practical way to plot the support of the bulk spectrum [\cref{sm:eq:bulk_approx}] of the FSRM $\vb J$. The proof is simple, taking
\begin{align}
    x(\theta) &= \sigma a\cos(\theta)\qty(1 - c\epsilon^2\sin(\theta)^2), \nonumber\\
    y(\theta) &= \sigma b\sin(\theta)\qty(1 - c\epsilon^2\cos(\theta)^2),
\end{align}
we compute (to second order in $\epsilon$)
\begin{align}
    \qty(\frac{x}{a})^2 + \qty(\frac{y}{b})^2 &= \sigma^2\cos(\theta)^2\qty(1 - 2c\epsilon^2\sin(\theta)^2) + \sigma^2\sin(\theta)^2\qty(1 - 2c\epsilon^2\cos(\theta)^2) + O\qty(\epsilon^4), \nonumber\\
    &= \sigma^2 - 4\sigma^2c\epsilon^2\sin(\theta)^2\cos(\theta)^2 + O\qty(\epsilon^4), \nonumber\\
    &= \sigma^2 - \frac{4c\epsilon^2}{\sigma^2}\qty(\frac{x}{a})^2\qty(\frac{y}{b})^2 + O\qty(\epsilon^4).
\end{align}
Hence, the expressions are equivalent to second order in $\epsilon$.
\subsection{Outlier eigenvalue}
\label{sm:section:approx_outlier}
As discussed in \cref{sm:section:outlier_expansion}, whilst a general FSRM can have multiple outlier eigenvalues, a FSRM with a small amount of fine structure only has a single outlier eigenvalue, much like in the usual elliptical law \cite{orourkeLowRankPerturbations2014}. The location of the outlier eigenvalue [\cref{eq:outlier_approx} in the main text] is obtained by substituting \cref{sm:eq:G_outlier_approx} into \cref{sm:eq:G_approx} and expanding to second order in $\epsilon$, we arrive at \cref{eq:outlier_approx} in the main text
\begin{align}
    \frac{\outlier}{\mu} = 1 + \frac{\gamma \sigma^2}{\mu^2} + \epsilon^2\qty[\frac{2\sigma^4}{\mu^4}T + \qty(1 - \frac{\gamma \sigma^2}{\mu^2})\qty(U + \frac{2\sigma^2}{\mu^2}V)].\label{sm:eq:outlier_approx}
\end{align}
We note that \cref{sm:eq:outlier_approx} is only valid for certain values of the model parameters. To check that $\outlier$ lies outside the bulk we must use \cref{sm:eq:G_outlier_ellimination_approx}, the fine-structure correction to \cref{sm:eq:outlier_ellimination_general}. Noting that $G(\outlier)$ is real, this condition is
\begin{align}
    |\mu| > \sigma\qty\bigg[1 + \epsilon^2\qty(U + \frac{2\mu^2}{\sigma^2}V) - \frac12\epsilon^2\qty\Big(S + T + 4R)],
\end{align}
which is equivalent to \cref{eq:outlier_condition} in the main text.
\section{Effect of fine structure on stability}
\label{sm:section:effect_f_fs_on_stability}

\subsection{Proof that \texorpdfstring{$\|\cdot\|$}{} is a semi-norm}
In \cref{section:small_fs_stability}, we make a number of conclusions about the effect of symmetry and antisymmetry in the structure of the statistics of the FSRM $\vb J$. Central to our analysis is the following function
\begin{align}
    \|\vb A\| = \sqrt{\frac{1}{N^3}\sum_i\qty(\sum_jA_{ij})^2},
\end{align}
which we claim is a matrix semi-norm. That is, the function $\|\cdot\|$ satisfies the triangle inequality, absolute homogeneity and positivity
\begin{align}
    \|\vb A + \vb B\|& \leq \|\vb A\| + \|\vb B\|, \nonumber\\
    \| a\vb A\| &= |a|\|\vb A\|, \nonumber\\
    \|\vb A\| &\geq 0. \label[plural_equation]{sm:eq:semi_norm}
\end{align}
The properties \cref{sm:eq:semi_norm} all follow from the observation that $\|\vb A\| = |\vb A \vb 1|/\sqrt{N^3}$, where $\vb 1$ is a vector of ones and $|\dot|$ is the usual vector norm $|\vb a| = \sqrt{\sum_ia_i^2}$. We also note that $\|\vb A\| = 0$ if and only if all column sums of $\vb A$ are equal to zero, which is straightforward to verify from the definition.
\subsection{Writing \texorpdfstring{$R, S, U$, and $V$}{} in terms of \texorpdfstring{$\rho_1, \rho_2, S_s, S_a, U_s, U_a$ and $T$.}{}}
Recalling the definitions in \cref{sm:eq:fs_params}, as well as the definitions of $S_s, S_a, U_s, U_a$ from \cref{eq:sym_antisym_params} in the main text, we can write 
\begin{align}
    S_s - S_a &= \frac{1}{4\sigma^4}\qty\Bigg(\left\|\vb s^{(1)} + \vb s^{(1)T}\right\|^2 - \left\|\vb s^{(1)} - \vb s^{(1)T}\right\|^2),\nonumber\\
    &= \frac{1}{4\sigma^4N^3}\sum_{j}\qty{\qty[\sum_i\qty(s^{(1)}_{ij} + s^{(1)}_{ji})]^2 - \qty[\sum_i\qty(s^{(1)}_{ij} - s^{(1)}_{ji})]^2}, \nonumber\\
    &= \frac{1}{\sigma^4N^3}\sum_{ijk}s^{(1)}_{ij}s^{(1)}_{jk} = S,
\end{align}
To confirm \cref{eq:fs_params_sym_antisym} in the main text, we first write the fine-structure parameters $R$ and $V$ as 
\begin{align}
    R &= \frac{1}{2\sigma^4N^3} \qty[\qty(\vb s^{(1)} + \vb s^{(1)T})\vb 1]\vdot \qty[\vb t^{(1)}\vb 1], \nonumber\\
    V &= \frac{1}{2\mu\sigma^2N^3} \qty[\qty(\vb u^{(1)} + \vb u^{(1)T})\vb 1]\vdot \qty[\vb t^{(1)}\vb 1],
\end{align}
where $\vdot$ denotes the vector dot product. Using the identity $\vb u \vdot \vb v = |\vb u||\vb v|\cos(\theta)$ for some arbitrary angle $\theta$, we can write
\begin{align}
    R &= \frac{1}{2N^3\sigma^4}\qty|\qty(\vb s^{(1)}  + \vb s^{(1)T})\vb 1|\qty|\vb t^{(1)}\vb 1|\cos(\theta_1) \equiv \frac{1}{2\sigma^4}\left\|\vb s^{(1)}  + \vb s^{(1)T}\right\|\left\|\vb t^{(1)}\right\|\cos(\theta_1), \nonumber\\
    V &= \frac{1}{2N^3\sigma^2\mu}\qty|\qty(\vb u^{(1)}  + \vb u^{(1)T})\vb 1|\qty|\vb t^{(1)}\vb 1|\cos(\theta_2) \equiv \frac{1}{2\sigma^2\mu}\left\|\vb u^{(1)}  + \vb u^{(1)T}\right\|\left\|\vb t^{(1)}\right\|\cos(\theta_2).\label[plural_equation]{sm:eq:rho_12_definition}
\end{align}
\cref{eq:fs_params_sym_antisym} from the main text follow with $\rho_1 = \cos(\theta_1)$ and $\rho_2 = \cos(\theta_2)$. In the following, we are not interested in the specific values of $\rho_1$ and $\rho_2$, only that they are between $-1$ and $1$.
\subsection{When fine structure is stabilizing and destabilizing}
Stability is determined from the rightmost eigenvalue of the spectrum of a FSRM $\vb J$. The rightmost eigenvalue is either equal to the edge of the bulk spectrum $\edge$, or is equal to the outlier eigenvalue $\outlier$, for which the fine-structure corrections are given in \cref{eq:edge_approx,eq:outlier_approx} in the main text. Written in terms of $S_s, S_a, U_s, U_a, T, \rho_1, \rho_2$, they are 
\begin{align}
    \frac{\edge}{\sigma} &= 1 + \gamma + \frac{1}{2}\qty\Big[(1 - \gamma)(S_s - S_a + T + 4\rho_1\sqrt{S_sT}) + 4T], \nonumber\\
    \frac{\outlier}{\mu} &= 1 + \frac{\gamma \sigma^2}{\mu^2} + \qty(1 - \frac{\gamma\sigma^2}{\mu^2})\qty(U_s - U_a + \frac{2\sigma^2\rho_2}{\mu^2}\sqrt{U_sT}) + \frac{2\sigma^2}{\mu^2}T.
\end{align}
Recall that $|\gamma| \leq 1$, and that for an outlier to emerge from the bulk we must have $\mu \geq \sigma$. Taking derivatives with respect to $S_a$ and $U_a$, we have
\begin{align}
    \pdv{\edge}{S_a} &= -\frac{\sigma}{2}(1 - \gamma) \leq 0, \nonumber\\
    \pdv{\outlier}{U_a} &= -\mu\qty(1 - \frac{\gamma\sigma^2}{\mu^2}) \leq 0,\label{sm:eq:stability_sym_antisym}
\end{align}
hence, as discussed in \cref{section:small_fs_stability} the main text, antisymmetric fine structure in the statistics of a FSRM is stabilizing. 

The effect of symmetric fine structure on stability is determined by the remainder of the fine-structure corrections in \cref{sm:eq:stability_sym_antisym}
\begin{align}
    \edge^s &\equiv \frac{\sigma}{2}\qty\Big[(1 - \gamma)\qty(S_s + T + 4\rho_1\sqrt{S_sT}) + 4T], \nonumber\\
    \outlier^s &\equiv \mu\qty[\qty(1 - \frac{\gamma\sigma^2}{\mu^2})\qty(U_s + \frac{2\sigma^2\rho_2}{\mu^2}\sqrt{U_sT}) + \frac{2\sigma^2}{\mu^2}T], \label[plural_equation]{sm:eq:stability_symmetric}
\end{align}
if $\edge^s$, $\outlier^s$ is positive/negative, then we say that symmetric fine structure in the statistics of the FSRM $\vb J$ is destabilizing/stabilizing. Noticing that \cref{sm:eq:stability_symmetric} are quadratics in the variables $\sqrt{T}$ and $\sqrt{S_s}$, completing the square for the variables $\sqrt{S_s}$ and $\sqrt{U_s}$ respectively gives
\begin{align}
    \edge^s &\equiv \frac{\sigma}{2}\qty\Big{[1 - \gamma]\qty[\sqrt{T} + 2\rho_1\sqrt{S_s}]^2 + \qty\Big[5 - \gamma - 4(1 - \gamma)\rho_1^2]T}, \label{sm:eq:stability_edge_symmetric_2}\\
    \outlier^s &\equiv \mu\qty{\qty[1 - \frac{\gamma\sigma^2}{\mu^2}]\qty[\sqrt{U_s} + \frac{\sigma^2\rho_2}{\mu^2}\sqrt{T}]^2 + \qty[\frac{2\sigma^2}{\mu^2} - \qty(1 - \frac{\gamma\sigma^2}{\mu^2})\frac{\sigma^4}{\mu^4}\rho_2^2]T}. \label{sm:eq:stability_outlier_symmetric_2}
 \end{align}
The first term in both expressions is always positive. Therefore, sufficient conditions for the positivity of $\edge^s$ and $\outlier^s$ can be found by analyzing when the second term in both expressions changes sign. As $T, S_s \geq 0$ and $|\gamma|\leq 1$, the second term in the curly braces on the RHS of \cref{sm:eq:stability_edge_symmetric_2} is positive
\begin{align}
    \rho_1^2 \leq \frac{5 - \gamma}{4(1 - \gamma)}. \label{sm:eq:rho_1_gamma_condition}
\end{align}
The LHS of \cref{sm:eq:rho_1_gamma_condition} takes values between $0$ and $1$, and the RHS takes values between $3/4$ and $\infty$. Therefore, if the LHS is smaller than $3/4$, or if the RHS is greater than $1$, then \cref{sm:eq:rho_1_gamma_condition} is automatically satisfied. Solving for $\rho_1$ and $\gamma$ in these two cases, we find that symmetric fine structure in the statistics of a FSRM $\vb J$ is destabilizing if $\gamma \geq -1/3$ or if $\rho_1 \geq -\sqrt{3}/2$. In the main text, we only quote the first of these conditions.

The situation is more straightforward for the outlier eigenvalue. As $T, U_s \geq 0$, $|\gamma|\leq 1$ and $\mu \geq \sigma$ for an outlier to emerge to the right of the bulk, the second term in \cref{sm:eq:stability_outlier_symmetric_2} is positive provided
\begin{align}
    \rho_2^2 \leq \frac{2\mu^2}{\sigma^2\qty(1 - \frac{\gamma\sigma^2}{\mu^2})}. \label{sm:eq:rho_2_gamma_condition}
\end{align}
The LHS of \cref{sm:eq:rho_2_gamma_condition} takes values between $0$ and $1$, and the RHS takes values between $1$ and $\infty$. Therefore, $\outlier^s$ is always positive, and we conclude that symmetric fine structure in the statistics of a FSRM is always destabilizing. 
\section{Explicit solutions for the support of the spectrum}
\label{sm:section:explicit_solutions_for}
In this section we detail exact solutions for the support of the bulk spectrum and outlier eigenvalues for some particular choices of the statistics $\vb u, \vb s$ and $\vb t$ [see \cref{eq:avg_params} in the main text]. We first look at the case where the fine-structure parameter $T = 0$. This is a restriction on the matrix $\vb t$ which includes many interesting examples, including the standard elliptical law, the cascade model of Ref.~\cite{allesinaPredictingStabilityLarge2015}, directed networks \cite{neriLinearStabilityAnalysis2020} and FSRMs for which the matrices $\vb u, \vb s$, and $\vb t$ are circulant. Then, we analyze the cascade model as presented in \cref{section:exact_solution} and find explicit expressions for the support of the bulk spectrum and outlier eigenvalues. Ultimately, these expressions are too complicated to provide insight into stability, but they do allow us to verify the results of \cref{section:spectrum_approximation} of the main text.
\subsection{The case \texorpdfstring{$T = 0$}{}.}
\label{sm:section:the_case_T_0}
If the fine-structure parameter $T = 0$, then the matrix $\vb t$ must have constant row sums. To see this, we note that $T$ can be written as [which follows from the symmetry of the matrix $\vb t$ and \cref{eq:fs_params} in the main text]
\begin{align}
    T = \frac{1}{N}\sum_i \qty[\qty(\frac{1}{\sigma^2N}\sum_jt_{ij}) - \gamma]^2.
\end{align}
Now, as the matrix $\vb t$ has constant row-sums, it follows that the diagonal elements of the resolvent matrix $G_i(z)$ do not depend on the index $i$. This is because the ansatz $G_i(z) = G(z)$ solves \cref{sm:eq:resolvent_general}, provided $G(z)$ satisfies
\begin{align}
    \frac{1}{G(z)} = z - \gamma\sigma^2G(z).\label{sm:eq:G_T_0}
\end{align}
To find the support of the bulk, we substitute $\vb G(z) = G(z)\vb I$ into \cref{sm:eq:outlier_ellimination_general} and obtain
\begin{align}
    |G(\outliers{k})|^2\PF\qty[\vb s/N] \leq 1,\label{sm:eq:outlier_elimination_T_0}
\end{align}
where $\PF[\vb s/N]$ is the PF eigenvalue of the matrix $\vb s/N$.  As discussed in \cref{sm:section:boundary_of_the_bulk_spectrum}, the bulk boundary is the collection of points $z = \bulk$ for which the above is an equality. Hence, to find the boundary of the bulk spectrum, we set $|G(\bulk)|^2 = 1/\PF\qty[\vb s/N]$ and substitute into \cref{sm:eq:G_T_0}. Writing $\bulk = x + i y$, we find that the boundary of the bulk spectrum is given by the ellipse 
\begin{align}
    \qty(\frac{x}{1 + \gamma\sigma^2/\PF[\vb s/N]})^2 + \qty(\frac{y}{1 - \gamma\sigma^2/\PF[\vb s/N]})^2 = \PF[\vb s/N]. \label{sm:eq:bulk_boundary_T_0}
\end{align}
To find the location of the outlier eigenvalues, we substitute $\vb G(\outliers{k}) = G(\outliers{k}) \vb I$ into \cref{sm:eq:outliers} and obtain 
\begin{align}
    \det[\vb I - \frac{1}{N}G\qty(\outliers{k})\vb u] = 0.
\end{align}
Hence, for each $k$, $\outliers{k}$ is an outlier eigenvalue if $1/G(\outliers{k})$ is equal to any of the eigenvalues of the matrix $\vb u/N$. By \cref{sm:eq:outlier_elimination_T_0}, the eigenvalue must have magnitude greater than $\sqrt{\PF[\vb s/N]}$. Supposing that there are some outlier eigenvalues $\outliers{k}$ indexed by $k$, we substitute $G(\outliers{k}) = 1/\eigval{k}[\vb u/N]$ into \cref{sm:eq:G_T_0} to obtain
\begin{align}
    \outliers{k} = \eigval{k}[\vb u/N] + \frac{\gamma\sigma^2}{\eigval{k}[\vb u/N]},\label{sm:eq:outliers_T_0}
\end{align}
which only corresponds to an outlier eigenvalue if $|\eigval{k}[\vb u/N]| \geq \sqrt{\PF[\vb s/N]}$, this result is a generalization of Theorem.~2.4 in Ref.~\cite{orourkeLowRankPerturbations2014}.

Hence, if the fine-structure parameter $T = 0$, or, equivalently, if the matrix $\vb t$ has constant row (or column) sums, then we only need to know the eigenvalues of the matrix $\vb u$ and the Perron-Frobenius eigenvalue of the matrix $\vb s$ to find the boundary of the spectrum. 

If we were to expand the statistics in small $\epsilon$ as we did to derive the fine-structure correction to the elliptical law and outlier eigenvalue, then we recover the same prediction as would be obtained by using the modified elliptical law directly. We first set 
\begin{align}
    u_{ij} &= \mu + \epsilon u^{(1)}_{ij}, \nonumber\\
    s_{ij} &= \sigma^2 + \epsilon s^{(1)}_{ij}.
\end{align}
To find the bulk spectrum and outlier eigenvalues to second order in $\epsilon$, we need the eigenvalues of $\vb u$ and $\vb s$ to second order in $\epsilon$. It is relatively straightforward (the derivations are very similar to those in \cref{sm:section:approximate_self_consistent_equations}) to show that, to second order in $\epsilon$, these eigenvalues are 
\begin{align}
    \PF\qty[\vb s/N] &= \sigma^2\qty(1 + \epsilon^2S), \nonumber\\
    \eigval{k}\qty[\vb u/N] &=
    \begin{cases}
        \mu\qty(1 + \epsilon^2U), & k = 1, \\
        0, & k > 1,
    \end{cases}
\end{align}
where $S$ and $U$ are as defined in \cref{sm:eq:fs_params}. Substituting into \cref{sm:eq:bulk_boundary_T_0,sm:eq:outliers_T_0}, and noting that if $T=0$ then both $R$ and $V$ are also equal to $0$, we recover the result that would be obtained from the modified elliptical law and outlier eigenvalue.
\subsubsection{The elliptical law}
\label{sm:section:elliptical_law}
An elliptical matrix ensemble is defined by the statistics
\begin{align}
    u_{ij} &= \mu,\nonumber\\
    s_{ij} &= \sigma^2,\nonumber\\
    t_{ij} &= \gamma\sigma^2.
\end{align}
These are the statistics of a FSRM with no fine structure. Therefore, it is clear that the fine-structure parameter $T = 0$, and we can use the exact solution detailed in \cref{sm:section:the_case_T_0}, of course we could also set all fine-structure parameters to zero in the modified elliptic law in \cref{eq:bulk_approx,eq:outlier_approx} of the main texts, the result is the same. We need to know the PF eigenvalue of $\vb s/N$ and all eigenvalues of $\vb u/N$, they are 
\begin{align}
    \PF\qty[\vb s/N] &= \sigma^2, \nonumber\\
    \eigval{k}[\vb u/N] &= \begin{cases}
        \mu, & k = 1, \\
        0, & k > 1
    \end{cases}.
\end{align}
Only the $k = 1$ outlier of $\vb u/N$ can correspond to an outlier eigenvalue as it is the only eigenvalue of $\vb u/N$ which satisfies $|\eigval{k}[\vb u/N]| > \PF[\vb s/N]$. Substituting into \cref{sm:eq:bulk_boundary_T_0,sm:eq:outliers_T_0}, we find expressions which agree with the modified elliptical law and outlier in \cref{eq:bulk_approx,eq:outlier_approx} of the main text.
\subsubsection{The cascade model of \texorpdfstring{Ref.~\cite{allesinaPredictingStabilityLarge2015}}{}}
The cascade model of Ref.~\cite{allesinaPredictingStabilityLarge2015} is equivalent (albeit with a different parametrization) to the cascade model we present in \cref{section:exact_solution} with the model parameter $\gamma_1$ set to zero. The statistics are  
\begin{align}
    u_{ij} &=
    \begin{cases}
        \mu_L, &i < j, \\
        \mu_U, &i > j
    \end{cases}, \nonumber\\
    s_{ij} &= 
    \begin{cases}
        \sigma_L^2, &i < j, \\
        \sigma_U^2, &i > j        
    \end{cases}, \nonumber\\
    t_{ij} &= \Gamma\sigma_L\sigma_U\label[plural_equation]{sm:eq:cascade_0_stats}.
\end{align}
As the matrix $\vb t$ is a constant matrix, the fine-structure parameter $T = 0$. Hence, we can again use the exact solution detailed in \cref{sm:section:the_case_T_0}, for which we need the PF eigenvalue of $\vb s/N$ and all eigenvalues of $\vb u/N$. In \cref{sm:section:cascade_model}, we show that they are given by 
\begin{align}
    \PF[\vb s/N] &= \frac{\sigma_L^2 - \sigma_U^2}{\ln \sigma_L^2/\sigma_U^2}, \nonumber\\
    \eigval{k}[\vb u/N] &= \frac{\mu_L - \mu_U}{\ln|\mu_L/\mu_U| + i\pi\qty[2k - 1 - \Theta(\mu_L\mu_U)]},
\end{align}
where $\Theta(x) = 1$ if $x > 0$ and is $0$ otherwise. Substituting into \cref{sm:eq:bulk_boundary_T_0,sm:eq:outliers_T_0} gives the exact support for the bulk spectrum, as well as the location of the outlier eigenvalues for the cascade model. In the case of the boundary of the bulk spectrum, the results here are the same as those obtained in \cite{allesinaPredictingStabilityLarge2015,aljadeffLowdimensionalDynamicsStructured2016}. As far we are aware, the exact location of the outlier eigenvalues has not been presented before.
\subsubsection{Directed networks}
In section \cref{section:directed_network} in the main text, and in \cref{sm:section:directed_networks}, we claim that our fine-structure approximation for the spectrum of network with directed and undirected links [see \cref{section:directed_network} of the main text] reduces to the exact solution when the underlying network contains only exclusively directed links. The FSRM of interest has statistics (obtained by setting $P_{i\leftrightarrow j} = 0$ in \cref{eq:network_stats} of the main text)
\begin{align}
    u_{ij} &= \frac{N}{p}P_{i\gets j}\mu, \nonumber\\
    s_{ij} &= \frac{N}{p}P_{i\gets j}\mu, \nonumber\\
    t_{ij} &= 0,
\end{align}
where $P_{i\gets j}$ is the probability of a link existing in the network going from node $i$ to node $j$. As $t_{ij} = 0$, clearly the fine-structure parameter $T$ vanishes, and therefore we can use \cref{sm:eq:bulk_boundary_T_0,sm:eq:outliers_T_0} to find the support of the spectrum. Hence, we only need to know the eigenvalues of $P_{i\gets j}$ to determine the spectrum, or equivalently, we need to know the eigenvalues of the adjacency matrix $A_{ij}$. The eigenvalues of undirected and directed networks are considered in Refs.~\cite{restrepoApproximatingLargestEigenvalue2007,chungSpectraGeneralRandom2011}. Provided the maximum degree of a node in the network is sufficiently small, the leading eigenvalue is equal to
\begin{align}
    \PF\qty[\vb A] = \frac{1}{Np}\sum_i\qty[k^\text{in}_i - p]\qty[k^\text{out}_i - p] = p\qty(1 + \rho),
\end{align}
where we recall the definition of $\rho$ from \cref{eq:network_stats} in the main text. All other eigenvalues are negligible (again, assuming the maximum degree is sufficiently small), hence, by \cref{sm:eq:outlier_elimination_T_0}, there is only one outlier eigenvalue. Substituting into \cref{sm:eq:bulk_boundary_T_0,sm:eq:outliers_T_0}, we find the bulk boundary and outlier eigenvalue
\begin{align}
    x^2 + y^2 &= \sigma\qty(1 + \rho), \nonumber\\
    \outlier &= \mu\qty(1 + \rho). \label[plural_equation]{sm:eq:network_exact_bulk_and_outlier}
\end{align}
\cref{sm:eq:network_exact_bulk_and_outlier} are equivalent to Eq.~(71) in \cite{neriLinearStabilityAnalysis2020}. 
\subsubsection{Circulant variances and correlations}
We now turn to an example FSRM ensemble for which the matrix $\vb t$ contains fine structure, but for which the value of $T$ is nonetheless equal to zero, we consider an FSRM for which the matrices $\vb t$ and $\vb s$ are circulant. That is, we suppose that $t_{ij} = t(i - j \mod N)$ and $s_{ij} = s(i - j \mod N)$. In this case the PF eigenvalue of the matrix $s_{ij}$ is simply the sum of any one of its rows $\frac{1}{N}\sum_js_{ij}$. 

For example, consider a FSRM with the following circulant statistics 
\begin{align}
    s_{ij} &= \sin\qty[\pi\frac{(i - j)\mod N}{N}]^2, \nonumber \\
    t_{ij} &= \cos\qty[2\pi\frac{(i - j)\mod N}{N}]\sin\qty[\pi\frac{(i - j)\mod N}{N}]^2, \label[plural_equation]{sm:eq:circulant_stats}
\end{align}
and with the matrix $\vb u$ chosen such that its only non-zero eigenvalues are $1, -3(1\pm i)/4$. In the large $N$ limit we approximate the row sums of $\vb s$ and $\vb t$ as the following integrals
\begin{align}
    \PF[\vb s/N] &= \frac{1}{N}\sum_{j}\sigma_{1j}^2 = \int_0^1\dd{\alpha}\sin(\pi \alpha)^2 = \frac{1}{2}, \nonumber\\
    \gamma &= \frac{1}{\sigma^2N}\sum_j\gamma_{1j}\sigma_{1j}\sigma_{j1} = \int_0^1\dd{\alpha}\cos(2\pi \alpha)\sin(\alpha)^2 = -\frac{1}{2}.
\end{align}
We therefore expect the bulk region to be bounded by the ellipse
\begin{align}
    \qty(\frac{x}{1 - \frac{1}{2}})^2 + \qty(\frac{y}{1 + \frac{1}{2}})^2 = \frac12,\label{sm:eq:circulant_bulk}
\end{align}
with three outlier eigenvalues located at $\frac{3}{4}, (-7 \pm 11 i)/12$ [\cref{sm:eq:outliers_T_0}]. \Cref{sm:eq:circulant_bulk}, and the prediction for the outliers are verified in \cref{sm:fig:circulant_statistics}
\begin{figure*}
    \includegraphics[width=0.7\textwidth]{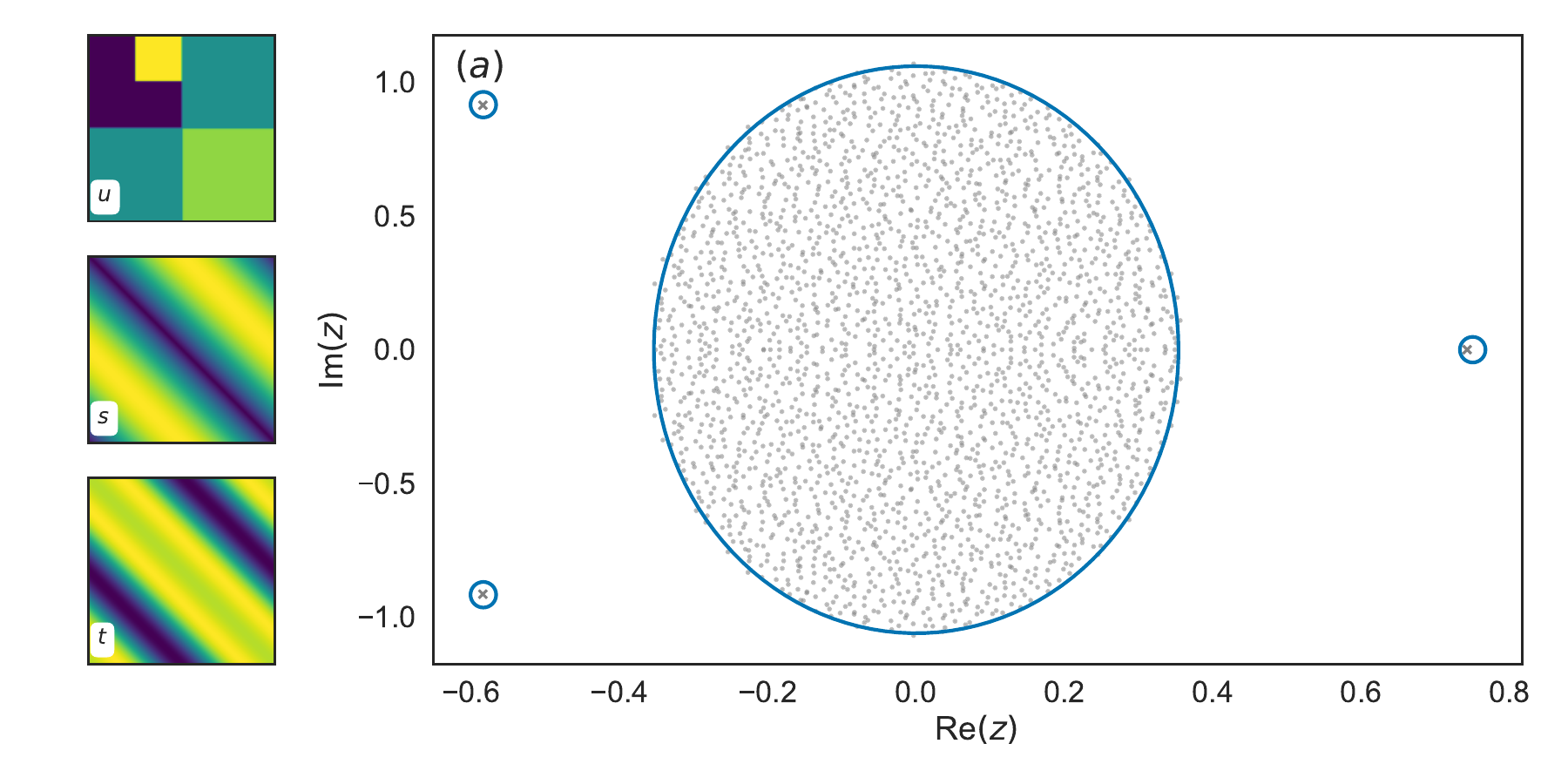}
    \caption{Eigenvalue spectrum of a FSRM with circulant statistics as described in \cref{sm:eq:circulant_stats}. $(u), (s), (t)$: heatmaps of the structure of the corresponding statistics of the random matrix. (a) Eigenvalues of a single $5000\times 5000$ realization of the random matrix (grey markers) with the analytic prediction for the bulk [\cref{sm:eq:circulant_bulk}] and outlier eigenvalues.}
    \label{sm:fig:circulant_statistics}
\end{figure*}

\subsection{The cascade model}
\label{sm:section:cascade_model}
\subsubsection{Statistics}
In \cref{section:exact_solution} of the main text, we compute the fine-structure correction for the spectrum of an FSRM with statistics that generalize the cascade model of Ref.~\cite{allesinaPredictingStabilityLarge2015}. We then claim that the correction we find [\cref{eq:cascade_fs_corrections} in the main text] agrees with a direct expansion of the exact solution in powers of the small parameter $\epsilon$, which controls the degree of fine structure present, therefore verifying our modified elliptic law [\cref{eq:bulk_approx,eq:outlier_approx}]. In this section, we fill in the details of this argument by first deriving the exact solution by solving the self-consistent equations \cref{sm:eq:resolvent_general,sm:eq:bulk_boundary,sm:eq:outliers} with $\vb u, \vb s$, and $\vb t$ as in \cref{eq:cascade_stats} in the main text. We then expand the exact solution in powers of $\epsilon$ and verify that it agrees with the fine-structure correction we find in \cref{section:exact_solution} of the main text.

For reference, we repeat \cref{eq:cascade_stats} here
\begin{align}
    u_{ij} &= \mu + \mu_1\sign(i - j), \nonumber\\
    s_{ij} &= \sigma^2 + \sigma_1^2\sign(i - j), \nonumber\\
    t_{ij} &= \gamma\sigma^2 + \gamma_1\sigma^2\sign(i + j - N),
    \label[plural_equation]{sm:eq:cascade_stats}
\end{align}
where $\sign(x) = 1$ if $x > 0$ and is $-1$ otherwise.
\subsubsection{Solving for the resolvent}
To find the support of a FSRM with statistics as in \cref{sm:eq:cascade_stats}, we must solve the self-consistent equations \cref{sm:eq:resolvent_general,sm:eq:bulk_boundary,sm:eq:outliers}. We will start by solving \cref{sm:eq:resolvent_general} for the diagonal elements of the resolvent matrix $G_i(z)$. For large $N$, the sums in \cref{sm:eq:resolvent_general} approach integrals over the continuous functions $G(i/N, z) \equiv G_i(z)$ and $t(i/N, j/N) \equiv t_{ij}$, giving
\begin{align}
    \frac{1}{G(\alpha, z)} &= z - \int_0^1\dd{\beta}t(\alpha, \beta)G(\beta, z),\nonumber \\
    &= z - \sigma^2\qty[\gamma \int_0^1\dd{\beta}G(\beta, z) + \gamma_1\int_0^1\dd{\beta}\sign(\alpha + \beta - 1)G(\beta, z)],\label{sm:eq:resolvent_cascade_integral_equation}
\end{align}
\cref{sm:eq:resolvent_cascade_integral_equation} can be exactly solved for $G(\alpha, z)$ with the following manipulations. First, we differentiate \cref{sm:eq:resolvent_cascade_integral_equation} with respect to $\alpha$ [noting that the distributional derivative of $\sign(\alpha)$ is $2\delta(\alpha)$], the result reads
\begin{align}
    \partial_\alpha G(\alpha, z) = 2\gamma_1\sigma^2 G(\alpha, z)^2G(1 - \alpha, z).\label{sm:eq:resolvent_cascade_integral_equation_1}
\end{align}
An ansatz of the form $G(\alpha, z) = B(z) \exp[2A(z) \alpha]$ solves \cref{sm:eq:resolvent_cascade_integral_equation_1} provided the functions $A(z)$ and $B(z)$ satisfy $A(z) = \gamma_1 \sigma^2 B(z)^2\exp[2A(z)]$. On substituting the resolvent ansatz back into \cref{sm:eq:resolvent_cascade_integral_equation} and carrying out the integrals, we obtain
\begin{align}
    z = \frac{\sigma}{\sqrt{A(z)\gamma_1}}\qty\Big{\gamma \sinh[A(z)] + \gamma_1\cosh[A(z)]}.\label{sm:eq:cascade_z}
\end{align}
Hence, the resolvent $G(\alpha, z)$ can be found by inverting \cref{sm:eq:cascade_z} for $A(z)$. Luckily, we do not need to do this to find the spectrum of $\vb J$, as \cref{sm:eq:bulk_boundary,sm:eq:outliers} can be manipulated into expressions involving the function $A(z)$ only with relative ease. 

% \begin{align}
%     \lambda_1(\vb s) \frac{1}{N}\Tr\qty\Big[\vb G(\bulk)\conj{\vb G}(\bulk)] &= 1, \nonumber\\
%     %
%     \det[\vb I - \frac{1}{N}\vb u \Tr\vb G\qty(\outliers{k})] &= 0.\label[plural_equation]{sm:eq:cascade_0_spectrum}
% \end{align}
% Both relationships follow from the following general fact, which we will prove at the end of this section. 
To find the bulk boundary and any outlier eigenvalues, we must solve \cref{sm:eq:bulk_boundary,sm:eq:outliers} with the resolvent $G(\alpha, z)$ determined from \cref{sm:eq:cascade_z} and $\vb u$ and $\vb s$ as in \cref{eq:cascade_stats}. In our solution of both equations, we will use the following useful information, which we prove in \cref{sm:section:eigenvalues_of_u_and_s}. For a matrix $[\vb A]_{ij} = a + a_1\sign(i - j)$ and any diagonal matrix $\vb D$, the following holds
\begin{align}
    \eigval{k}\qty\big[\vb A \vb D/N] = \eigval{k}\qty\big[\vb A/N]\frac1N\Tr \vb D.\label{sm:eq:general_cascade_eigvals_0}
\end{align}
where the eigenvalues of $\vb A$ are given by
\begin{align}
    \eigval{k}\qty\big[\vb A/N] = \frac{2a_1}{\ln\qty|\frac{a + a_1}{a - a_1}| + i\pi q_k},\label{sm:eq:general_cascade_eigvals_1}
\end{align}
where $q_k$ runs through the even integers if $a > a_1$ and runs through the odd integers otherwise.
\subsubsection{Bulk Boundary}
To solve \cref{sm:eq:bulk_boundary}, we first use \cref{sm:eq:general_cascade_eigvals_0}, which on writing the trace as an integral over the continuous variable $\alpha$ gives
\begin{align}
    \PF\qty\big[\vb s/N]\int_0^1\dd{\beta}|G(\beta, \bulk)|^2 = 1.\label{sm:eq:bulk_cascade_3}
\end{align}
Substituting $G(\alpha, z) = B(z)\exp[2A(z)\alpha]$, recalling $A(z) = \gamma_1 \sigma^2 B(z)^2\exp[2A(z)]$, we can carry out the integral. On writing $A(z)$ in polar form $A(z) = |A(z)|\exp[i\varphi(z)]$, we find
\begin{align}
    A(\bulk) &= \frac{1}{2\cos[\varphi(\bulk)]}\sinh^{-1}\qty{\frac{2|\gamma_1|\sigma^2}{\PF[\vb s]}\cos[\varphi(\bulk)]}e^{i\varphi(\bulk)}. \label{sm:eq:bulk_cascade_A}
\end{align} 
Finally, we substitute \cref{sm:eq:bulk_cascade_A} into \cref{sm:eq:cascade_z} to find an expression for $\bulk$ in terms of $\varphi$, the argument of $A(z)$. We therefore have an explicit parameterization for the bulk of the spectrum of a FSRM with cascade statistics in terms of the variable $\varphi \in [0, 4\pi]$
\begin{align}
    \bulk(\varphi) 
    &= \frac{\sigma}{\sqrt{\gamma_1 |A(\varphi)|}}\qty\bigg{\gamma \sinh\qty\Big[|A(\varphi)|e^{i\varphi}] + \gamma_1\cosh\qty\Big[|A(\varphi)|e^{i\varphi}]}e^{-\frac{1}{2}i\varphi},\label{sm:eq:bulk_boundary_cascade_final}
\end{align}
with 
\addtocounter{equation}{-1}
\begin{subequations}
\begin{align}
    |A(\varphi)| = \frac{1}{2\cos(\varphi)}\sinh^{-1}\qty{|\gamma_1|\ln(\frac{\sigma^2 + \sigma_1^2}{\sigma^2 - \sigma_1^2})\frac{\sigma^2}{\sigma_1^2}\cos(\varphi)}.\label{sm:eq:bulk_boundary_cascade_final_1}
\end{align}
\end{subequations}
This explicit solution is used to produce the star shaped boundary of the bulk spectrum shown in \cref{fig:HT_spectrum}(a) of the main text.
\subsubsection{Outliers}
For the outliers we follow the same steps as in the derivation of the bulk boundary. First, we recognise \cref{sm:eq:outliers} as an eigenvalue equation, $\outliers{k}$ is an outlier eigenvalue if there is an eigenvalue of the matrix $\vb G(\outliers{k})\vb u$ equal to $1$. That is, $\outliers{k}$ is an outlier eigenvalue if there is some $\ell_k$ such that the $\ell_k$-th eigenvalue of $\vb G(\outliers{k})\vb u$ satisfies
\begin{align}
    \eigval{\ell_k}\qty[\vb G\qty(\outliers{k})\vb u/N] = 1.
\end{align}
Using \cref{sm:eq:general_cascade_eigvals_0} and writing the trace of $\vb G$ as an integral over the continuous variable $\alpha$ gives
\begin{align}
    \eigval{\ell_k}\qty\big[\vb u/N]\int_0^1\dd{\beta} G\qty(\beta, \outliers{k}) = 1, \label{sm:eq:outlier_cascade_1}
\end{align}
where we assume that the ordering of the eigenvalues of $\vb u$, dictated by the sequence $\{\ell_k\}$, is such that the real parts of the outlier eigenvalues are descending. On substituting $G(\alpha, z) = B(z)\exp[2A(z)\alpha]$ and carrying out the integral, we find 
\begin{align}
    \frac{\sinh\qty(A_k)}{\sqrt{A_k}} = \frac{\sqrt{|\gamma_1|\sigma^2}}{\eigval{\ell_k}\qty\big[\vb u/N]}, \label{sm:eq:outlier_cascade_2}
\end{align}
where $A_k \equiv A\qty(\outliers{k})$. Substituting \cref{sm:eq:outlier_cascade_2} into \cref{sm:eq:cascade_z}, we find the location of outlier eigenvalues in terms of $A_k$ [which must be computed numerically from \cref{sm:eq:outlier_cascade_2}]
\begin{align}
    \outliers{k} = \frac{\sinh(A_k)\cosh(A_k)}{A_k}\eigval{\ell_k}\qty\big[\vb u/N] + \frac{\gamma\sigma^2}{\eigval{\ell_k}\qty\big[\vb u/N]}. \label{sm:eq:outlier_cascade_final}
\end{align}
Recalling \cref{sm:eq:outlier_ellimination_general} and applying \cref{sm:eq:general_cascade_eigvals_0,sm:eq:general_cascade_eigvals_1}, we find that \cref{sm:eq:outlier_cascade_final} is only valid if $|\eigval{\ell_k}[\vb u/N]|\geq \PF[\vb s/N]$.
\subsubsection{Eigenvalues of \texorpdfstring{$\vb u$ and $\vb s$}{}}\label{sm:section:eigenvalues_of_u_and_s}
Here we derive \cref{sm:eq:general_cascade_eigvals_0,sm:eq:general_cascade_eigvals_1} for the matrix $\vb A$ with elements $A_{ij} = a + a_1\sign(i - j)$ assuming that $N$ is large. We wish to compute the eigenvalues of the matrix $\vb A \vb D/N$. Writing $A(i/N, j/N) \equiv A_{ij}$ for the matrix $\vb A$ and $D(i/N) = D_i$ for the diagonal elements of the matrix $\vb D$, the relevant eigenvalue equation [with eigenvalues $\lambda_k$ and eigenvectors (eigenfunctions) $B_k(\alpha)$] is
\begin{align}
    \lambda_k B_k(\alpha) = D(\alpha)\int_0^1\dd{\beta} \qty\big[a + a_1\sign(\alpha - \beta)]B_k(\beta).\label{sm:eq:general_cascade_eigvals_2}
\end{align}
We can find the eigenvalues $\lambda_k$ by first differentiating \cref{sm:eq:general_cascade_eigvals_2} with respect to $\alpha$, and then multiplying both sides by $D(\alpha)$, which gives
\begin{align}
    \lambda_k \frac{D(\alpha)}{B_k(\alpha)}\dv{\alpha}\qty[\frac{B_k(\alpha)}{D(\alpha)}] = 2a_1D(\alpha).
\end{align}
Recognizing the LHS side of this expression as the logarithmic derivative of $B_k(\alpha)/D(\alpha)$, we can integrate the above with respect to $\alpha$ to get 
\begin{align}
    B_k(\alpha) = \frac{B_k(0)}{D(0)}D(\alpha)\exp[\frac{2a_1}{\lambda_k}\int_0^\alpha\dd{\beta}D(\beta)].\label{sm:eq:general_cascade_eigvals_3}
\end{align}
Finally, we can find $\lambda_k$ by computing the value of $B_k(1)D(0)/D(1)B_k(0)$ in two different ways, first by substituting $\alpha = 0$ and $\alpha = 1$ into \cref{sm:eq:general_cascade_eigvals_2}, and then by substituting into \cref{sm:eq:general_cascade_eigvals_3}. The result is 
\begin{align}
    \lambda_k = \frac{2a_1}{\ln\qty|\frac{a + a_1}{a - a_1}| + i\pi q_k}\int_0^1 \dd{\beta}D(\beta),\label{sm:eq:general_cascade_eigvals_4}
\end{align}
where $q_k$ runs through the even integers if $a > a_1$ and runs through the odd integers otherwise. The factor of $q_k$ in the denominator comes from taking a complex logarithm when re-arranging \cref{sm:eq:general_cascade_eigvals_3} for $\lambda_k$.
\section{Analytical verification of the modified elliptical law: the cascade model}
\label{sm:section:verification_of_MEL}
In this section we verify our fine-structure correction to the elliptical law and outlier eigenvalue [\cref{eq:bulk_approx,eq:outlier_approx} in the main text] by showing that they correctly predict the small fine-structure approximation to the support of the spectrum for the cascade model [see \cref{section:exact_solution} in the main text], for which we know the support of the spectrum exactly [\cref{sm:eq:outlier_cascade_final,sm:eq:bulk_boundary_cascade_final}]. 
\subsection{Approximate statistics}
As discussed in \cref{section:exact_solution} of the main text, we can expand the statistics of the cascade model in a small parameter $\epsilon$ such that when $\epsilon = 0$ we recover an elliptical random matrix. The statistics we are interested in are \cref{eq:cascade_stats} in the main text (or \cref{sm:eq:cascade_stats} in the SM), with the replacements $\mu_1 \to \epsilon \mu_1$, $\sigma_1^2\to\epsilon\sigma_1^2$, and $\gamma_1\to \epsilon\gamma_1$. Keeping only terms to second order in $\epsilon$, the statistics we are interested in are therefore 
\begin{align}
    u_{ij} &= \mu + \epsilon\mu_1\sign(i - j), \nonumber\\
    s_{ij} &= \sigma^2 + \epsilon\sigma_1^2\sign(i - j), \nonumber\\
    t_{ij} &= \gamma\sigma^2 + \epsilon\gamma_1\sigma^2\sign(i + j - N).\label[plural_equation]{sm:eq:cascade_stats_perturbative}
\end{align} 
To find the support of a FSRM with statistics as in \cref{sm:eq:cascade_stats_perturbative} to second order in the small parameter $\epsilon$ we have two options. Firstly, we can substitute these statistics into the exact solution [\cref{sm:eq:outlier_cascade_final,sm:eq:bulk_boundary_cascade_final}] and expand the result up to second order in $\epsilon$. Secondly, we can compute the zeroth and first order fine-structure parameters ($\mu, \sigma, \gamma, R, S, T, U, V$) using \cref{eq:fs_params} in the main text (or \cref{sm:eq:fs_params} in the SM), then substitute the correction into the modified elliptical law [\cref{eq:bulk_approx,eq:outlier_approx} in the main text, or \cref{sm:eq:bulk_approx,sm:eq:outlier_approx} in the SM]. 
\subsection{Small parameter expansion of the exact spectrum}
\subsubsection{Bulk boundary}
In this section we detail our first option, substituting the statistics into the exact solution for the support of an FSRM with cascade statistics and expanding to second order in the small parameter $\epsilon$. We recall \cref{sm:eq:bulk_boundary_cascade_final,sm:eq:bulk_boundary_cascade_final_1}, the exact solution for the boundary of the bulk spectrum. For a FSRM with statistics as in \cref{sm:eq:cascade_stats_perturbative}, we have 
\begin{align}
    \bulk(\varphi) 
    &= \frac{\sigma}{\sqrt{\epsilon\gamma_1 |A(\varphi)|}}\qty\bigg{\gamma \sinh\qty\Big[|A(\varphi)|e^{i\varphi}] + \epsilon\gamma_1\cosh\qty\Big[|A(\varphi)|e^{i\varphi}]}e^{-\frac{1}{2}i\varphi},\label{sm:eq:bulk_boundary_cascade_final_perturbative}
\end{align}
with 
\begin{align}
    |A(\varphi)| = \frac{1}{2\cos(\varphi)}\sinh^{-1}\qty{\epsilon|\gamma_1|\ln(\frac{\sigma^2 + \epsilon\sigma_1^2}{\sigma^2 - \epsilon\sigma_1^2})\frac{\sigma^2}{\epsilon \sigma_1^2}\cos(\varphi)}.\label{sm:eq:bulk_boundary_cascade_final_1_perturbative}
\end{align}
In order to expand \cref{sm:eq:bulk_boundary_cascade_final_perturbative} to second order in $\epsilon$, we must expand \cref{sm:eq:bulk_boundary_cascade_final_1_perturbative} to third order. This is because of the factor of $1/\sqrt{\epsilon|A(z)|}$ in \cref{sm:eq:bulk_boundary_cascade_final_perturbative}. Expanding $|A(\varphi)|$ to third order in $\epsilon$, we have 
\begin{align}
    A(\varphi) &= \gamma_1 \epsilon - \frac{\gamma_1\epsilon^3}{6}\qty[\frac{\sigma_1^4}{\sigma^4} + 4\gamma_1^2\cos[2](\varphi)] + \order{\epsilon^5}.
\end{align}
Substituting into our exact expression for the bulk boundary [\cref{sm:eq:bulk_boundary_cascade_final_perturbative}] and expanding to second order in $\epsilon$, we find 
\begin{align}
    \frac{\bulk(\varphi)}{\sigma} = e^{-i\varphi} + \gamma e^{i\varphi} + 
    \frac16 \epsilon^2 \qty[\gamma  \gamma_1^2 e^{5i\varphi} + 2\gamma_1^2 \qty(e^{-i\varphi} - \gamma e^{i\varphi}) \cos[2](2\varphi) + 3\gamma_1^2e^{3i\varphi} + \frac{\sigma_1^4}{\sigma^4}\qty(e^{-i\varphi} - \gamma e^{i\varphi})] + \order{\epsilon^4},\label{sm:eq:cascade_bulk_exact_approx_complex}
\end{align}
where we have also replaced $\varphi\to2\varphi$ for convenience. As $\varphi$ parametrizes the bulk, this has no effect on the final result. Note that we cannot compare \cref{sm:eq:cascade_bulk_exact_approx_complex} directly with \cref{sm:eq:bulk_boundary_complex} as the parameter $\varphi$ is different in the two cases. We must therefore eliminate the parameter $\varphi$, which we do my manipulating \cref{sm:eq:cascade_bulk_exact_approx_complex} into the form of the modified elliptical law. The calculation follows very similar steps to the derivation of the modified elliptical law in \cref{sm:section:approx_bulk}. 

First, we write $\bulk(\varphi) = x(\varphi) + iy(\varphi)$ and take the real and imaginary part of \cref{sm:eq:cascade_bulk_exact_approx_complex}. We obtain
\begin{align}
    \frac{x(\varphi)}{\sigma} &= (1 + \gamma)\cos(\varphi) - \frac{1}{6}\qty(2\gamma_1^2 - 2\gamma_1^2(3 - \gamma)\cos(2\varphi) - \gamma_1^2(1 + \gamma)\cos(4\varphi) + (1 - \gamma)\frac{\sigma_1^4}{\sigma^4})\cos(\varphi)\epsilon^2, \nonumber\\
    \frac{y(\varphi)}{\sigma} &= (-1 + \gamma)\sin(\varphi) + \frac{1}{6}\qty(2\gamma_1^2 - 2\gamma_1^2(3 + \gamma)\cos(2\varphi) - \gamma_1^2(1 - \gamma)\cos(4\varphi) + (1 + \gamma)\frac{\sigma_1^4}{\sigma^4})\sin(\varphi)\epsilon^2.   
\end{align}
Following \cref{sm:section:approx_bulk}, we re-arrange the above into the following form
\begin{align}
    \frac{x(\varphi)^2}{\sigma^2(1 + \gamma)^2} =& \cos[2](\varphi) + \frac{\epsilon^2}{12(1 + \gamma)}\Bigg{\{} \gamma_1^2(1 + \gamma)\nonumber\\
    &+ 2\qty[3\gamma_1^2(3 - \gamma) - 2\frac{\sigma_1^4}{\sigma^4}(1 - \gamma)]\cos[2](\varphi) - 16\gamma_1^2\sin[2](2\varphi) + \gamma_1^2(1 + \gamma)\cos(6\varphi)\Bigg{\}},\nonumber\\
    \frac{y(\varphi)^2}{\sigma^2(1 - \gamma)^2} =& \sin[2](\varphi) + \frac{\epsilon^2}{12(1 - \gamma)}\Bigg{\{} \gamma_1^2(1 - \gamma) \nonumber\\
    &+ 2\qty[3\gamma_1^2(3 + \gamma) - 2\frac{\sigma_1^4}{\sigma^4}(1 + \gamma)]\sin[2](\varphi) - 16\gamma_1^2\sin[2](2\varphi) - \gamma_1^2(1 - \gamma)\cos(6\varphi)\Bigg{\}},\label[plural_equation]{sm:eq:xy_sq_cascade}
\end{align}
by considering the product of the above expressions, we find
\begin{align}
    \sin(2\varphi)^2 &= \frac{4x(\varphi)^2y(\varphi)^2}{\sigma^4(1 - \Gamma^2)^2} + O\qty(\epsilon^2). \label{sm:eq:cascade_trig_3}
\end{align}
Hence, we can sum \cref{sm:eq:xy_sq_cascade} to get
\begin{multline}
    \frac{x^2}{\sigma^2(1 + \gamma)^2}\qty{1 - \frac{\epsilon^2}{3(1 + \gamma)}\qty[\gamma_1^2(5 - \gamma) - \frac{\sigma_1^4}{\sigma^4}(1 - \gamma)]} + \frac{y^2}{\sigma^2(1 - \gamma)^2}\qty{1 - \frac{\epsilon^2}{3(1 - \gamma)}\qty[\gamma_1^2(5 + \gamma) - \frac{\sigma_1^4}{\sigma^4}(1 + \gamma)]} \\
    = 1 - \frac{32\gamma_1^2\epsilon^2}{3}\frac{x^2y^2}{\sigma^4(1 - \gamma^2)^3}.\label{sm:eq:approx_from_exact_cascade}
\end{multline}
Using the binomial identity $(1 + A\epsilon^2/2)^{-2} = 1 - A\epsilon^2 + O\qty(\epsilon^4)$ to move the terms of order $\epsilon^2$ on the left-hand side of \cref{sm:eq:approx_from_exact_cascade} to the denominator, we arrive at an expression in the form of the modified elliptical law
\begin{align}
    \frac{x^2}{\qty{1 + \gamma + \frac{\epsilon^2}{6}\qty[\gamma_1^2(5 - \gamma) - \frac{\sigma_1^4}{\sigma^4}(1 - \gamma)]}^2} + \frac{y^2}{\qty{1 - \gamma + \frac{\epsilon^2}{6}\qty[\gamma_1^2(5 + \gamma) - \frac{\sigma_1^4}{\sigma^4}(1 + \gamma)]}^2} = \sigma^2 - \frac{32\gamma_1^2\epsilon^2}{3}\frac{x^2y^2}{\sigma^2(1 - \gamma^2)^3}.\label{sm:eq:approx_from_exact_cascade_bulk}
\end{align}
\subsubsection{Outliers}
The exact $k$th outlier eigenvalues $\outliers{k}$ for the cascade model with statistics as described in \cref{eq:cascade_stats} of the main text are given by \cref{sm:eq:outlier_cascade_final,sm:eq:outlier_cascade_2}. We repeat the results here with the small parameter $\epsilon$ included
\begin{align}
    \outliers{k} &= \frac{\sinh(A_k)\cosh(A_k)}{A_k}\eigval{\ell_k}[\vb u] + \frac{\gamma\sigma^2}{\eigval{\ell_k}[\vb u]}, \label{sm:eq:hankel_outlier}
\end{align}
where the complex number $A_k$ is a solution to
\begin{align}
    \frac{\sinh(A_k)}{\sqrt{A_k}} = \frac{\sqrt{\epsilon \gamma_1}\sigma}{2\epsilon\mu_1}\qty(\ln\qty|\frac{\mu + \epsilon\mu_1}{\mu - \epsilon\mu_1}| + i\pi q_{\ell_k}). \label{sm:eq:hankel_outliers_k}
\end{align}
As we are assuming that $\epsilon$ is small, we see that $q_{\ell_k}$, which is even if $\epsilon\mu_1\leq \mu$ and is odd otherwise, must run through the even integers. In fact, $q_{\ell_k}$ must be equal to $0$. For all other values of $q_{\ell_k}$, the eigenvalue $\eigval{k}[\vb u]$ is of order $\epsilon$ and therefore violates the inequality $|\eigval{k}[\vb u]| \geq \PF[\vb s]$. Explicitly, we see this by expanding $\eigval{k}[\vb u]$ [given by \cref{sm:eq:general_cascade_eigvals_1}] to second order in $\epsilon$, finding
\begin{align}
    \eigval{k}[\vb u] = \begin{cases} \frac{2 i \mu_1 \epsilon}{\pi q_{\ell_k}} + \frac{4 \mu_1^2 \epsilon^2}{\pi^2\mu q_{\ell_k}^2} + \order{\epsilon^3}, & q_{\ell_k} \neq 0 \\[3mm]
    \mu - \frac{\mu_1^2\epsilon^2}{3\mu} + \order{\epsilon^4}, & q_{\ell_k} = 0
    \end{cases}.\label{sm:eq:mu_eigval_approx}
\end{align}
Hence, as $\PF[\vb s] = \order{1}$, we need only solve \cref{sm:eq:hankel_outliers_k} for $A_1$, and not for each $A_k$.

Assuming that the unknown quantity $A_1$ can be written $A_1 = A_0 + \epsilon A_1 + \epsilon^2 A_2$, we expand \cref{sm:eq:hankel_outliers_k} and equate powers of $\epsilon$, finding
\begin{align}
    A_0 = 0, \hspace{1cm} A_1 = \frac{\gamma_1\sigma^2}{\mu^2}, \hspace{1cm} A_2 = 0.
\end{align}
On substituting $A_1 = \epsilon \gamma_1\sigma^2/\mu^2$ into \cref{sm:eq:hankel_outlier} and expanding to second order in $\epsilon$, we arrive at \cref{eq:outlier_approx} in the main text
\begin{align}
    \frac{\outlier}{\mu} = 1 + \frac{\gamma\sigma^2}{\mu^2} + \frac{\epsilon^2}{3}\qty[\frac{2\gamma_1^2\sigma^4}{\mu^4} - \qty(1 - \frac{\gamma\sigma^2}{\mu^2})\frac{\mu_1^2}{\mu^2}]. \label{sm:eq:approx_from_exact_cascade_outlier}
\end{align}
\subsection{fine-structure correction from the modified elliptical Law}
\label{sm:section:MEL_cascade_boundary}
We will now reproduce \cref{sm:eq:approx_from_exact_cascade_bulk} using our modified elliptical law, thereby verifying its correctness. Computation of the zeroth order fine-structure parameters from \cref{sm:eq:cascade_stats_perturbative} is straightforward, the zeroth order parameters are simply $\mu, \sigma^2$ and $\gamma$.

Calculation of the parameters $R, S, T, U$ and $V$ is also relatively straightforward [see \cref{sm:eq:fs_params} for their definitions]. The fine-structure parts of the statistics can also be read off from \cref{sm:eq:cascade_stats_perturbative}, we have 
\begin{align}
    u^{(1)}_{ij} &= \epsilon \mu_1 \sign(i - j), \nonumber\\
    s^{(1)}_{ij} &= \epsilon \sigma_1^2 \sign(i - j), \nonumber\\
    t^{(1)}_{ij} &= \epsilon \gamma_1\sigma^2 \sign(i + j - N),
\end{align}
where $\sign(x)$ is equal to $1$ if $x > 0$ and is $-1$ otherwise. For $N$ large, we can compute the fine-structure parameters by integrating. For example, the first two fine-structure parameters are
\begin{align}
    R &= \frac{\epsilon^2\gamma_1\sigma_1^4}{\sigma^4}\int_0^1 \dd[3]{x}\qty\big[\sign(x_1 - x_2) + \sign(x_2 - x_1)]\sign(x_2 + x_3 - 1) = \frac{\epsilon^2\gamma_1\sigma_1^4}{\sigma^4}\qty(\frac13 - \frac13) = 0, \nonumber\\
    S &= \frac{\epsilon^2\sigma_1^4}{\sigma^4}\int_0^1\dd[3]{x}\sign(x_1 - x_2)\sign(x_2 - x_3) = -\frac{\epsilon^2\sigma_1^4}{3\sigma^4}
\end{align}
The remaining quantities $S, T, U, V$ are computed similarly. We find
\begin{align}
    T = \frac{\epsilon^2\gamma_1^2}{3}, \hspace{1cm} U = -\frac{\epsilon^2\mu_1^2}{3\mu^2}, \hspace{1cm} V = 0.
\end{align}
With the fine-structure parameters in hand, we simply plug them into \cref{sm:eq:bulk_approx,sm:eq:outlier_approx} [or \cref{eq:bulk_approx,eq:outlier_approx} in the main text]. The result is \cref{sm:eq:approx_from_exact_cascade_bulk,sm:eq:approx_from_exact_cascade_outlier}, expressions for the support of the spectrum of an FSRM with cascade statistics, derived from the exact solution. We have therefore succeeded in analytically verifying the fine-structure correction to the elliptical law and outlier eigenvalue in the case of the cascade model.
\section{Approximate solutions for the support of the spectrum}
\label{sm:section:approximate_solutions_for}
\cref{section:directed_network,section:neural_networks,section:niche_model} in the main text all follow a similar format. A FSRM model is defined via some choice of the statistics $\vb u$, $\vb s$ and $\vb t$. Then the zeroth order parameters $\mu, \sigma$ and $\gamma$, as well as the fine-structure parameters $R, S, T, U, V$ are all computed from these statistics. In this section we perform these calculations in detail, as well as any other details which are omitted from \cref{section:directed_network,section:neural_networks,section:niche_model} in the main text.
\subsection{Directed networks}
\label{sm:section:directed_networks}
In section \cref{section:directed_network} of the main text, we consider a network built such that each pair of nodes $i$ and $j$ is joined with a directed edge ($i\gets j$ or $j\gets i$), an undirected edge ($i\leftrightarrow j$) or no link ($i \not\leftrightarrow j$) with mutually exclusive probabilities $P_{i\gets j}$, $P_{j\gets i}$, $P_{i\leftrightarrow j}$ and $P_{i\not\leftrightarrow j}$ respectively. If we call the adjacency matrix of this network $\vb A$, then we are interested in the random matrix $\vb A \circ \vb K$, where $\circ$ denotes an element wise product and where $\vb K$ is an elliptical random matrix with statistics given in \cref{eq:dr_non_zero_stats} of the main text. The network structure implies that diagonally opposite pairs of elements of $\vb K$ are drawn from the following joint probability distribution
\begin{align}
    \Pr(K_{ij}, K_{ji}) = \qty[1 - P_{i\leftrightarrow j} - P_{i\leftarrow j}]\delta(K_{ij})\delta(K_{ji}) + P_{i\leftrightarrow j}\pi_{\leftrightarrow}(K_{ij}, K_{ji}) + P_{i\leftarrow j}\pi_{\leftarrow}(K_{ij}, K_{ji}).\label{sm:eq:dr_adjacency_statistics}
\end{align}
The distributions $\pi_\leftrightarrow(K_{ij}, K_{ji}), \pi_\leftarrow(K_{ij}, K_{ji})$ of the non-zero elements of $\vb A \circ \vb K$ are the statistics of $\vb K$ 
\begin{align}
    \begin{aligned}
        \avg{K_{ij}}_{\leftrightarrow} &= \frac{\mu}{p}, \\
        \variance_{\leftrightarrow}(K_{ij}) &= \frac{\sigma^2}{p}, \\
        \cov_{\leftrightarrow}(K_{ij}, K_{ji}) &= \frac{\Gamma\sigma^2}{p}, 
    \end{aligned}
    \hspace{1cm}
    \begin{aligned}
        \avg{K_{ij}}_{\leftarrow} &= \frac{\mu}{p}, \\
        \variance_{\leftarrow}(K_{ij}) &= \frac{\sigma^2}{p}, \\
        \cov_{\leftarrow}(K_{ij}, K_{ji}) &= 0,        
    \end{aligned}\label[plural_equation]{sm:eq:dr_non_zero_statistics}
\end{align}
where the covariance of $K_{ij}$ and $K_{ji}$ is zero for the directed distribution because one of $K_{ij}$ or $K_{ji}$ must be zero if there is a directed link $j\gets i$ or $i\gets j$. We have excluded the possibility of a link $i\gets j$ and $j\gets i$ by definition of the network.

To derive the statistics of $\vb A \circ \vb K$ over the whole ensemble, rather than just over the distribution of non-zero elements, we show that the statistics of a random matrix constructed according to \cref{sm:eq:dr_adjacency_statistics,sm:eq:dr_non_zero_statistics} have the same statistics as a fully connected FSRM $\vb J$ with statistics [\cref{eq:dr_stats} in the main text]
\begin{align}
    \avg{J_{ij}}_J &= \qty(P_{i\leftrightarrow j} + P_{i\gets j})\frac{\mu}{p}, \nonumber\\
    \variance_J(J_{ij}) &= \qty(P_{i\leftrightarrow j} + P_{i\gets j})\frac{\sigma^2}{p}, \nonumber\\
    \cov_J(J_{ij}, J_{ji}) &= P_{i\leftrightarrow j}\frac{\Gamma\sigma^2}{p}.\label{sm:eq:network_stats}
\end{align}
To show this, we consider the eigenvalue potential associated with the matrix $\vb A\circ \vb K$ with statistics described in \cref{sm:eq:dr_adjacency_statistics,sm:eq:dr_non_zero_statistics}, it is given by 
\begin{align}
    \exp[N\Phi(z)]
    &= \avg{\int \prod_{ij} \frac{\dd[2]{p_i}\dd[2]{q_j}}{\pi^2} \exp[- \sum_{i}\conj{q_i} q_i + i\sum_{i} \conj{p_i} (\conj z \delta_{ij} - A_{ji}K_{ji})q_j + \conj{q_i}(z \delta_{ij} - A_{ij}K_{ij})p_j]}_{A\circ K}.
\end{align}
Focusing only on the parts which contain factors of $A_{ij}K_{ij}$ for the disorder average, we can write the average over realizations of $\vb A\circ \vb K$ in terms of the averages $\avg{\cdot}_\leftrightarrow$ and $\avg{\cdot}_\gets$ as 
\begin{multline}
    \avg{\exp[-i\sum_{ij} \conj{p_i} A_{ji}K_{ji}q_j + \conj{q_i}A_{ij}K_{ij}p_j]}_{A\circ K} = 1 - P_{i\leftrightarrow j} - P_{i \leftarrow j} \\
    + P_{i\leftrightarrow j}\avg{\exp[-i\sum_{ij} \conj{p_i} K_{ji}q_j + \conj{q_i}K_{ij}p_j]}_\leftrightarrow + P_{i\leftarrow j}\avg{\exp[-i\sum_{ij} \conj{p_i} K_{ji}q_j + \conj{q_i}K_{ij}p_j]}_\leftarrow.\label{sm:eq:disorder_avergae_dr0}
\end{multline}
Assuming that the average node degree $p$ is large, we can expand the exponentials inside the averages in powers of $1/p$ and average the terms in the expansion. We have
\begin{multline}
    \avg{\exp[-i\sum_{ij} \conj{p_i} K_{ji}q_j + \conj{q_i}K_{ij}p_j]}_\leftrightarrow = 1 - \frac{i}{p}\sum_{ij}\mu\qty(q_i\conj{p_i} + \conj{q_i}p_j)\\
    - \frac{1}{2p} \sum_{ij}\sigma^2 \qty(q_i\conj{p_j} + \conj{q_i}p_j)^2 + \Gamma\sigma^2\qty(p_i\conj{q_j} + \conj{p_i}q_j)\qty(p_j\conj{q_i} + \conj{p_j}q_i),
\end{multline}
for the disorder average over reciprocated links and 
\begin{align}
    \avg{\exp[-i\sum_{ij} \conj{q_i} K_{ji}q_j + \conj{q_i}K_{ij}q_j]}_\leftarrow = 1 - \frac{i}{p}\sum_{ij}\mu\qty(q_i\conj{q_j} + \conj{q_i}q_j)
    - \frac{1}{2p} \sum_{ij}\sigma^2 \qty(q_i\conj{q_j} + \conj{q_i}q_j)^2,
\end{align}
for the disorder average over directed links. Substituting these results back into \cref{sm:eq:disorder_avergae_dr0} and using the relation $1 + A/p \approx \exp(A/p)$, which is valid for large $p$, we can re-exponentiate the averaged statistics to obtain
\begin{align}
    \avg{\exp[-i\sum_{ij} \conj{q_i} A_{ji}K_{ji}q_j + \conj{q_i}A_{ij}K_{ij}q_j]}_{A\circ K} = \exp[- \frac{i}{p}\sum_{ij}\mu\qty{P_{i\leftrightarrow j} + P_{i \leftarrow j}}\qty(q_i\conj{q_i} + \conj{q_i}q_j)] \nonumber\\
    \times\exp[- \frac{1}{2p} \sum_{ij}\sigma^2 \qty{P_{i\leftrightarrow j} + P_{i \leftarrow j}}\qty(q_i\conj{q_j} + \conj{q_i}q_j)^2 + \Gamma\sigma^2P_{i\leftrightarrow j}\qty(q_i\conj{q_j} + \conj{q_i}q_j)\qty(q_j\conj{q_i} + \conj{q_j}q_i)].
\end{align}
Finally, we recognize that one would obtain the exact same expression for the eigenvalue potential from a fully connected FSRM with statistics as in \cref{sm:eq:network_stats}. Note that it would also be straightforward to modify this argument to allow for the matrix $\vb K$ to be an FSRM, rather than an elliptical matrix.
\subsubsection{fine-structure parameters}
To find the elliptical and fine-structure parameters for a FSRM with statistics as in \cref{sm:eq:network_stats}, we first recall the definitions of the undirected, in and out degree distributions from the main text [\cref{eq:dr_degree_distributions}]
\begin{align}
    k^\leftrightarrow_i = \sum_j P_{i\leftrightarrow j}, \hspace{5mm}
    k^\text{in}_i = \sum_j P_{i\gets j},\hspace{5mm}
    k^\text{out}_i = \sum_j P_{j\gets i}.\label[plural_equation]{sm:eq:degree_distributions}
\end{align}
We also recall the definition of the reciprocity of the network from the main text [\cref{eq:reciprocity}]
\begin{align}
    r = \frac{1}{pN}\sum_{ij}P_{i\leftrightarrow j},\label{sm:eq:reciprocity}
\end{align}
where $p$ is the average degree of a node in the network, which can be expressed in terms of the probabilities $P_{i\leftrightarrow j}$ and $P_{i\gets j}$
\begin{align}
    p = \frac{1}{N}\sum_{ij}\qty\Big(P_{i\leftrightarrow j} + P_{i\gets j}).
\end{align}

We first use \cref{eq:avg_params} from the main text to compute the zeroth order parameters in the fine structure. For the zeroth order mean interaction strength, we have
\begin{align}
    \frac{1}{N^2}\sum_{ij}\frac{N}{p}\qty\Big[P_{i\leftrightarrow j} + P_{i\gets j}]\mu = \mu,
\end{align}
similarly, the remaining two zeroth order parameters are $\sigma^2$ and $\gamma = \Gamma r$.

The first order fine-structure parameters are similarly straightforward to compute from the definitions given in \cref{eq:fs_params} in the main text. For example, we compute the parameter $R$
\begin{align}
    R &= \frac{1}{2N^3\sigma^4}\sum_{ijk}\qty[\frac{N\Gamma\sigma^2}{p}P_{i\leftrightarrow j} - \Gamma r\sigma^2]\qty[\frac{N\sigma^2}{p}\qty\Big(2P_{i\leftrightarrow j} + P_{i\gets j} + P_{j\gets i}) - 2\sigma^2], \nonumber\\
    &= \frac{\Gamma}{2Np^2}\sum_j\qty\Big[k^\leftrightarrow_j - rp]\qty\Big[2k^\leftrightarrow_j + k^\text{in}_j + k^\text{out}_j - 2p], \nonumber\\
    &= \Gamma\qty(h^2 + \tau),
\end{align}
where we have recognized $h^2$ and $\tau$ from their definitions in \cref{eq:network_stats} of the main text, they are the degree heterogeneity and the correlation coefficient between directed and exclusively undirected links respectively. Recalling the definition of the degree correlation coefficient $\rho$ from \cref{eq:network_stats} in the main text, we can compute the remaining fine-structure parameters following the same procedure as that used for $R$. Ultimately, all the fine-structure parameters can be computed from the degree distributions $k^\text{in}_i, k^\text{out}_i$ and $k^\leftrightarrow_i$, we find
\begin{align}
    R &= \Gamma \qty(\tau + h^2), \nonumber\\
    S &= \rho + 2\tau + h^2, \nonumber\\
    T &= \Gamma^2 h^2, \nonumber\\
    U &= \rho + 2\tau + h^2, \nonumber\\
    V &= \Gamma \qty(\tau + h^2).\label[plural_equation]{sm:eq:network_fs_params}
\end{align}
It is now straightforward to compute the fine-structure correction to the elliptical law and outlier eigenvalue using \cref{eq:bulk_approx,eq:outlier_approx} due to the network structure in $\vb A\circ \vb K$. 

\subsubsection{Test of the outlier condition with dichotomous degree distributions}
\label{sm:section:dichotomous_deg_dist}
To test our fine-structure correction to the spectrum of a directed network we use a network with dichotomous degree distributions. Specifically, the network has degree distributions 
\begin{align}
    k^\text{in}_i &= \begin{cases}
        (1 - r)(p + \epsilon p^\text{in}) &\text{if  }i < \frac{N}{2}, \\
        (1 - r)(p - \epsilon p^\text{in}) &\text{if  }i \geq \frac{N}{2},
    \end{cases}\nonumber\\
    k^\text{out}_i &= \begin{cases}
        (1 - r)(p + \epsilon p^\text{out}) &\text{if  }i < \frac{N}{2}, \\
        (1 - r)(p - \epsilon p^\text{out}) &\text{if  }i \geq \frac{N}{2},
    \end{cases}\nonumber\\
    k^\leftrightarrow_i &= \begin{cases}
        r(p + \epsilon p^\leftrightarrow) &\text{if  }i < \frac{N}{2}, \\
        r(p - \epsilon p^\leftrightarrow) &\text{if  }i \geq \frac{N}{2}.
    \end{cases}\label[plural_equation]{sm:eq:dichotomous_degree_distributions}
\end{align}
We can compute the relevant statistics for the fine-structure correction, they are 
\begin{align}
    h^2 &= \qty(\frac{r p^\leftrightarrow}{p})^2\epsilon^2, \nonumber\\
    \tau &= r(1 - r)\frac{p^\leftrightarrow\qty( p^\text{in} + p^\text{out})}{2p^2}\epsilon^2, \nonumber\\
    \rho &= (1 - r)^2\frac{p^\text{in} p^\text{out}}{p^2}\epsilon^2.
\end{align}
On substituting into \cref{eq:bulk_approx,eq:outlier_approx} in the main text, we find the fine-structure correction to the spectrum of random matrices with a dichotomous degree distribution.

\begin{figure}
    \includegraphics[width=0.7\textwidth]{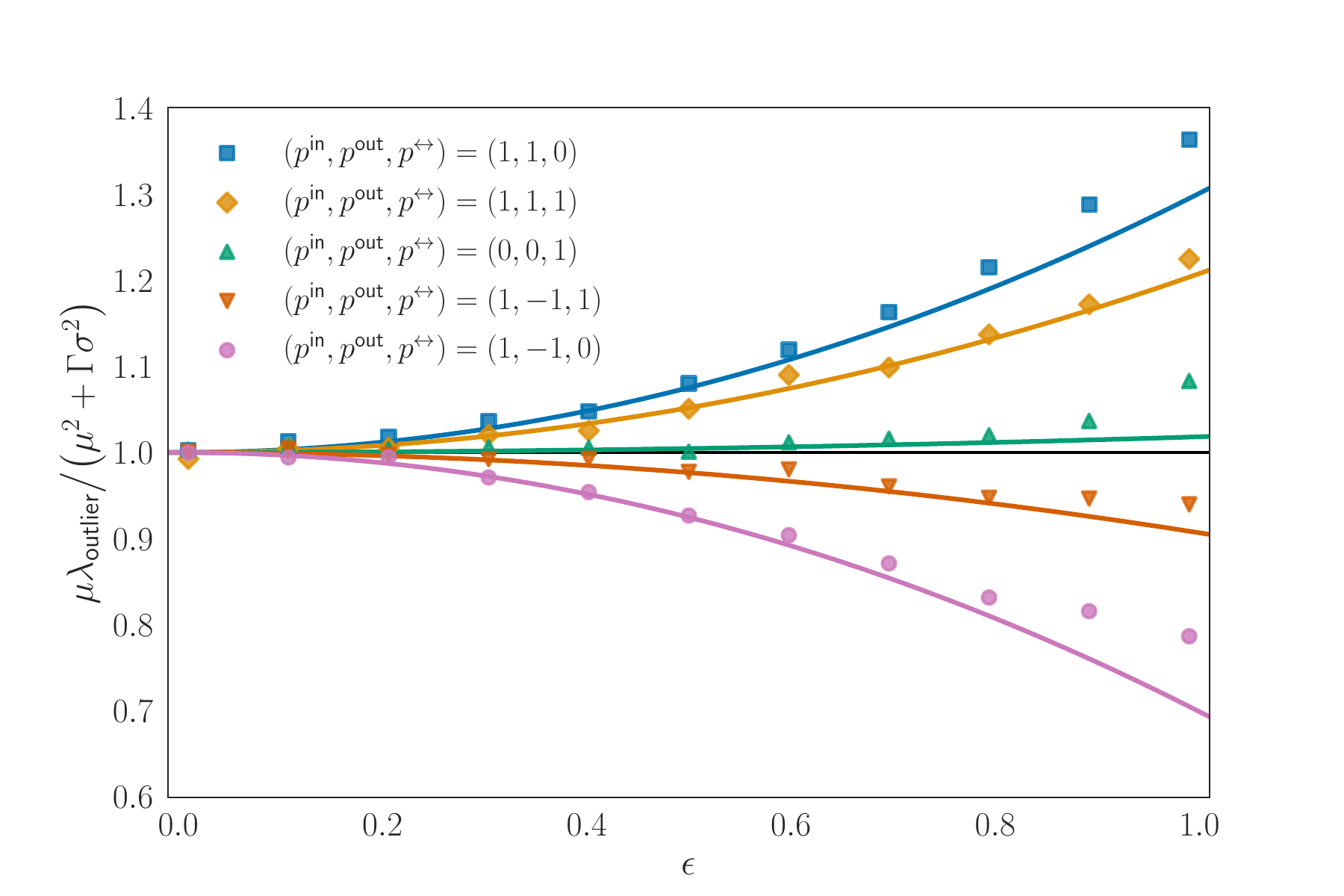}
    \caption{Test of the fine-structure correction to the outlier of a directed network [\cref{eq:network_outlier}]. Parameters for all curves are $\mu=1.5, \sigma=1.0, \gamma=0.5, r=0.4$ and $p=0.1$. The parameters $p^\text{in}$, $p^\text{out}$ and $p^\leftrightarrow$ are set to $1$ or $0$ as indicated. Curves are generated from \cref{eq:outlier_approx} in the main text and markers are the maximum eigenvalue of random matrices constructed according the description in \cref{sm:section:directed_networks} with the statistics of non-zero elements given by \cref{sm:eq:dr_non_zero_statistics} and on a network with degree distributions given by \cref{sm:eq:dichotomous_degree_distributions}. Markers are the average of $10$ realizations of the matrix with $N = 4000$.}
    \label{sm:fig:network_outlier}
\end{figure}

In the main text, we claim that if the underlying network is either completely directed ($P_{i\gets j} = 0$) or completely undirected ($P_{i\leftrightarrow j} = 0$), then our results reduce to known cases \cite{neriLinearStabilityAnalysis2020,baron_directed_2022}. In the following two sections we demonstrate this.
\subsubsection{Undirected underlying network}
\label{sm:section:undirected_underlying_network}
When the underlying network is undirected, we have $k^\text{in}_i = k^\text{out}_i = 0$. In this case we find that the reciprocity of the network $r = 1$ and that the network statistics $\rho$ and $\tau$ both vanish. In this case, we can substitute the values of $\mu, \sigma^2, \gamma, R, S, T, U, V$ from \cref{sm:eq:network_fs_params} [\cref{eq:network_fs_params} in the main text] into the modified elliptic law [\cref{eq:bulk_approx} in the main text] to find the fine-structure correction to the boundary of the bulk spectrum 
\begin{align}
    \frac{x^2}{\qty(1 + \Gamma)^2\qty[1 + \frac12h^2\qty(1 + 2\Gamma - \Gamma^2)]^2} + \frac{y^2}{\qty(1 - \Gamma)^2\qty[1 + \frac12h^2\qty(1 - 2\Gamma - \Gamma^2)]^2} = \sigma^2,
\end{align}
as well as the location of the outlier eigenvalue
\begin{align}
    \outlier &= \qty(\mu + \frac{\Gamma \sigma^2}{\mu})\qty(1 + h^2).
\end{align}
These results are equivalent to Eqs.~(30) and (32) of Ref \cite{baron_directed_2022}, verifying our fine-structure corrections to the elliptical law and outlier eigenvalue.
\subsubsection{Directed underlying network}
\label{sm:section:directed_underlying_network}
When the underlying network is directed, we have $k^\leftrightarrow_i = 0$ and find $r = h^2 = \tau = 0$. In this case we can substitute the fine-structure parameters into the modified elliptical law [\cref{eq:bulk_approx}]
\begin{align}
    \frac{x^2}{\qty(1 + \rho/2)^2} + \frac{y^2}{\qty(1 + \rho/2)^2} = \sigma^2,\label{sm:eq:directed_network_bulk}
\end{align}
and we can find the fine-structure correction to the outlier eigenvalue [\cref{eq:outlier_approx}]
\begin{align}
    \outlier = \mu\qty(1 + \rho).\label{sm:eq:directed_netowork_outlier}
\end{align}
In fact, if we re-arrange \cref{sm:eq:directed_network_bulk} using the binomial expansion $1/(1 + \rho/2)^2 = 1 - \rho + \order{\rho^2}$ into the form $x^2 + y^2 = \sigma^2(1 + \rho)$, then \cref{sm:eq:directed_network_bulk,sm:eq:directed_netowork_outlier} are the exact results for the support of the FSRM $\vb J$ found in Ref.~\cite{neriLinearStabilityAnalysis2020}. In \cref{sm:section:the_case_T_0}, we re-derive the exact solution found in Ref.~\cite{neriLinearStabilityAnalysis2020} using \cref{sm:eq:bulk_boundary_general,sm:eq:outliers_general,sm:eq:outlier_ellimination_general,sm:eq:resolvent_block}, our self-consistent equations for the support of the spectrum of a general FSRM.
\subsection{Neural network}
\label{sm:section:neural_network}
In \cref{section:neural_networks} of the main text, we consider a FSRM model for which the fine structure is due to spatial dependence of the interactions between neurons in a grid. The statistics of the FSRM we consider are given by \cref{eq:neural_stats} in the main text, we repeat them here 
\begin{align}
    u_{ij} &= \mu, \nonumber\\
    s_{ij} &= \sigma^2, \nonumber\\
    t_{ij} &= \Gamma\sigma^2\cos(k|\vb r_i - \vb r_j|). \label[plural_equation]{sm:eq:neural_stats}   
\end{align}
The quantity $\vb r_i$ is the position of the $i^\text{th}$ neuron in the grid. As discussed in the main text, for $N = n^2$ neurons regularly spaced in a $1\times 1$ grid, we observe that the location of each neuron is of the form $\vb r_i = (x_i/n, y_i/n)$ for some natural numbers $x_i, y_i$. One particular choice of the ordering of neurons is shown in \cref{fig:neural_networks} (t) in the main text.

The absence of fine structure in the matrices $\vb u$ and $\vb s$ implies that the only non-zero fine-structure parameter is $T$. The first non-trivial calculation is the derivation of \cref{eq:neural_gamma_calc} in the main text, that is, the calculation of the correlation coefficient 
\begin{align}
    \gamma &= \frac{\Gamma}{N^2}\sum_{ij}\cos(k|\vb r_i - \vb r_j|), \nonumber\\
    &= \frac{\Gamma}{n^4}\sum_{i,j=1}^N\cos\qty[k\sqrt{\qty(\frac{x_i}{n} - \frac{x_j}{n})^2 + \qty(\frac{y_i}{n} - \frac{y_j}{n})^2}].\label{sm:eq:neural_gamma_calc}
\end{align}
The exact functional form of $x_i$ and $y_i$ depends on the specifics of how the $N$ neurons are arranged and labelled in the $n\times n$ grid. However, as the summation is over all neurons in the grid, the specific ordering, $x_i$ and $y_i$, is unimportant. We only need to ensure that the sum runs over all possible interneuron distances. This is achieved by writing the double sum in \cref{sm:eq:neural_gamma_calc} as an equivalent quadruple sum 
\begin{align}
    \gamma = \frac{\Gamma}{n^4}\sum_{ijkl}^{n}\cos\qty[k\sqrt{\qty(\frac{i}{n} - \frac{j}{n})^2 + \qty(\frac{k}{n} - \frac{l}{n})^2}].
\end{align}
If the number of neurons $n$ is large, then the quadruple sum approaches the following integral
\begin{align}
    \gamma = \Gamma\int_0^1\dd{x}\dd{x'}\dd{y}\dd{y'}\cos(k\sqrt{(x - x')^2 + (y - y')^2}),
\end{align}
which is equivalent to \cref{eq:neural_gamma_calc} in the main text. We can also follow a very similar line of reasoning to derive \cref{eq:neural_T_calc} in the main text. 
\subsection{The niche model}
\label{sm:section:niche_model}
\begin{figure}
    \includegraphics[width=0.6\textwidth]{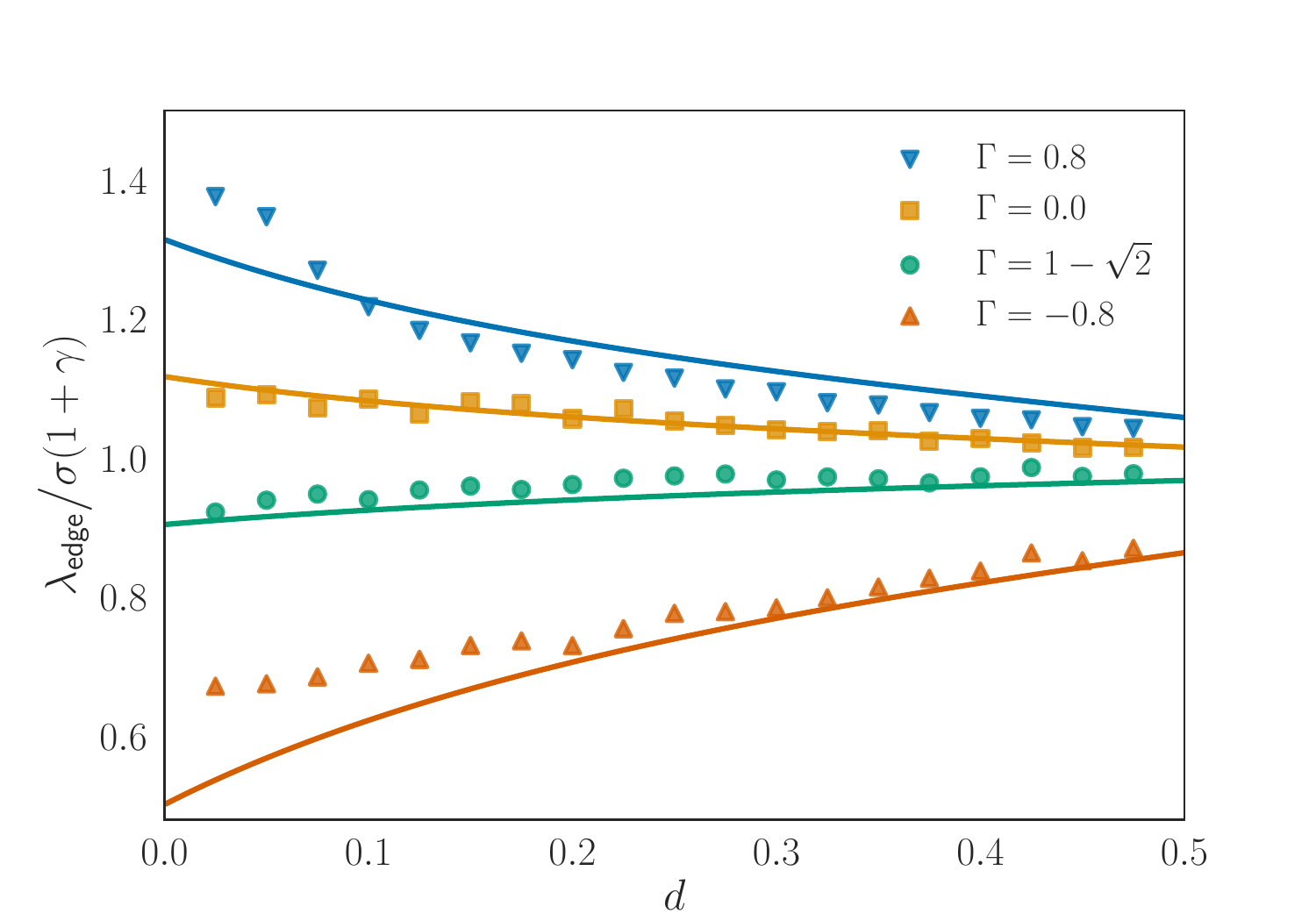}
    \caption{Ratio of the bulk edge of a niche model matrix to an elliptical matrix with no fine-structure correction. Lines indicate the result of the fine-structure correction [\cref{eq:outlier_approx,eq:bulk_approx}] and markers are computed from a single realization of a $10000\times 10000$ matrix, with $\mu_L=0, \mu_U=0, \sigma_L=1.2, \sigma_U=0.8$ and $\Gamma, d$ as indicated.}
    \label{sm:fig:niche_varying_d}
\end{figure}
In this section, we focus on the more technical parts of the derivation of \cref{eq:niche_fs_params,eq:niche_network_stats_expression} in the main text. In particular, we derive the leading-order fine-structure correction parameters for a FSRM constructed according to the niche model. Given a particular realization of the matrix $\vb E$, the statistics of such a matrix are [see \cref{eq:niche_stats} in the main text]
\begin{align}
    u_{ij} &= E_{ij}\mu_L + E_{ji}\mu_U, \nonumber\\
    s_{ij} &= E_{ij}\sigma^2_L + E_{ji}\sigma^2_U, \nonumber\\
    t_{ij} &= \Gamma\sigma_L\sigma_U\qty(E_{ij} + E_{ji}). \label[plural_equation]{sm:eq:niche_stats}
\end{align}
We emphasize that the statistics given are those of a random matrix constructed according to the niche model for a fixed instance of the matrix $\vb E$. However, as we will show, the elliptical and fine-structure parameters computed from \cref{sm:eq:niche_stats} ultimately depend on statistical properties of $\vb E$, not on any particular instance. 

To find the fine-structure correction, we first plug the statistics \cref{sm:eq:niche_stats} into \cref{sm:eq:avg_params} to find expressions for the elliptical and fine-structure parameters in terms of $\mu_L, \mu_U, \sigma_L, \sigma_U, \Gamma$, and the matrix $\vb E$. We will then show that these parameters do not depend on the specific instance of $\vb E$ that we use, but on the statistics of the elements of $\vb E$, which are ultimately functions of the average niche radius $d$ only. By approximating these statistics, we finally find expressions for the elliptical and fine-structure parameters in terms of the model parameters $\mu_L, \mu_U, \sigma_L, \sigma_U, \Gamma$, and $d$.
\subsubsection{fine-structure parameters}
\label{sm:section:eijeji_0}
We can find the elliptical parameters in a straightforward manner by substituting the statistics \cref{sm:eq:niche_stats} into \cref{sm:eq:avg_params} [\cref{eq:avg_params} in the main text], which gives
\begin{align}
    \mu &= \tilde d\qty(\mu_L + \mu_U), \nonumber\\
    \sigma^2 &= \tilde d\qty(\sigma^2_L + \sigma^2_U), \nonumber\\
    \gamma &= \frac{2\Gamma\sigma_L\sigma_U}{\sigma_L^2 + \sigma_U^2},
\end{align}
where $\tilde d = N^{-2}\sum_{ij}E_{ij}$. We can also compute the fine-structure parameters similarly, by plugging the statistics into \cref{sm:eq:fs_params} [\cref{eq:fs_params} in the main text]. For example, we can compute $S$ as follows
\begin{align}
    S &= \frac{1}{\tilde d^2\qty(\sigma_L^2 + \sigma_U^2)^2N^3}\sum_{ijk}\qty\Big(\qty\big[E_{ij}\sigma_L^2 + E_{ji}\sigma_U^2]\qty\big[E_{jk}\sigma_L^2 + E_{jk}\sigma_U^2] - \tilde d^2\qty\big[\sigma_L^2 + \sigma_U^2]^2), \nonumber \\
    &= \frac{1}{\tilde d^2\qty(\sigma_L^2 + \sigma_U^2)^2N^3}\sum_{ijk}\qty\Big(\qty\big[\sigma_L^2 - \sigma_U^2]^2E_{ij}E_{jk} + \sigma_L^2\sigma_U^2\qty\big[E_{ij} + E_{ji}]\qty\big[E_{jk} + E_{kj}] - \tilde d^2\qty\big[\sigma_L^2 + \sigma_U^2]^2), \nonumber \\
    &= \frac{\sigma_L^2\sigma_U^2}{\qty(\sigma_L^2 + \sigma_U^2)^2}\frac{1}{\tilde d^2N^3}\sum_{ijk}\qty\Big(\qty\big[E_{ij} + E_{ji}]\qty\big[E_{jk} + E_{kj}] - 4\tilde d^2) + \qty(\frac{\sigma_L^2 - \sigma_U^2}{\sigma_L^2 + \sigma_U^2})^2\frac{1}{\tilde d^2N^3}\sum_{ijk}\qty\Big(E_{ij}E_{jk} - \tilde d^2).
\end{align}
Using the identity $\sigma_L^2\sigma_U^2/(\sigma_L^2 + \sigma_U^2)^2 = [1 - (\sigma_L^2 - \sigma_U^2)^2/(\sigma_L^2 + \sigma_U^2)^2]/4$, we arrive at
\begin{align}
    S &= h^2 + \qty(\frac{\sigma_L^2 - \sigma_U^2}{\sigma_L^2 + \sigma_U^2})^2f,
\end{align}
where we have defined the following quantities (which are statistical properties of the matrix $\vb E$),
\begin{align}
    h^2 &= \frac{1}{4\tilde d^2N^3}\sum_{ijk}\qty\Big(\qty\big[E_{ij} + E_{ji}]\qty\big[E_{jk} + E_{kj}] - 4\tilde d^2),\nonumber\\
    f &= \frac{1}{4\tilde d^2N^3}\sum_{ijk}\qty\Big(E_{ij} - E_{ji})\qty\Big(E_{jk} - E_{kj}). \label[plural_equation]{sm:eq:niche_hf}
\end{align}
The remaining fine-structure parameters can be calculated similarly, and eventually we end up with 
\begin{align}
    R &= \gamma h^2, \nonumber\\
    S &= h^2 + \qty(\frac{\sigma_L^2 - \sigma_U^2}{\sigma_L^2 + \sigma_U^2})^2f\nonumber\\
    T &= \gamma^2 h^2, \nonumber\\
    U &= h^2 + \qty(\frac{\mu_L - \mu_U}{\mu_L + \mu_U})^2f\nonumber\\
    V &= \gamma h^2.\label[plural_equation]{sm:eq:niche_fs_params}
\end{align}
Whilst we have derived the elliptical and fine-structure parameters for a specific realization of the matrix $\vb E$, \cref{sm:eq:niche_fs_params} reveal that our correction only depends on the network structure through $d, h^2$, and $f$, which, for large $N$, are properties of the ensemble from which $\vb E$ is drawn, and not of any particular instance of the network. That is, for large $N$, we have 
\begin{align}
    \tilde d &= \avg{\frac{1}{N^2}\sum_{ij}E_{ij}}_E, \nonumber\\
    h^2 &= \avg{\frac{1}{4\tilde d^2N^3}\sum_{ijk}\qty\Big(\qty\big[E_{ij} + E_{ji}]\qty\big[E_{jk} + E_{kj}] - 4\tilde d^2)}_E, \nonumber\\
    f &= \avg{\frac{1}{4\tilde d^2N^3}\sum_{ijk}\qty\Big(E_{ij} - E_{ji})\qty\Big(E_{jk} - E_{kj})}_E. \label[plural_equation]{sm:eq:niche_averaged_network_stats}    
\end{align}
Hence, $\tilde d, h^2$, and $f$ are non-random functions of $d$ only, as that is the only parameter used to construct realizations of the network. If we can determine these functions, then we have the fine-structure correction to the spectrum of a random matrix constructed according to the niche model.
\subsubsection{Approximating the statistics of the network}
First, we observe that $\tilde d, h^2$, and $f$ can all be written in terms of the following quantities, which are simpler
\begin{align}
    E^\text{in}_i &= \frac{1}{N}\sum_jE_{ij}, \nonumber\\
    E^\text{out}_i &= \frac{1}{N}\sum_jE_{ji}. \label[plural_equation]{sm:eq:niche_ein_eout}
\end{align}
 We also repeat the definition of $E_{ij}$ from \cref{section:niche_model_definition} in the main text with the dependence of $E_{ij}$ on the niche position $\eta_i$, range $d_i$, and center $c_i$ (which are all random variables, also defined in \cref{section:niche_model_definition} of the main text) made explicit
\begin{align}
    E_{ij}\qty(\eta_j, \eta_i, d_i, c_i) &= 
    \begin{cases}
        1&\qc c_i - d_i/2 < \eta_j < c_i + d_i/2,\\
        0&\qc \text{otherwise}.
    \end{cases}
\end{align}
For the definitions of the random variables $\eta_i, d_i$, and $c_i$, see \cref{section:niche_model_definition} of the main text.

It turns out that $E^\text{in}_i$ is much easier to compute than $E^\text{out}_i$, so we start with $E^\text{in}_i$. As $\eta_j$ is a uniformly distributed random variable on the interval $[0, 1]$, and as both of $c_i - d_i/2$ and $c_i + d_i/2$ are, by definition, confined to the interval $[0, 1]$, the sum $\sum_jE_{ij}/N$ is simply equal to the length of the interval $[c_i - d_i/2, c_i + d_i/2]$ when $N$ is large. That is, we have 
\begin{align}
    E^\text{in}_i = d_i. \label{sm:eq:niche_e_in}
\end{align}
As an immediate consequence, we also have $\tilde d = \sum_i d_i/N = d$. The  expression for $\avg{E^\text{in}_i}_E$ is compared to numerical simulations in \cref{sm:fig:niche_e_inout}(b).

On the other hand, $E^\text{out}_i$ is more difficult to evaluate, so we resort to an approximation. We write
\begin{align}
    E^\text{out}_i &= \frac{1}{N}\sum_j E^\text{approx}_{ji}
    + \delta_i, \label{sm:eq:eout_approx}
\end{align}
where 
\begin{align}
    E^\text{approx}_{ij} &= E_{ij}\qty\Big(\avg{\eta_j}_E, \avg{\eta_i}_E, \avg{d_i}_E, \avg{c_i}_E),
\end{align}
and where $\delta_i$ is the error in the approximation. The error depends on the higher central moments of $\eta_i, d_i$, and $c_i$. Whilst $\delta_i$ itself is generally not small [see \cref{sm:fig:niche_e_inout}(c)], we will see that the sum $\sum_i\delta_i/N$ vanishes for large $N$.

The average of $\eta_i, d_i$, and $c_i$ over realizations of $\vb E$ are straightforward to compute using their definitions [see \cref{section:niche_model_definition} of the main text]. The variables $\eta_i$ are uniformly distributed random variables on the interval $[0, 1]$, and they are ordered so that $\eta_1 < \eta_2 < \dots < \eta_N$. The fact that these variables are ordered implies that $\avg{\eta_i}_E = i/N$ and that the variance of $\eta_i$ is sub-leading in $N$ \cite{mostellerUsefulInefficientStatistics1946,jonesKumaraswamyDistributionBetatype2009}. That is, each $\eta_i$ is equal to its average, up to a term sub-leading in $N$. The niche range of each species is given by $d_i = X\eta_i$, where $X$ is a beta distributed random variable drawn independently for each species with PDF $p_X(x) = \qty(\frac{1}{2r} - 1) (1 - x)^{\frac{1}{2r} - 2}$. The average value of $d_i$ over realizations of $\vb E$ is therefore 
\begin{align}
    \avg{d_i}_E = 2d\eta_i.\label{sm:eq:avg_di}
\end{align}
The distribution of the niche center of species $i$, $c_i$, is
\begin{align}
    c_i = 
    \begin{cases}
        c^{(1)}_i\qc &\eta_i + \frac12 d_i < 1\\
        c^{(2)}_i\qc &\eta_i + \frac12 d_i > 1
    \end{cases}\label{sm:eq:ci}
\end{align}
where $c^{(1)}_i$ is a uniform random variable on the interval $[d_i/2, \eta_i]$ and $c^{(2)}_i$ is uniform on the interval $[d_i/2, 1 - d_i/2]$. To compute $\avg{c_i}_E$, we have to average over both the distribution of $d_i$ and of $c_i$, this can be done using the law of total expectation. Recalling $d_i = X\eta_i$, we have 
\begin{align}
    \avg{c_i}_E &= \int_0^{2\qty(\frac{1}{\eta_i}-1)} \frac12\eta_i\qty(1 + \frac12 x)p_X(x)\dd{x} + \int_{2\qty(\frac{1}{\eta_i}-1)}^1 \frac12 p_X(x) \dd{x} \nonumber\\
    &= \frac12\eta_i\qty(1 + d) - \frac12 \eta_i d\qty(3 - \frac{2}{\eta_i})^{\frac{1}{2d}}H\qty(\eta_i - \frac{2}{3}),\label{sm:eq:ci_exact}
\end{align}
where $H(x) = \max(0, x)$. 

By inspection, we can simplify $\avg{c_i}_E$ with the following approximation
\begin{align}
    \avg{c_i}_E \approx \frac{1}{2}\max\qty[\eta_i(1 + d), 1]. \label{sm:eq:ci_approx}
\end{align}
The expressions in Eqs.~(\ref{sm:eq:ci_exact}) and (\ref{sm:eq:ci_approx}) are compared with one another in \cref{sm:fig:niche_e_inout}(a). Overall, we find that there is a maximum error of $\sim 5\%$. The two expression coincide with each other if any of the following conditions is fulfilled: $\eta_i < 2/3$, $\eta_i=1$, $d = 1/2$, or $d = 0$. We will use the approximation in \cref{sm:eq:ci_approx} to compute the fine-structure parameters. 

Substituting for $\avg{\eta_i}_E, \avg{d_i}_E$, and our approximation to $\avg{c_i}_E$ [$\avg{\eta_i}_E = i/N$, \cref{sm:eq:avg_di,sm:eq:ci_approx}] for the summand in \cref{sm:eq:eout_approx}, we arrive at
\begin{align}
    E_{ij}^\text{approx} \approx 
    \begin{cases}
        1, & \text{if }\quad i\qty(1 - d) < 2j < i\qty(1 + 3d) \text{ and } i(1 + d) < N \\
        \hspace{4mm}&\text{or }\quad N - 2di < 2j < N + 2di\ \text{ and } i(1 + d) > N, \\
        0, &\text{otherwise.}
    \end{cases}\label{sm:eq:niche_E_approx}
\end{align}
To determine $E^\text{out}_i$ from $E_{ij}^\text{approx}$, we note that, in the large $N$ limit, the summations in \cref{sm:eq:niche_ein_eout} can be approximated by integrals over the variable $\alpha \equiv i/N$ from $0$ to $1$. Defining $E^\text{approx}(i/N, j/N) = E_{ij}^\text{approx}$, and similar for $E_{ij}, E^\text{in}_i,E^\text{out}_i$, and the error term $\delta_i$, we have
\begin{align}
    E^\text{out}(\alpha) &= \int_0^1\dd{\beta}E^\text{approx}(\beta, \alpha) + \delta(\alpha)
    = 
    \left.
    \begin{cases}
        \frac{8d\alpha}{(1 - d)(1 + 3d)}\qc &0 < \alpha < \frac12 - d,\\
        1 - \frac{1}{2d} + \frac{\alpha}{d} + \frac{8d\alpha}{(1 - d)(1 + 3d)}\qc &\frac12 - d < \alpha < \frac12\frac{1 - d}{1 + d}, \\
        1 - \frac{2\alpha}{1 + 3d}\qc &\frac12\frac{1 - d}{1 + d} < \alpha < \frac12\frac{1 + 3d}{1 + d}, \\
        1 + \frac{1}{2d} - \frac{\alpha}{d}\qc &\frac12\frac{1 + 3d}{1 + d} < \alpha < \frac12 + d, \\
        0\qc & \text{otherwise}
    \end{cases}
    \right\} + \delta(\alpha)\label{sm:eq:niche_e_out}.
\end{align}
Noting that $\int_0^1\dd{\alpha}E^\text{out}(\alpha) = d$ exactly, we can compute the integral of the RHS of \cref{sm:eq:niche_e_out} and conclude that $\int_0^1\dd{\alpha}\delta(\alpha) = 0$. We compare our expressions for \cref{sm:eq:niche_e_in,sm:eq:niche_e_out} to the results of numerical simulations in \cref{sm:fig:niche_e_inout}(b) and (c).
\begin{figure}
    \includegraphics[width=\textwidth]{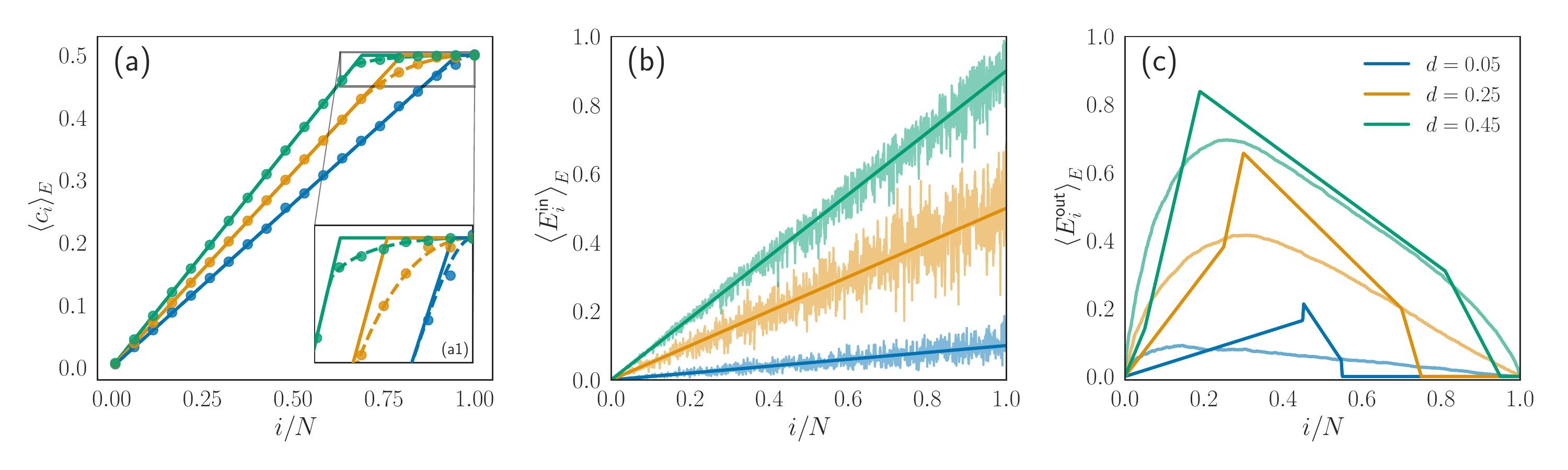}
    \caption{(a) Comparison of the exact (dashed lines) expression for $\avg{c_i}_E$ [\cref{sm:eq:ci_exact}] and its approximation (solid lines) [\cref{sm:eq:ci_approx}]. The functions coincide when $d = 0$ or $d = 1/2$ and only differ in a small range for intermediate values of $d$, shown zoomed in as indicated in (a1). Markers are the result of averaging $10000$ realizations of the niche center $c_i$. (b) and (c) Comparison of numerical (faded wiggley lines) values of $\avg{E^\text{in}_i}_E$ and $\avg{E^\text{out}_i}_E$ [\cref{sm:eq:niche_ein_eout}, averaged over $10$ realizations] and their approximations [(a) solid straight lines, (b) solid piecewise straight lines] as given in \cref{sm:eq:niche_e_in,sm:eq:niche_e_out}.}
    \label{sm:fig:niche_e_inout}
\end{figure}

Let us now use $E^\text{in}(\alpha)$ and $E^\text{out}(\alpha)$ [\cref{sm:eq:niche_e_in,sm:eq:niche_e_out}] to find expressions for $h^2$ and $f$ in terms of the average niche range $d$ only. Neglecting terms proportional to $\delta(\alpha)$, we have 
\begin{align}
    h^2 &\approx \frac{1}{4d^2}\int_0^1\qty[E^\text{in}(\alpha) + E^\text{out}(\alpha)]^2\dd{\alpha}  - 1, \nonumber \\
    f &\approx \frac{1}{4d^2}\int_0^1 \qty[E^\text{in}(\alpha) - E^\text{out}(\alpha)]^2\dd{\alpha}. \label[plural_equation]{sm:eq:niche_h_and_f_integrals} 
\end{align}
On carrying out the integrals, we find 
\begin{align}
    h^2 &\approx \frac{2 + d - 5d^2 - 3d^3 + 3d^4}{6(1 - d)(1 + d)^2(1 + 3d)},\nonumber\\
    f &\approx -\qty(\frac{1}{3(1 + d)^2} + h^2). \label{sm:eq:niche_h_and_f}
\end{align}
To obtain \cref{eq:niche_network_stats_expression} in the main text from \cref{sm:eq:niche_h_and_f}, we take the order $[2/2]$ Pad\'e approximant of $h^2$ around $d = 0$, which incurs a maximum relative error of approximately $0.004$ when $d = 1/2$, or approximately $5\%$ of the value of $h^2$.

On substituting $h^2$ and $f$ into \cref{sm:eq:niche_fs_params}, we have successfully found an explicit fine-structure correction for FSRMs constructed according to the niche model, in terms of only the model parameters $\mu_L, \mu_U, \sigma_L, \sigma_U, \Gamma$ and $d$. We compare our approximation $h^2$ and $f$ in \cref{fig:NM_adjacency_matrices}(b) and (c) in the main text. The fine-structure correction to the outlier eigenvalue is verified in \cref{fig:NM_spectrum}(b) and the correction to the bulk of the spectrum is verified in \cref{sm:fig:niche_varying_d}.
\end{widetext}

\end{document}